\documentclass[english,aip, numerical,jcp,reprint,twocolumn,floatfix,prd]{revtex4-1}
\usepackage{natbib,hyperref}
\usepackage{amsmath}
\usepackage{stmaryrd}
\usepackage{braket}
\usepackage{graphicx}
\usepackage{dcolumn}
\usepackage{tabularx}
\usepackage{bm}
\usepackage{epsfig,color,xspace,multirow,xr,bbold}
\usepackage[all]{xy}
\usepackage{setspace}
\usepackage{url}
\usepackage{threeparttable, tablefootnote}
 
\newcommand{\RRefs}[1]{\,Refs.\nocite{#1}\citenum{#1}}
\usepackage{natbib,hyperref}
\usepackage{longtable}
\usepackage{float}
\usepackage{placeins}
\usepackage[utf8]{inputenc}

\newcommand{\SGonefivesix}{{\it P}3{\it m}1\,}
\newcommand{\SGonesixzero}{{\it R}3{\it m}\,}
\newcommand{\SGonesixfour}{{\it P}$\overline{3}${\it m}1\,}
\newcommand{\SGonesixsix}{{\it R}$\overline{3}${\it m}\,}

\newcommand{\SGoneeightsix}{{\it P}$6_3${\it mc}\,}
\newcommand{\SGoneeightseven}{{\it P}$\overline{6}${\it m}$2$\,}
\newcommand{\SGoneninefour}{{\it P}$6_3$/{\it mmc}\,}

\newcommand{\SGtwoonesix}{{\it F}$\overline{4}$3{\it m}\,}
\newcommand{\SGtwotwoseven}{{\it Fd}$\overline{3}${\it m}\,}

\usepackage[none]{hyphenat}
\usepackage{calligra}
\usepackage{xtab,afterpage,longtable}

\begin{document}

\title[]{
Band gaps of long-period   
polytypes of IV, IV-IV, and III-V  
semiconductors estimated with an Ising-type additivity model
}
\author{Raghunathan Ramakrishnan}
\email{ramakrishnan@tifrh.res.in}
\affiliation{Tata Institute of Fundamental Research Hyderabad, Hyderabad 500046, India.}

\author{Shruti Jain}
\affiliation{Indian Institute of Science Education and Research Mohali,
Mohali 140306, Punjab, India.}
\affiliation{VSRP fellow at Tata Institute of Fundamental Research Hyderabad, Hyderabad 500046, India.}

\date{\today}

\begin{abstract}
We apply an Ising-type model to estimate the
band gaps of the polytypes of group IV elements (C, Si, and Ge)
and binary compounds of groups:
IV-IV (SiC, GeC, and GeSi), and III-V 
(nitride, phosphide, and arsenide of B, Al, and Ga). 
The models use reference band gaps 
of the simplest polytypes comprising 2--6 bilayers
calculated with the hybrid density functional approximation, HSE06. 
We report four models capable of
estimating band gaps of  nine polytypes containing 7 and 8 bilayers with 
an average error of $\lesssim0.05$ eV. We apply the best model
with an error of $<0.04$ eV
to predict 
the band gaps of 497 polytypes with up to 15 bilayers in 
the unit cell, providing a comprehensive view of the variation
in the electronic structure with the degree of hexagonality 
of the crystal structure. Within our enumeration, we identify 
four rhombohedral polytypes of SiC---9$R$, 12$R$, 15$R$(1), and 15$R$(2)---and perform
detailed stability and band structure analysis. 
Of these, 15$R$(1) that has not been experimentally characterized has the widest band gap ($>3.4$ eV); phonon analysis and cohesive energy reveal  15$R$(1)-SiC to be metastable. Additionally, we model the energies of valence and conduction bands of the rhombohedral SiC phases at the high-symmetry points of the Brillouin zone and predict band structure characteristics around the Fermi level.
The models presented in this study may aid in identifying 
polytypic phases suitable for various applications, such as the design of wide-gap materials, that are relevant to high-voltage applications. In particular, the method holds promise for forecasting electronic properties of long-period and ultra-long-period polytypes for which accurate first-principles modeling is computationally challenging.
\end{abstract}


\maketitle

\section{Introduction}\label{sec_intro}
The fundamental gap (or the transport gap) of a material 
is the difference between its ionization energy and electron 
affinity: $E_g={\rm IP}-{\rm EA}$. For an $N$-electron system, $E_g$ can be formally expressed as $E_g = E\left( N + 1 \right) + E\left( N - 1 \right) - 2 E\left( N  \right)$, where $E$ is the total energy. Within the Kohn--Sham (KS) formalism of density functional theory (DFT), $E_g$ is approximated by the KS gap, which is the difference between the energy of the 
conduction band minimum (CBM) and that of the valence band maximum (VBM): 
$\varepsilon_g^{\rm KS}=\varepsilon_{\rm CMB}-\varepsilon_{\rm VBM}$\cite{perdew2017understanding}. The difference $E_g-\varepsilon_g^{\rm KS}=\Delta_{\rm XC}$  is called the derivative discontinuity 
of the exchange-correlation (XC) contribution to the total energy\cite{perdew1983physical}. Typically,
$\Delta_{\rm XC} >0$ for semi-local KS-DFT methods---local-density approximation (LDA), generalized gradient approximation (GGA), and meta-generalized gradient approximation (mGGA)---resulting in the underestimation of a material's fundamental gap\cite{perdew2018density}. As stated by John P. Perdew {\it et al.}, even with the exact KS band structure, $\varepsilon_g$ will underestimate $E_g$,\cite{perdew2010fourteen} as the former approximates the first exciton energy, while the latter, the asymptotic limit of a Rydberg series of exciton energies\cite{perdew2009some}.
The Hartree--Fock method, with 100\% exact exchange, has been known to overestimate $E_g$, often predicting 
paramagnetic metals as antiferromagnetic insulators\cite{mizokawa1996electronic}. 
Hence, KS-DFT with a hybrid-XC treatment\cite{heyd2003hybrid,krukau2006influence}, 
which includes a fraction of the exact exchange energy,
offers a more realistic description of $E_g$ due to favorable error cancellations\cite{perdew1996rationale}.

\begin{table*}[ht]
\centering
\caption{Definitions of small polytypes
considered in this study: Stacking sequence
in close-packing, $hk$, and H\"agg notations are given. Polytype names (in Ramsdell notation) and space group
details are given separately for compounds and elements. Names of binary compounds are used 
in the text for all systems. Also given is a crystal structure's degree of
hexagonality (\%hex). 
For 3$C$, space group details for a non-primitive rhombohedral
unit cell in the hexagonal lattice are also
given.}
\small\addtolength{\tabcolsep}{1.2pt}
\begin{tabular}[t]{llll lll lll l}
\hline
\multicolumn{4}{l}{Stacking sequence} & 
\multicolumn{3}{l}{Binary compounds} &
\multicolumn{3}{l}{Elements} &
\multicolumn{1}{l}{\%hex}\\
\cline{1-3} \cline{5-6} \cline{8-9}   
\multicolumn{1}{l}{Close-packing} & 
\multicolumn{1}{l}{$hk$} & 
\multicolumn{2}{l}{H\"agg} & 
\multicolumn{1}{l}{Name} &
\multicolumn{2}{l}{Space group} &
\multicolumn{1}{l}{Name} &
\multicolumn{2}{l}{Space group} &
\multicolumn{1}{l}{}\\
\hline 
 {\it AB}         & {\it hh        }& $+-$        &&  2$H$       &  \SGoneeightsix (186) &&   2$H$       & \SGoneninefour (194)   && 100 \\
 {\it ABC}        & {\it kkk       }& $+++$       &&  3$C$ / 3$R$       &  \SGtwoonesix  (216) / \SGonesixsix  (166)   &&   3$C$ /3$R$       & \SGtwotwoseven (227) /\SGonesixzero  (160)   &&  0\\
 {\it ABAC}       & {\it hkhk      }& $+--+$      &&  4$H$       &  \SGoneeightsix (186) &&   4$H$       & \SGoneninefour (194)   && 50 \\
 {\it ABABC}      & {\it hhkkk     }& $+-+++$     &&  5$T$       &  \SGonefivesix  (156) &&   5$T$       & \SGonesixfour (164)    && 40 \\
 {\it ABABAC}     & {\it hhhkhk    }& $+-+--+$    &&  6$T$       &  \SGonefivesix  (156) &&   6$H$(1)    & \SGoneeightseven (187) && 66.67\\
 {\it ABACBC }    & {\it hkkhkk    }& $+---++$    &&  6$H$       &  \SGoneeightsix (186) &&   6$H$(2)    & \SGoneninefour (194)   && 33.33\\
 {\it ABABABC }   & {\it hhhhkkk   }& $+-+-+++$   &&  7$T$(1)    &  \SGonefivesix  (156) &&   7$T$(1)    & \SGonesixfour (164)    && 57.14\\
 {\it ABABCAC }   & {\it hhkhhkk   }& $+-+++-+$   &&  7$T$(2)    &  \SGonefivesix  (156) &&   7$T$(2)    & \SGonesixfour (164)    && 57.14 \\
 {\it ABACABC }   & {\it hkhkkkk   }& $+--++++$   &&  7$T$(3)    &  \SGonefivesix  (156) &&   7$T$(3)    & \SGonesixfour (164)    && 28.57\\
 {\it ABABABAC}   & {\it hhhhhkhk  }& $+-+-+--+$  &&  8$T$(1)    &  \SGonefivesix  (156) &&   8$H$(1)    & \SGoneeightseven (187) && 75 \\
 {\it ABABACAC }  & {\it hhhkhhhk  }& $+-+--+-+$  &&  8$H$(1)    &  \SGoneeightsix (186) &&   8$H$(2)    & \SGoneninefour (194)   && 75\\
 {\it ABABACBC }  & {\it hhhkkhkk  }& $+-+---++$  &&  8$T$(2)    &  \SGonefivesix  (156) &&   8$H$(3)    & \SGoneeightseven (187) && 50\\
 {\it ABABCABC}   & {\it hhkkkkkk  }& $+-++---+$  &&  8$T$(3)    &  \SGonefivesix  (156) &&   8$T$(1)    & \SGonesixfour (164)    && 25\\
 {\it ABABCBAC}   & {\it hhkhkkhk  }& $+-++++++$  &&  8$T$(4)    &  \SGonefivesix  (156) &&   8$T$(2)    & \SGonesixfour (164)    && 50 \\
 {\it ABACBABC}   & {\it hkkkhkkk  }& $+----+++$  &&  8$H$(2)    &  \SGoneeightsix (186) &&   8$H$(4)    & \SGoneninefour (194)   && 25\\
\hline
\end{tabular}
\label{tab:one}
\end{table*}%
\begin{figure*}[!hbtp]
    \centering
    \includegraphics[width=\linewidth]{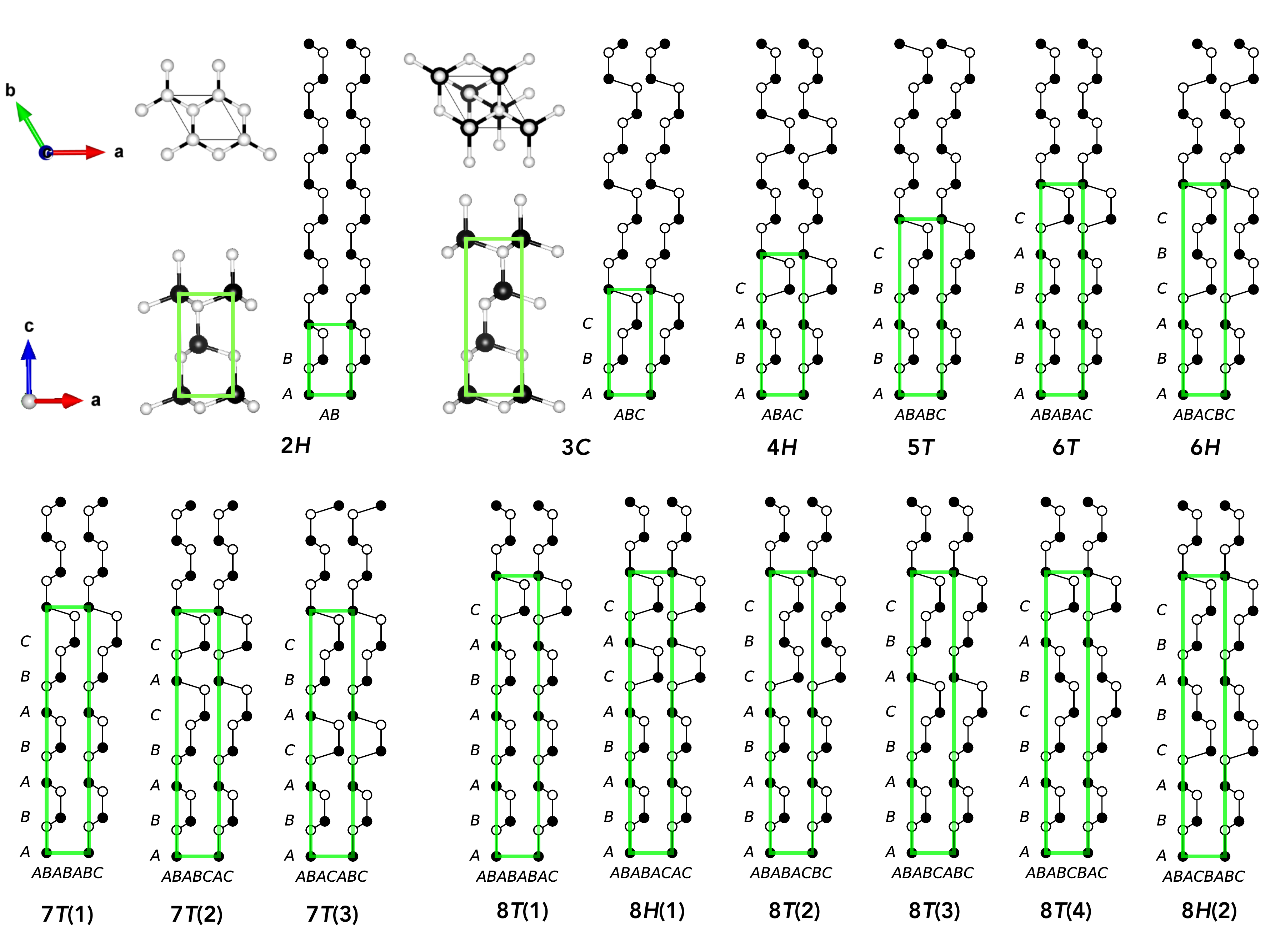}
    \caption{
    Stacking arrangement of polytypes with up to 8 bilayers
    as viewed from the [10$\overline{1}$0] plane
    of the hexagonal unit cell. 
    For 2$H$ and 3$C$, side and top views of a ball-and-stick representation are shown for comparison. 
    For binary compounds, white and black circles correspond to different atom types, while for elements, they correspond to the same atom type. The unit cell is shown in a green rectangle.
    }
    \label{fig:one}
\end{figure*}

Range-separated hybrid XC functionals are highly effective in accurately predicting the experimental band gaps of semiconductors, with minimal systematic errors, but they are not suitable for modeling metals\cite{perdew2018density}. Furthermore, applying hybrid XC methods for large unit cell systems becomes impractical due to their high computational cost. As a result, alternative surrogate methods 
incorporating empirical parameters to estimate hybrid DFT-level band gaps are finding increased applications. The $\Delta-$sol method estimates $E_g$ as the finite derivative of the energy ($E$) with respect to an empirically determined fraction of valence electrons in the unit cell\cite{chan2010efficient}. A popular empirical approach is the DFT+U method, where a Hubbard-type energy correction term improves the accuracy of the band gap ($\varepsilon_g$), particularly for strongly localized bands\cite{verma2016does,ramakrishnan2009dftpu,kulik2015perspective}. Morales {\it et al.}\cite{morales2017empirical} employed linear regression using reference results from the many-body perturbation theory (GW) for a representative ``training set'' to enhance the accuracy of GGA level $\varepsilon_g$ for a ``test set'' of semiconductors. 
More rigorous methods based on machine learning (ML) use
GGA-level $\varepsilon_g$ as input features to 
predict GW-level results\cite{lee2016prediction}. For the same task, 
 non-linear regressions based on neural networks
offer better performances\cite{na2020tuplewise}. ML models can also predict hybrid-DFT level $\varepsilon_g$ using the GGA-level electron density as a feature\cite{moreno2021machine,lentz2020predicting}. Models have also been trained on the difference between $\varepsilon_g$ from hybrid-DFT and GGA methods\cite{zhang2022accurate} using the $\Delta-$ML approach\cite{ramakrishnan2015big}, which effectively provide non-systematic corrections to GGA with increased training. ML models of $\varepsilon_g$ show high accuracy when trained on narrow property ranges\cite{borlido2020exchange}, indicating
that for a given composition, electronic properties such as $\varepsilon_g$ exhibit slight variations from a mean value with gradual changes of configuration variables in the materials space. Consequently, empirical models of $\varepsilon_g$ designed for specific classes of materials hold significant potential for various applications.

In this study, we explore the application of an additivity model
inspired by the cluster expansion of the Ising model\cite{wu2016cluster}. 
This approach can very accurately determine the energies 
of compounds within a consistent compound space\cite{laks1992efficient}. The Ising model exploits near-sightedness in chemical interactions and is very efficient for modeling the energetics of spatially modulated phases\cite{elliott1961phenomenological}. 
Individual interacting units can be atoms or molecules in a lattice or
sub-units of the bulk crystal, such as layers. Specifically, when the model focuses on phases that exhibit modulation of charge density
along an axis perpendicular to a layered structure, it is referred to 
as the axial next-nearest-neighbor-interaction (ANNNI) model or the ANNNI-Ising model\cite{selke1988annni}, referred hereafter as the Ising-type model. 

Using reference energies of small unit cell phases calculated using first-principles methods, the Ising-type model has demonstrated high accuracy in predicting the relative energies of larger unit cell phases, 
achieving less than milli-electron volt (meV) accuracy \cite{limpijumnong1998total,panse2011polytypism,raffy2002properties,moriguchi2021annni,keller2023first,kobayashi2012first,moriguchi2013comparative,bechstedt2013structure}. This approach aids in understanding the phase diagram or the relative thermodynamic stabilities of the polytypes of layered compounds\cite{smith1984new,trigunayat1991survey}, particularly silicon carbide (SiC)\cite{heine1985computation,cheng1989theory,cheng1990silicon}. 
These phenomenological models do not account for the dynamic stability of a long-period polytype 
based on phonon band structure diagnostics\cite{kayastha2021high,pallikara2022physical}.

In our study, we aim to assess the applicability of the Ising {\it Ansatz} as a mathematical representation for modeling $\varepsilon_g$
across the polytypes of group IV elements (C, Si, and Ge), 
IV-IV (SiC, GeC, and GeSi), and III-V binary compounds (BN, BP, BAs, AlN, AlP, AlAs, GN, GaP, GaAs). The model is applied for each composition to estimate the band gaps of 497 polytypes containing up to 15 bilayers in the unit cell. Furthermore, we analyze the trends in band gaps with the hexagonality of the crystal structure and conduct a more comprehensive analysis focusing on the rhombohedral phases of SiC more commonly encountered in experimental studies. Finally, we explore the scope to develop models of the energies of valence and conduction bands at the high-symmetry points in the Brillouin zone to rapidly generate the band structure characteristics of polytypes around the Fermi level.

\section{Theory}\label{subsec:C}
Polytypism is a particular case of polymorphism where structural variations arise from the stacking sequence of layered units along an axis perpendicular to the layers. 
Three-dimensional structural information, such as lattice parameters and Wyckoff positions, of many polytypes are well documented\cite{rohrer2001structure,tilley2020crystals,bechstedt2013structure,halac2002static}. However, for a complete enumeration of structures, 
an algorithmic approach to cover the entire compound space spanned by polytypes is necessary. P{\'o}lya enumeration is a powerful algebraic technique used to derive closed-form expressions for the number of equivalent patterns, eliminating the need for explicit structure construction\cite{polya2012combinatorial}. This method has been successfully employed for various problems, including the generation of a comprehensive catalogue of heteroatom-substituted benzenoid compounds \cite{chakraborty2019chemical} and other compounds \cite{douglas2012symmetry,muller2013symmetry}. However, explicit construction and de-duplication are necessary to identify unique structures when dealing with three-dimensional structures. For rapid enumeration using combinatorial techniques, structures can also be represented as string-based fingerprints.

A previous study\cite{mclarnan1981numbers} has enumerated distinct close-packings and polytypes. To generate the corresponding sequence\cite{lothaire1997combinatorics}, Jagodzinski's $hk$-notation is employed \cite{guinier1984nomenclature}. In this notation, a letter $h$ denotes layers sandwiched between similar layers, as in hexagonal close-packing, while $k$ denotes layers sandwiched between different neighbors, representing a local cubic structure. From the $hk$ sequence, we obtained the close-packing sequence and also
calculated the percentage of the locally hexagonal layers,  
$\%{\rm hex}=100 \times \#h/\left( \#h + \#k \right)$, where \# indicates the count. For the simplest 15 polytypes of elements and binary compounds containing up to 8 bilayers ({\it i.e.} formula units), Table~\ref{tab:one} provides the stacking sequence in various notations, space group details, and \%hex. FIG.~\ref{fig:one} presents the schematic representations of the corresponding stacking arrangements.

In the case of binary semiconductors, where the formula unit comprises two different types of atoms, the corresponding space group becomes a ``Klassengleiche'' subgroup of the elemental phase's space group (see Table~\ref{tab:one}). 
The polytypes are named using the Ramsdell notation\cite{ramsdell1947studies}, denoted as $pY(z)$. Here, $p$ represents the number of bilayers in the unit cell, and $Y$ signifies the crystal class ($C$: cubic, $H$: hexagonal, $T$: trigonal, or $R$: rhombohedral). The number $z$ in parentheses indicates the index. In this study, we arranged the polytypes in lexicographic order based on their close-packing sequence. For instance, the 7$H$ polytypes, namely $ABABABC$, $ABABCAC$, and $ABACABC$, are assigned indices 1, 2, and 3, respectively (see Table~\ref{tab:one}). The
3$C$ polytype has the cubic zincblende structure and 2$H$ 
polytype corresponds to the hexagonal wurtzite structure. 
We provide the details of long-period polytypes with many bilayers at the relevant places in the text.

\begin{figure}[!hbtp]
    \centering
    \includegraphics[width=\linewidth]{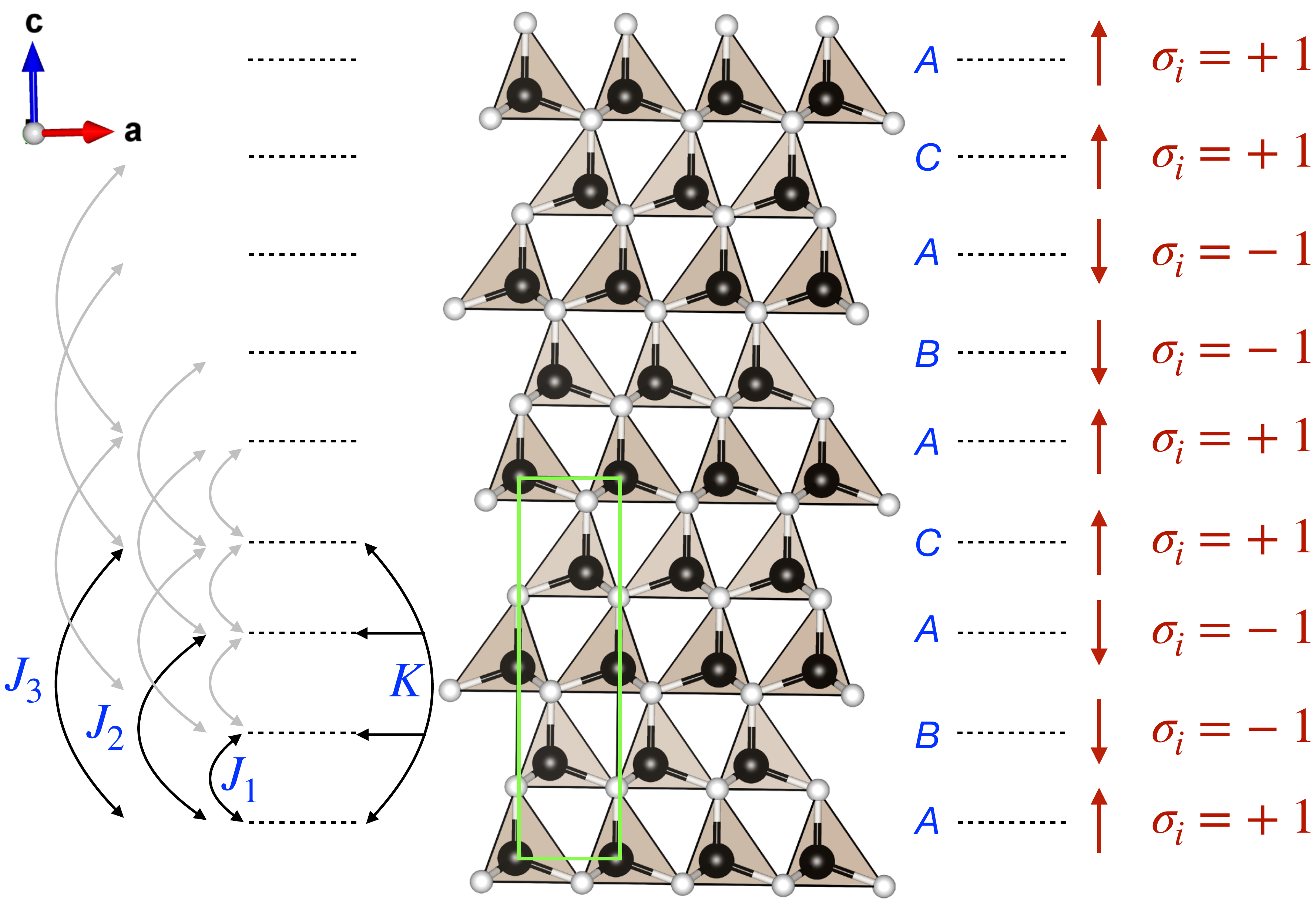}
    \caption{
    ABAC stacking sequence of 4$H$ polytype mapped to an arrangement 
    of spins.
    Each row of pyramids corresponds to a bilayer in the ${\bf a}$-${\bf b}$
    plane. The pseudo-spin value of a bilayer is 
    $\sigma_i=+1$ (symbolically mapped to the
    `up' spin) when 
    a layer and the one above follows the order 
    $A\rightarrow B$, 
    $B\rightarrow C$, or 
    $C\rightarrow A$. 
    When the sequence is reversed as
    $B\rightarrow A$, 
    $C\rightarrow B$, or 
    $A\rightarrow C$, the
    pseudo-spin value of a bilayer is $\sigma_i=-1$ ({\it i.e.} `down' spin). 
    First, second and third-neighbor
    2-body and 4-body interactions are shown. 
    }
    \label{fig:pseudospoin}
\end{figure}

For Ising-type models, we require a numerical representation that forms a
bijective mapping to the close-packing sequence. The H\"agg representation\cite{guinier1984nomenclature,guinier1984nomenclature,ortiz2013prolific} achieves this, where
a layer (or bilayer when the formula unit has two atoms) 
is assigned a {\it pseudo-spin} value of $\sigma_i=+1$ when the layer above is in the order $A\rightarrow B$, $B\rightarrow C$, or 
$C\rightarrow A$. When the sequence moves as $B\rightarrow A$,
$C\rightarrow B$, or $A\rightarrow C$, the assignment is $\sigma_i=-1$. As a shorthand notation, only the signs of $\sigma_i$ are given in TABLE~\ref{tab:one}. The mapping of a bilayer's local structure to a pseudospin value is shown for the 4$H$ polytype in FIG.~\ref{fig:pseudospoin}. The simplest model studied here contains four parameters:
\begin{eqnarray}
\varepsilon_g & = & J_0 -
J_1  \langle \sigma_i \sigma_{i+1} \rangle  -
J_2  \langle  \sigma_i \sigma_{i+2}  \rangle - J_3 \langle  \sigma_i \sigma_{i+3}  \rangle
\label{eq:annni1}
\end{eqnarray}
While there are no formal constraints on increasing the
number of interaction terms in Eq.~(\ref{eq:annni1}), we aim to develop
a simple model suitable for a rapid analysis of its performance. 
The first coefficient, $J_0$, corresponds to an intercept of a hyperplane in the model space. When the inter-layer interactions are negligible in the polytypes, the model will predict $\varepsilon_g$ to be $J_0$ for all polytypes. The `2-body' interaction terms $\langle \sigma_i \sigma_{i+1} \rangle$, $\langle \sigma_i \sigma_{i+2} \rangle$, and $\langle \sigma_i \sigma_{i+3} \rangle$ correspond to the interaction of layer-$i$ with
the first, second, and third layers, respectively. 

The notation $\langle x \rangle$ implies averaging over all layers in the unit cell, $(1/p)\sum_{i=1}^{p}x_i$, for a given polytype comprising
$p$ bilayers. For example, for the 4$H$ polytype (with four bilayers, $p=4$), the first neighbor interaction term is evaluated as 
\begin{eqnarray}
    \langle \sigma_i \sigma_{i+1} \rangle = 
    \frac{1}{4} 
    \left[ 
    \sigma_1 \sigma_{2} +
    \sigma_2 \sigma_{3}  + 
    \sigma_3 \sigma_{4} + 
    \sigma_4 \sigma_{5} 
    \right]. 
    \label{eq:firstnn}
\end{eqnarray}
For the 4$H$ polytype, $i=5$ corresponds to $i=1$ by periodicity. Inserting the pseudo-spin values shown in  FIG.~\ref{fig:pseudospoin} in 
Eq.~(\ref{eq:firstnn}), we find
\begin{eqnarray}
    \frac{1}{4} 
    \left[ 
    (1)(-1) +
    (-1)(-1) + 
    (-1)(1) + 
    (1)(1) 
    \right] = 0.
\end{eqnarray}

For SiC, the unknown coefficients $J_{0-3}$ were obtained by a linear regression using the HSE06 band gaps of 2$H$, 3$C$, 4$H$, and 6$H$ polytypes. We denote the resulting model as Model-1.  Since we aim to identify structure-property correlations across the compound space of polytypes, we select the reference systems for determining the models' parameters without any experimental bias. To this end, we evaluate the model's performance by applying it to nine long-period polytypes: 7$H$(1-3), 8$H$(1-2), and 8$T$(1-4) listed in TABLE~\ref{tab:one}. We also explore replacing 6$H$ with 6$T$, resulting in Model-2. We note in passing that using the 5$T$ polytype in a 4-parameter model with only 2-body terms along with 2$H$, 3$C$, and 4$H$ resulted in linear dependency because the 2-body interaction
terms of 3$C$ are proportional to that of 5$T$ (see TABLE~\ref{tab:two}). 

We have also tested the role of a `4-body' term  
$\langle  \sigma_i \sigma_{i+1} \sigma_{i+2} \sigma_{i+3}  \rangle$. As previously discussed\cite{cheng1988inter}, for polytypes of 
elements and binary compounds, odd terms such as the 3-body terms vanish due to permutation symmetry. However, these terms play a role in modeling the relative energies of polytypes of other compounds\cite{price1984application,plumer1988ising}. We augment both Model-1 and Model-2 with a 4-body term by including the 5$T$ polytype to arrive at Model-3 and Model-4 containing five unknown parameters:
\begin{eqnarray}
\varepsilon_g & = & J_0 -
J_1  \langle \sigma_i \sigma_{i+1} \rangle  -
J_2  \langle  \sigma_i \sigma_{i+2}  \rangle - J_3 \langle  \sigma_i \sigma_{i+3}  \rangle \nonumber \\
& &  - K \langle  \sigma_i \sigma_{i+1} \sigma_{i+2} \sigma_{i+3}  \rangle
\label{eq:annni2}
\end{eqnarray}

Table~\ref{tab:two} provides numerical values of the 
interaction terms for several polytypes. These terms are property-independent; hence, for SiC, we have applied the same interaction terms to determine $J_{0-3}$ and $K$ for modeling the energies of VB and CB at the high-symmetry points of the Brillouin zone. In this case, the model parameters will be implicit functions of these $ k$ points.

\begin{table}[ht]
\centering
\caption{Terms in the model of $\varepsilon_g$ given in Eq.~(\ref{eq:annni1})
and Eq.~(\ref{eq:annni2})
for various polytypes. 
 }
\small\addtolength{\tabcolsep}{1.2pt}
\begin{tabular}[t]{l c c c c}
\hline 
\multicolumn{1}{l}{Polytype} & 
\multicolumn{1}{l}{$\langle \sigma_i\sigma_{i+1} \rangle$} & 
\multicolumn{1}{l}{$\langle \sigma_i\sigma_{i+2} \rangle$} & 
\multicolumn{1}{l}{$\langle \sigma_i\sigma_{i+3} \rangle$} & 
\multicolumn{1}{l}{$\langle \sigma_i\sigma_{i+1}\sigma_{i+2}\sigma_{i+3} \rangle$} \\
\hline 
2$H$     & -1    &   1   & -1    &   1\\
3$C$     & 1     &   1   &   1   &   1\\
4$H$     & 0     &  -1   &  0    &   1\\
5$T$     & 1/5   &   1/5 &   1/5 & -3/5\\
6$T$     & -1/3  & -1/3  &   1/3 &   1/3\\
6$H$     & 1/3   & -1/3  & -1    & -1/3\\
7$H$(1)  & -1/7  &  3/7 & -1/7  & -1/7\\
7$H$(2)  & -1/7  & -1/7  &   3/7 & -5/7\\
7$H$(3)  & 3/7   & -1/7  & -1/7  &   3/7\\
8$H$(1)  & -1/2  &  0    &  0    &   1/2\\
8$T$(1)  & -1/2  &  0    &   1/2 &  0\\
8$T$(2)  & 0     &  0    & -1/2  & -1/2\\
8$T$(3)  & 1/2   &   1/2 &   1/2 &  0\\
8$T$(4)  & 0     & -1/2  &  0    &  0\\
8$H$(2)  & 1/2   &  0    & -1/2  &  0\\
9$R$     & -1/3  & -1/3  &   1   & -1/3\\
12$R$    & 0     &  0    &  0    & -1\\
15$R$(1) & -3/5  &   1/5 &   1/5 &   1/5\\
15$R$(2) & 1/5   & -3/5  & -3/5  &   1/5\\
\hline
\end{tabular}
\label{tab:two}
\end{table}%

\section{Results and Discussions}

\subsection{Structural and electronic properties}\label{ref:IIIA}
GGA-PBE minimum energy crystal structures of the simplest 
15 polytypes of all semiconductors studied in this work
are given in Table~S1 and Table~S2 in the 
Supporting Information (SI). 
For all structures, the $x$ and $y$ fractional coordinates ({\it i.e.} $x$ and $y$ Wyckoff coordinates) remain in their ideal positions, as in the corresponding composition's wurtzite phase. In our investigation of 225 systems, encompassing 15 polytypes and 15 compositions, we observed minimal deviations from the ideal positions ($u_i$ and $v_i$ stated in \ref{subsec:A}), measuring less than $10^{-3}$ \AA{}. It is worth noting that such minor deviations have proven to be significant for accurately refining experimental crystal structures \cite{gomes1967refinement,bauer1998high}. However, given that these deviations are 2 or 3 orders of magnitude smaller than the uncertainty in the GGA-PBE lattice constants compared to experimental ones, we have not focused on $u_i$ and $v_i$. 
\begin{table*}[!htpb]
\centering
\caption{
Band gaps (in eV) of 15 polytypes of semiconductors calculated with
PBE, SCAN, and HSE06 DFT methods using PBE equilibrium crystal structure. 
The band gap character is denoted as $d$ for direct and $i$ for indirect.
}
\small\addtolength{\tabcolsep}{1.2pt}
\begin{tabular}[t]{ll ll ll ll l lll lll ll}
\hline
\multicolumn{1}{l}{Sys.}   &
\multicolumn{1}{l}{XC}   &
\multicolumn{1}{l}{2$H$} & 
\multicolumn{1}{l}{3$C$} & 
\multicolumn{1}{l}{4$H$} & 
\multicolumn{1}{l}{5$T$} & 
\multicolumn{1}{l}{6$T$} & 
\multicolumn{1}{l}{6$H$} &
\multicolumn{1}{l}{7$T$(1) } & 
\multicolumn{1}{l}{7$T$(2) } & 
\multicolumn{1}{l}{7$T$(3) } & 
\multicolumn{1}{l}{ 8$T$(1)} & 
\multicolumn{1}{l}{ 8$H$(1)} & 
\multicolumn{1}{l}{ 8$T$(2)} & 
\multicolumn{1}{l}{ 8$T$(3)} & 
\multicolumn{1}{l}{ 8$T$(4)} & 
\multicolumn{1}{l}{ 8$H$(2)} \\
\hline 
C&PBE&     3.58  $i$&  4.18  $i$&  4.58  $i$&  4.35  $i$&  4.51  $i$&  4.42  $i$ &  4.29  $i$&  4.45  $i$&  4.37  $i$&  4.13  $i$&  4.42  $i$&  4.47  $i$&  4.24  $i$&  4.46  $i$&  4.39  $i$\\
&SCAN&     4.00  $i$&  4.59  $i$&  4.99  $i$&  4.75  $i$&  4.95  $i$&  4.82  $i$ &  4.73  $i$&  4.85  $i$&  4.78  $i$&  4.57  $i$&  4.87  $i$&  4.87  $i$&  4.64  $i$&  4.86  $i$&  4.79  $i$\\
&HSE06&    4.74  $i$&  5.41  $i$&  5.82  $i$&  5.59  $i$&  5.73  $i$&  5.66  $i$ &  5.49  $i$&  5.69  $i$&  5.61  $i$&  5.33  $i$&  5.64  $i$&  5.71  $i$&  5.48  $i$&  5.70  $i$&  5.62  $i$\\
Si&PBE&    0.48  $i$&  0.66  $i$&  0.63  $i$&  0.61  $i$&  0.56  $i$&  0.64  $i$ &  0.57  $i$&  0.58  $i$&  0.67  $i$&  0.54  $i$&  0.54  $i$&  0.60  $i$&  0.64  $i$&  0.60  $i$&  0.68  $i$\\
&SCAN&     0.72  $i$&  0.90  $i$&  0.87  $i$&  0.85  $i$&  0.80  $i$&  0.88  $i$ &  0.81  $i$&  0.82  $i$&  0.91  $i$&  0.77  $i$&  0.78  $i$&  0.83  $i$&  0.88  $i$&  0.84  $i$&  0.92  $i$\\
&HSE06&    1.05  $i$&  1.24  $i$&  1.21  $i$&  1.19  $i$&  1.14  $i$&  1.22  $i$ &  1.15  $i$&  1.15  $i$&  1.26  $i$&  1.11  $i$&  1.12  $i$&  1.17  $i$&  1.22  $i$&  1.18  $i$&  1.26  $i$\\
Ge&PBE&    0.06  $i$&  0.03  $i$&  0.08  $i$&  0.03  $i$&  0.05  $i$&  0.06  $i$ &  0.04  $i$&  0.07  $i$&  0.08  $i$&  0.05  $i$&  0.02  $i$&  0.01  $i$&  0.02  $i$&  0.01  $i$&  0.03  $i$\\
&SCAN&     0.13  $d$&  0.04  $d$&  0.12  $d$&  0.04  $d$&  0.07  $d$&  0.06  $d$ &  0.08  $d$&  0.03  $d$&  0.06  $d$&  0.04  $d$&  0.08  $d$&  0.06  $d$&  0.03  $d$&  0.02  $d$&  0.00  $d$\\
&HSE06&    0.12  $d$&  0.15  $d$&  0.18  $d$&  0.17  $d$&  0.16  $d$&  0.18  $d$ &  0.15  $d$&  0.16  $d$&  0.18  $d$&  0.15  $d$&  0.15  $d$&  0.16  $d$&  0.18  $d$&  0.17  $d$&  0.19  $d$\\
SiC&PBE&   2.33  $i$&  1.54  $i$&  2.35  $i$&  1.87  $i$&  2.38  $i$&  2.13  $i$ &  1.90  $i$&  2.10  $i$&  1.81  $i$&  2.42  $i$&  2.33  $i$&  2.06  $i$&  1.65  $i$&  2.14  $i$&  1.90  $i$\\
&SCAN&     2.70  $i$&  1.88  $i$&  2.70  $i$&  2.22  $i$&  2.75  $i$&  2.48  $i$ &  2.25  $i$&  2.46  $i$&  2.15  $i$&  2.80  $i$&  2.70  $i$&  2.41  $i$&  2.00  $i$&  2.50  $i$&  2.25  $i$\\
&HSE06&    3.29  $i$&  2.50  $i$&  3.34  $i$&  2.84  $i$&  3.38  $i$&  3.10  $i$ &  2.87  $i$&  3.09  $i$&  2.77  $i$&  3.42  $i$&  3.33  $i$&  3.04  $i$&  2.62  $i$&  3.12  $i$&  2.87  $i$\\
GeC&PBE&   2.42  $i$&  1.80  $i$&  2.16  $i$&  1.96  $i$&  2.18  $i$&  2.13  $i$ &  2.00  $i$&  2.09  $i$&  1.95  $i$&  2.22  $i$&  2.17  $i$&  2.08  $i$&  1.88  $i$&  2.13  $i$&  2.02  $i$\\
&SCAN&     2.66  $i$&  2.03  $i$&  2.40  $i$&  2.20  $i$&  2.41  $i$&  2.37  $i$ &  2.24  $i$&  2.33  $i$&  2.19  $i$&  2.46  $i$&  2.40  $i$&  2.31  $i$&  2.12  $i$&  2.37  $i$&  2.25  $i$\\
&HSE06&    3.31  $i$&  2.63  $i$&  3.03  $i$&  2.81  $i$&  3.05  $i$&  3.00  $i$ &  2.85  $i$&  2.95  $i$&  2.80  $i$&  3.10  $i$&  3.04  $i$&  2.93  $i$&  2.72  $i$&  2.99  $i$&  2.87  $i$\\
GeSi&PBE&  0.40  $i$&  0.66  $i$&  0.59  $i$&  0.56  $i$&  0.51  $i$&  0.58  $i$ &  0.50  $i$&  0.54  $i$&  0.61  $i$&  0.47  $i$&  0.48  $i$&  0.54  $i$&  0.61  $i$&  0.54  $i$&  0.66  $i$\\
&SCAN&     0.69  $i$&  0.88  $i$&  0.86  $i$&  0.78  $i$&  0.77  $i$&  0.80  $i$ &  0.76  $i$&  0.76  $i$&  0.84  $i$&  0.74  $i$&  0.76  $i$&  0.77  $i$&  0.83  $i$&  0.76  $i$&  0.89  $i$\\
&HSE06&    1.00  $i$&  1.20  $i$&  1.18  $i$&  1.10  $i$&  1.09  $i$&  1.12  $i$ &  1.07  $i$&  1.08  $i$&  1.16  $i$&  1.06  $i$&  1.08  $i$&  1.09  $i$&  1.15  $i$&  1.08  $i$&  1.21  $i$\\
BN&PBE&    5.39  $i$&  4.57  $i$&  5.58  $i$&  5.01  $i$&  5.67  $i$&  5.26  $i$ &  5.06  $i$&  5.30  $i$&  4.92  $i$&  5.70  $i$&  5.64  $i$&  5.25  $i$&  4.74  $i$&  5.34  $i$&  5.01  $i$\\
&SCAN&     5.93  $i$&  5.08  $i$&  6.10  $i$&  5.53  $i$&  6.20  $i$&  5.78  $i$ &  5.58  $i$&  5.82  $i$&  5.44  $i$&  6.23  $i$&  6.17  $i$&  5.77  $i$&  5.25  $i$&  5.87  $i$&  5.53  $i$\\
&HSE06&    6.81  $i$&  6.00  $i$&  7.05  $i$&  6.45  $i$&  7.14  $i$&  6.71  $i$ &  6.50  $i$&  6.75  $i$&  6.36  $i$&  7.17  $i$&  7.11  $i$&  6.70  $i$&  6.17  $i$&  6.80  $i$&  6.45  $i$\\
BP&PBE&    1.08  $i$&  1.37  $i$&  1.35  $i$&  1.33  $i$&  1.31  $i$&  1.37  $i$ &  1.30  $i$&  1.32  $i$&  1.36  $i$&  1.29  $i$&  1.29  $i$&  1.33  $i$&  1.30  $i$&  1.34  $i$&  1.37  $i$\\
&SCAN&     1.33  $i$&  1.65  $i$&  1.63  $i$&  1.61  $i$&  1.58  $i$&  1.64  $i$ &  1.58  $i$&  1.59  $i$&  1.63  $i$&  1.56  $i$&  1.57  $i$&  1.61  $i$&  1.58  $i$&  1.61  $i$&  1.65  $i$\\
&HSE06&    1.80  $i$&  2.15  $i$&  2.13  $i$&  2.10  $i$&  2.08  $i$&  2.14  $i$ &  2.07  $i$&  2.09  $i$&  2.13  $i$&  2.05  $i$&  2.06  $i$&  2.10  $i$&  2.07  $i$&  2.11  $i$&  2.15  $i$\\
BAs&PBE&   1.19  $i$&  1.31  $i$&  1.29  $i$&  1.26  $i$&  1.24  $i$&  1.29  $i$ &  1.26  $i$&  1.25  $i$&  1.29  $i$&  1.23  $i$&  1.23  $i$&  1.28  $i$&  1.26  $i$&  1.26  $i$&  1.32  $i$\\
&SCAN&     1.42  $i$&  1.56  $i$&  1.52  $i$&  1.49  $i$&  1.47  $i$&  1.52  $i$ &  1.49  $i$&  1.48  $i$&  1.53  $i$&  1.46  $i$&  1.46  $i$&  1.51  $i$&  1.49  $i$&  1.50  $i$&  1.55  $i$\\
&HSE06&    1.87  $i$&  2.00  $i$&  1.97  $i$&  1.94  $i$&  1.92  $i$&  1.97  $i$ &  1.94  $i$&  1.93  $i$&  1.98  $i$&  1.91  $i$&  1.91  $i$&  1.96  $i$&  1.94  $i$&  1.95  $i$&  2.00  $i$\\
AlN&PBE&   4.03  $i$&  3.45  $i$&  4.00  $i$&  3.80  $i$&  4.01  $i$&  3.91  $i$ &  3.79  $i$&  4.00  $i$&  3.69  $i$&  4.00  $i$&  4.01  $i$&  3.92  $i$&  3.56  $i$&  3.99  $i$&  3.75  $i$\\
&SCAN&     4.67  $i$&  4.07  $i$&  4.65  $i$&  4.43  $i$&  4.66  $i$&  4.54  $i$ &  4.41  $i$&  4.65  $i$&  4.31  $i$&  4.65  $i$&  4.66  $i$&  4.54  $i$&  4.18  $i$&  4.62  $i$&  4.37  $i$\\
&HSE06&    5.43  $i$&  4.74  $i$&  5.40  $i$&  5.11  $i$&  5.40  $i$&  5.22  $i$ &  5.09  $i$&  5.34  $i$&  4.99  $i$&  5.40  $i$&  5.41  $i$&  5.23  $i$&  4.86  $i$&  5.31  $i$&  5.05  $i$\\
AlP&PBE&   1.96  $i$&  1.63  $i$&  1.91  $i$&  1.80  $i$&  1.93  $i$&  1.90  $i$ &  1.79  $i$&  1.85  $i$&  1.74  $i$&  1.93  $i$&  1.88  $i$&  1.85  $i$&  1.69  $i$&  1.86  $i$&  1.80  $i$\\
&SCAN&     2.31  $i$&  1.95  $i$&  2.26  $i$&  2.14  $i$&  2.28  $i$&  2.24  $i$ &  2.13  $i$&  2.19  $i$&  2.07  $i$&  2.28  $i$&  2.23  $i$&  2.18  $i$&  2.02  $i$&  2.21  $i$&  2.13  $i$\\
&HSE06&    2.69  $i$&  2.35  $i$&  2.64  $i$&  2.53  $i$&  2.66  $i$&  2.63  $i$ &  2.52  $i$&  2.57  $i$&  2.46  $i$&  2.66  $i$&  2.61  $i$&  2.57  $i$&  2.41  $i$&  2.59  $i$&  2.52  $i$\\
AlAs&PBE&  1.69  $i$&  1.62  $i$&  1.68  $i$&  1.63  $i$&  1.67  $i$&  1.70  $i$ &  1.64  $i$&  1.68  $i$&  1.65  $i$&  1.69  $i$&  1.67  $i$&  1.67  $i$&  1.63  $i$&  1.69  $i$&  1.68  $i$\\
&SCAN&     2.10  $i$&  1.92  $i$&  2.01  $i$&  1.95  $i$&  2.00  $i$&  2.03  $i$ &  1.96  $i$&  2.01  $i$&  1.97  $i$&  2.02  $i$&  2.00  $i$&  1.99  $i$&  1.94  $i$&  2.02  $i$&  1.99  $i$\\
&HSE06&    2.42  $i$&  2.27  $i$&  2.34  $i$&  2.29  $i$&  2.33  $i$&  2.36  $i$ &  2.29  $i$&  2.34  $i$&  2.30  $i$&  2.35  $i$&  2.33  $i$&  2.32  $i$&  2.28  $i$&  2.34  $i$&  2.33  $i$\\
GaN&PBE&   1.72  $d$&  1.55  $d$&  1.64  $d$&  1.61  $d$&  1.66  $d$&  1.61  $d$ &  1.64  $d$&  1.65  $d$&  1.60  $d$&  1.67  $d$&  1.68  $d$&  1.63  $d$&  1.59  $d$&  1.63  $d$&  1.59  $d$\\
&SCAN&     2.04  $d$&  1.87  $d$&  1.95  $d$&  1.93  $d$&  1.98  $d$&  1.93  $d$ &  1.95  $d$&  1.96  $d$&  1.91  $d$&  1.99  $d$&  1.99  $d$&  1.95  $d$&  1.90  $d$&  1.95  $d$&  1.91  $d$\\
&HSE06&    2.92  $d$&  2.74  $d$&  2.83  $d$&  2.80  $d$&  2.85  $d$&  2.80  $d$ &  2.83  $d$&  2.84  $d$&  2.79  $d$&  2.87  $d$&  2.87  $d$&  2.82  $d$&  2.77  $d$&  2.82  $d$&  2.78  $d$\\
GaP&PBE&   1.28  $d$&  1.58  $d$&  1.43  $d$&  1.44  $d$&  1.37  $d$&  1.47  $d$ &  1.36  $d$&  1.39  $d$&  1.47  $d$&  1.34  $d$&  1.35  $d$&  1.39  $d$&  1.45  $d$&  1.41  $d$&  1.49  $d$\\
&SCAN&     1.59  $d$&  1.88  $d$&  1.74  $d$&  1.74  $d$&  1.68  $d$&  1.77  $d$ &  1.67  $d$&  1.70  $d$&  1.78  $d$&  1.64  $d$&  1.65  $d$&  1.70  $d$&  1.75  $d$&  1.72  $d$&  1.79  $d$\\
&HSE06&    2.04  $d$&  2.35  $d$&  2.19  $d$&  2.20  $d$&  2.13  $d$&  2.24  $d$ &  2.12  $d$&  2.15  $d$&  2.24  $d$&  2.09  $d$&  2.11  $d$&  2.15  $d$&  2.21  $d$&  2.17  $d$&  2.26  $d$\\
GaAs&PBE&  0.17  $d$&  0.15  $d$&  0.17  $d$&  0.16  $d$&  0.16  $d$&  0.16  $d$ &  0.16  $d$&  0.16  $d$&  0.16  $d$&  0.16  $d$&  0.16  $d$&  0.16  $d$&  0.15  $d$&  0.16  $d$&  0.16  $d$\\
&SCAN&     0.41  $d$&  0.40  $d$&  0.41  $d$&  0.40  $d$&  0.41  $d$&  0.41  $d$ &  0.40  $d$&  0.41  $d$&  0.41  $d$&  0.41  $d$&  0.41  $d$&  0.40  $d$&  0.40  $d$&  0.41  $d$&  0.41  $d$\\
&HSE06&    0.88  $d$&  0.87  $d$&  0.89  $d$&  0.88  $d$&  0.88  $d$&  0.88  $d$ &  0.88  $d$&  0.88  $d$&  0.88  $d$&  0.88  $d$&  0.88  $d$&  0.88  $d$&  0.88  $d$&  0.88  $d$&  0.88  $d$\\
\hline
\end{tabular}
\label{tab:sthree}
\end{table*}%

\begin{table*}[hpt]
\centering
\caption{Accuracy of band gap models based on polytypes
with up to 6 layers, when applied to larger unit cell
polytypes 7$T$(1-3), 8$H$(1-2), and 8$T$(1-4). For each composition,
mean absolute deviation (in eV) in model-predicted $\varepsilon_g$ with respect to
HSE06 values are given for four models. 
Standard deviations (in eV) are given in parentheses. For comparison, similar 
error metrics for 7 and 8 bilayer systems
are given for PBE and SCAN.
 }
\small\addtolength{\tabcolsep}{1.2pt}
\begin{tabular}[t]{l cc cc cc cc cc cc }
\hline
\multicolumn{2}{l}{Composition} &
\multicolumn{2}{l}{Model-1}&
\multicolumn{2}{l}{Model-2}&
\multicolumn{2}{l}{Model-3}&
\multicolumn{2}{l}{Model-4}&
\multicolumn{2}{l}{PBE} &
\multicolumn{1}{l}{SCAN} \\
  \hline      
  C       &&  0.143 (0.126) &&  0.184 (0.204) &&  0.075 (0.145)    &&  0.125 (0.218)     &&  1.230 (0.015) &&  0.812 (0.037)\\
  Si      &&  0.016 (0.025) &&  0.009 (0.015) &&  0.015 (0.024)    &&  0.009 (0.014)     &&  0.578 (0.004) &&  0.339 (0.002)\\
  Ge      &&  0.007 (0.007) &&  0.008 (0.009) &&  0.011 (0.013)    &&  0.007 (0.009)     &&  0.128 (0.026) &&  0.121 (0.037)\\
  SiC     &&  0.130 (0.249) &&  0.116 (0.229) &&  0.120 (0.193)    &&  0.092 (0.147)     &&  0.978 (0.011) &&  0.623 (0.003)\\
  GeC     &&  0.077 (0.161) &&  0.077 (0.163) &&  0.063 (0.094)    &&  0.043 (0.065)     &&  0.856 (0.011) &&  0.620 (0.011)\\
  GeSi    &&  0.048 (0.092) &&  0.040 (0.079) &&  0.020 (0.039)    &&  0.021 (0.036)     &&  0.558 (0.020) &&  0.318 (0.001)\\
  BN      &&  0.142 (0.241) &&  0.135 (0.226) &&  0.147 (0.261)    &&  0.123 (0.226)     &&  1.452 (0.014) &&  0.931 (0.009)\\
  BP      &&  0.047 (0.043) &&  0.045 (0.041) &&  0.031 (0.061)    &&  0.034 (0.063)     &&  0.769 (0.002) &&  0.493 (0.001)\\
  BAs     &&  0.017 (0.031) &&  0.013 (0.026) &&  0.013 (0.022)    &&  0.013 (0.022)     &&  0.683 (0.001) &&  0.450 (0.001)\\
  AlN     &&  0.063 (0.130) &&  0.054 (0.119) &&  0.068 (0.089)    &&  0.049 (0.065)     &&  1.331 (0.037) &&  0.701 (0.027)\\
  AlP     &&  0.038 (0.077) &&  0.038 (0.075) &&  0.038 (0.040)    &&  0.035 (0.035)     &&  0.727 (0.004) &&  0.387 (0.003)\\
  AlAs    &&  0.022 (0.046) &&  0.027 (0.053) &&  0.024 (0.042)    &&  0.017 (0.032)     &&  0.653 (0.003) &&  0.329 (0.005)\\
  GaN     &&  0.007 (0.014) &&  0.008 (0.015) &&  0.005 (0.006)    &&  0.004 (0.005)     &&  1.190 (0.002) &&  0.874 (0.001)\\
  GaP     &&  0.028 (0.058) &&  0.030 (0.062) &&  0.019 (0.019)    &&  0.017 (0.018)     &&  0.763 (0.003) &&  0.457 (0.004)\\
  GaAs    &&  0.003 (0.006) &&  0.003 (0.006) &&  0.002 (0.002)    &&  0.002 (0.003)     &&  0.722 (0.002) &&  0.476 (0.000)\\
  \hline 
  All     &&  0.053 (0.083) &&  0.052 (0.089) &&  0.043 (0.075)    &&  0.039 (0.070)     &&  0.841 (0.333) &&  0.529 (0.221)\\
\hline
\end{tabular}
\label{tab:four}
\end{table*}%

Band gaps of 15 polytypes of various semiconductors are collected in TABLE~\ref{tab:sthree}. In all cases, the GGA-PBE band gaps underestimate the hybrid-DFT-HSE06 counterparts, while
the mGGA-SCAN values fall in between. All the polytypes of Ge and Ga pnictides are found to be direct gap materials by all three DFT methods. For elemental Ge, HSE06 predicts all polytypes as direct gap semiconductors. Despite underestimating the gap, SCAN predicts the direct nature correctly, while PBE wrongly predicts them
as indirect gap semiconductors. 

The discussion then shifts to the accuracy of HSE06 $\varepsilon_g$ determined using GGA-PBE crystal structures. For the diamond phase of C and Si, the obtained results of 5.41 eV and 1.24 eV, respectively, are in agreement with experimental values (5.50 eV and 1.17 eV)\cite{yang2016more}  with a deviation of less than 0.1 eV. 
However, for 3$C$-Ge, the PBE lattice constant $a=5.768$ \AA{}
overestimates the experimental value 5.66 \AA{} resulting in a diminished
HSE06 band gap of 0.15 eV (see TABLE~\ref{tab:sthree}). 
We performed a separate lattice relaxation of 3$C$-Ge
with the HSE06 method and obtained $a=5.698$ \AA{} resulting in an
improved gap of 0.54 eV approaching the 
experimental value of 0.74 eV\cite{heyd2005energy}. 

In the case of SiC, specifically the 2$H$, 3$C$, 4$H$, 6$H$, and 8$H$(2) phases, the HSE06 band gaps 
determined in this study are 3.29/2.50/3.34/3.10/2.87 eV closely matching the experimental values 3.330/2.390/3.263/3.023/2.80 eV\cite{backes1994energy}. Additionally, for the 3$C$ polytype of BN,
BP, and BAs, the HSE06 band gaps are determined to be 6.00 eV, 2.15 eV, and 2.00 eV, respectively, which closely align with previously reported HSE values (5.98 eV, 2.16 eV, and 1.92 eV)\cite{heyd2005energy}. Both sets reasonably approximate the experimental values of 6.22 eV, 2.4 eV, and 1.46 eV, respectively\cite{zhuang2012electronic}.

Regarding the pnictides of Al and Ga, the HSE06 band gaps obtained using PBE crystal structure slightly underestimate the HSE06 band gaps obtained using HSE06 lattice parameters\cite{heyd2005energy}. For GaAs, the HSE06 band gap of the 3$C$ phase (0.87 eV) underestimates the experimental value of 1.43 eV\cite{kusch2012band}. Again, the discrepancy is attributed to the crystal structure used. GGA-PBE predicts the lattice constant of 3$C$-GaAs as 5.749 \AA{}(experimental value is 5.65 \AA), resulting in an HSE06 band gap of 0.87 eV. However, for the HSE06 lattice constant of 5.692 \AA{} determined through a separate calculation, 
the HSE06 band gap widens to 1.14 eV, 
 a value close to 1.21 eV at the HSE06 level reported previously\cite{zhao2009calculation}. 

The main objective of our study is to explore the polytypes of SiC for which HSE06 band gaps determined using PBE lattice constants show good agreement with experimental results, with a mean error of less than 0.1 eV. Additionally, several past studies have utilized GGA-PBE for investigating SiC polytypes. Therefore, our study chooses to proceed with the PBE-level crystal structure, and we do not investigate the impact of lattice constants on the band gaps for other materials. 
The HSE06 values of $\varepsilon_g$ determined using PBE equilibrium crystal structures of polytypes with 2--6 bilayers are used for developing the pseudo-spin models of $\varepsilon_g$. The performances of these models are then evaluated for 7- and 8-bilayer polytypes. Finally, the best model is applied to estimate the band gaps of 497 polytypes, spanning up to 15 bilayers across all 15 compositions.

\begin{table}[hpb]
\centering
\caption{Parameters involved in Eq.~\ref{eq:annni2} in the main text obtained 
by fitting to $\varepsilon_g$ of the polytypes: 2$H$, 3$C$, 4$H$,
5$T$, and 6$T$. All values are in eV. 
 }
\small\addtolength{\tabcolsep}{1.2pt}
\begin{tabular}[t]{l rr rr r}
\hline 
\multicolumn{1}{l}{Composition} & 
\multicolumn{1}{l}{$J_0$} & 
\multicolumn{1}{l}{$J_1$} & 
\multicolumn{1}{l}{$J_2$} & 
\multicolumn{1}{l}{$J_3$} &
\multicolumn{1}{l}{$K$}  \\
\hline 
C    & 5.543 & -0.021 &  0.372 & -0.312 &  0.094  \\
Si   & 1.179 & -0.124 &  0.031 &  0.029 &  0.000  \\
Ge   & 0.167 & -0.026 &  0.024 &  0.013 &  0.009  \\
SiC  & 3.022 &  0.575 &  0.222 & -0.179 & -0.096  \\
GeC  & 2.928 &  0.302 &  0.029 &  0.040 & -0.072  \\
GeSi & 1.104 & -0.113 &  0.038 &  0.016 & -0.035  \\
BN   & 6.647 &  0.743 &  0.322 & -0.339 & -0.078  \\
BP   & 2.071 & -0.105 &  0.075 & -0.068 &  0.020  \\
BAs  & 1.942 & -0.074 &  0.018 &  0.012 & -0.012  \\
AlN  & 5.223 &  0.356 &  0.158 & -0.012 & -0.020  \\
AlP  & 2.576 &  0.183 &  0.062 & -0.011 & -0.005  \\
AlAs & 2.315 &  0.042 & -0.001 &  0.033 & -0.024  \\
GaN  & 2.824 &  0.088 & -0.001 &  0.002 & -0.004  \\
GaP  & 2.180 & -0.160 &  0.001 &  0.004 & -0.014  \\
GaAs & 0.883 & -0.000 &  0.007 &  0.005 & -0.001  \\
\hline
\end{tabular}
\label{tab:five}
\end{table}%

\subsection{Performance of the band gap models}
The accuracies of the models of $\varepsilon_g$ are 
quantified using the error metrics: mean absolute deviation (MAD)
and standard deviation (SD). 
In Table~\ref{tab:four}, we have compiled the prediction errors for $\varepsilon_g$ in polytypes comprising 7 and 8 bilayers. 
Model-1 and Model-2, utilizing four parameters while lacking the four-body term, exhibit similar accuracies across different compositions. However, including the $K$-term significantly impacts the models' performances by reducing the MADs. Specifically, for compound C, the error of 
Model-1 is almost halved with the addition of the four-body term. 
Model-4 achieves the highest accuracy among the four models, with an MAD of 0.039 eV and an SD of 0.07 eV.
Comparing the performance of semi-local methods to the reference HSE06 results reveals an intriguing trend. In the case of both PBE and SCAN, the MAD values across compositions are approximately an order of magnitude greater than those of the models. The total MAD for PBE is 0.84 eV. For SCAN, the MAD drops to 0.53 eV. However, when assessing the SD, both semi-local methods yield significantly lower values for individual compositions, indicating that the errors in PBE and SCAN are predominantly systematic for each specific composition.
All the parameters involved in the best-performing model (Model-4) 
are listed in Table~\ref{tab:five}. Throughout the rest of the text, the term `model' specifically refers to Model-4.

We observe $J_0$ for all compositions to consistently align closely 
with the magnitude of the target property, $\varepsilon_g$ (Table~\ref{tab:five}). Additionally, the second and third nearest interaction contributions generally exhibit smaller magnitudes compared to the first nearest neighbor contribution quantified by
$J_1$. In general, the $K$-term is also smaller than the $J_3$ term, 
although there are a few exceptions. It is important to note that the coefficients' signs do not have a direct physical interpretation but are determined through an optimal fit. However, for modeling total or relative energies, there is no physically motivated explanation or quantum mechanical basis to justify why the interaction term $\langle \sigma_i \sigma_{i+1} \rangle$ between first neighbors is -1 for the 2$H$ polytype, while it is +1 for the 3$C$ structure, as indicated in TABLE~\ref{tab:two}.

As exemplary examples to showcase the application of the parameters listed in Table~\ref{tab:five}, let us consider 4$H$ and 12$R$ polytypes of SiC. Table~\ref{tab:two} provides the values of $\langle \sigma_i \sigma_{i+1} \rangle$, $\langle \sigma_i \sigma_{i+2} \rangle$, $\langle\sigma_i \sigma_{i+3} \rangle$, and $\langle\sigma_i \sigma_{i+1} \sigma_{i+2} \sigma_{i+3} \rangle$ as 0, -1, 0, and 1, respectively. The values of $J_{1-3}$ and $K$ are given in TABLE~\ref{tab:five} as 3.022, 0.575, 0.222, -0.179, and -0.096, respectively. These values, when used in \ref{eq:annni2}, provide the HSE06 level band gap as $J_0+J_1-K=3.022+0.222+0.096=3.34$ eV, which agrees with the value from the HSE06 calculation reported in Table~\ref{tab:sthree}. The estimation is exact for the 4$H$ polytype used to parameterize Model-4. For the 12$R$ polytype, all the 2-body interaction terms vanish, and the only non-vanishing term is the 4-body term,$\langle \sigma_i \sigma_{i+1} \sigma_{i+2} \sigma_{i+3}\rangle=-1$  (see Table~\ref{tab:two}). So, the band of 12$R$-SiC is $J_0+K)=3.022-0.096=2.926$ eV, deviating by $<0.1$ eV from the actual HSE06 value that is discussed later in the article.

\begin{figure}[hpb]
    \centering
    \includegraphics[width=\linewidth]{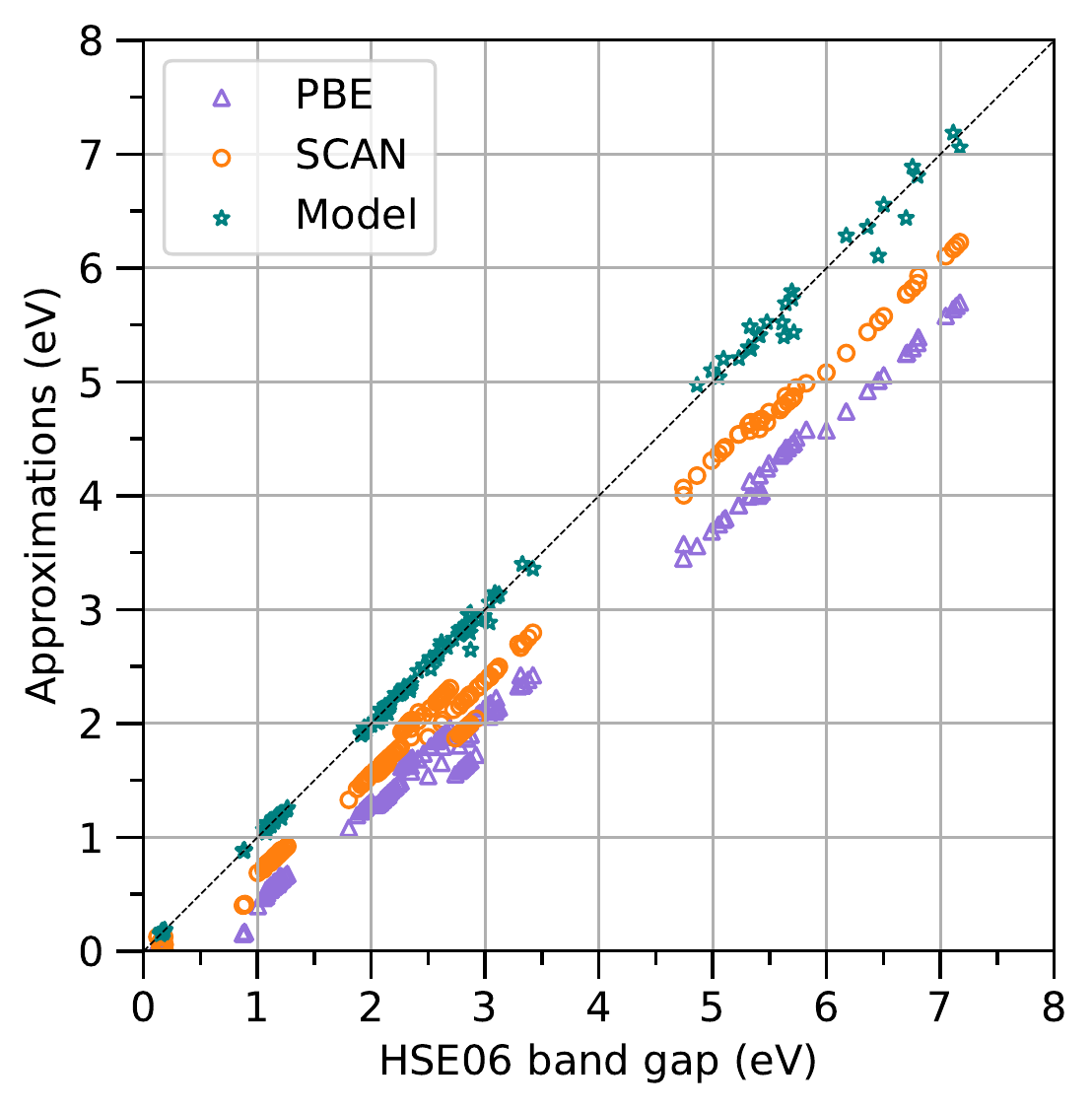}
    \caption{Scatterplot of the band gaps predicted with PBE, SCAN, and a model (Model-4 in TABLE~\ref{tab:four}) shown alongside the HSE06 values. For
    PBE and SCAN, we have plotted band gaps of 15 polytypes shown in FIG.~\ref{fig:one} 
    for the 15 compositions: C, Si, Ge, SiC, GeC, GeSi, BN, BP, BAs, AlN, AlP, AlAs, GaN,
    GaP, and GaAs
    For Model-4, we have shown results for the 7- and 8-bilayered 
    polytypes not used in the construction of the model.
    }
    \label{fig:two}
\end{figure}
\begin{figure*}
    \centering
    \includegraphics[width=\linewidth]{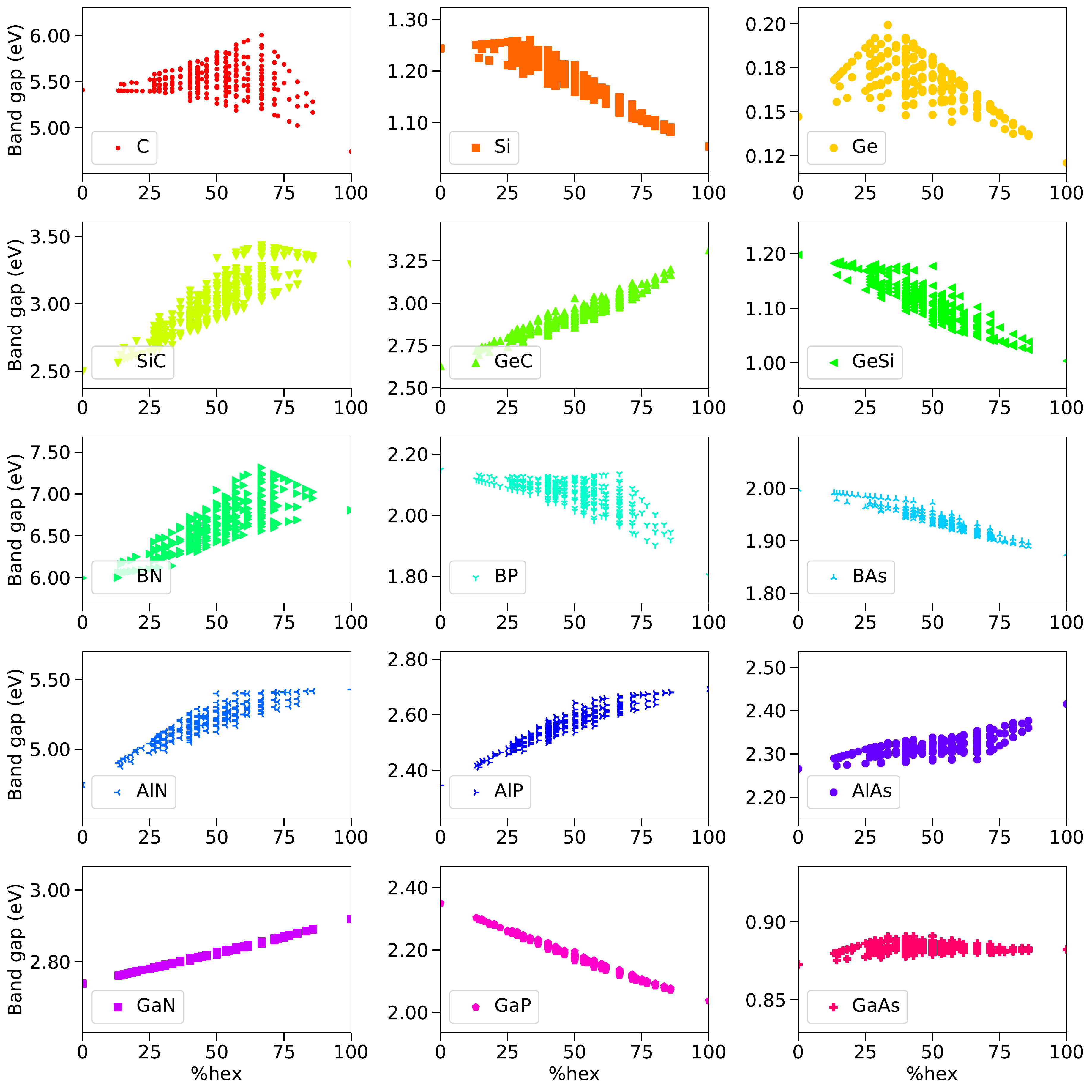}
    \caption{Model-predicted band gaps of 497 polytypes with up to
    15 layers in the unit cell. For various compositions,
    band gaps are plotted along with the 
    percentage hexagonality (\%hex) of the crystal structure. Extreme
    values of \%hex corresponds to 3$C$ (\%hex = 0) and 2$H$ (\%hex = 100) polytypes. 
    }
    \label{fig:four}
\end{figure*}

\subsection{Accuracy of the best model}
The pseudo-spin model (Model-4) developed using $\varepsilon_g$ of small polytypes with up to 6 bilayers in the unit cell, 
models the $\varepsilon_g$  of 7- and 8-bilayered polytypes not used in the construction of the model very efficiently, resulting in prediction errors centred around zero. FIG.~\ref{fig:two} displays a scatterplot comparing the predictions of Model-4 and the semi-local DFT methods with HSE06 values. The corresponding error metrics are listed in TABLE~\ref{tab:errormetrics} for a quantitative understanding of the methods' performances. 

GGA-PBE and mGGA-SCAN methods tend to underestimate the HSE06 values on an average by 0.84 and 0.53 eV, respectively. However, the agreement between their predictions and that of HSE06 shows a systematic linearity as indicated by their corresponding $R^2$ values greater than 0.99. This trend suggests that a linear adjustment of PBE and SCAN values can agree well with the HSE06 predictions. The corresponding slope and intercept can calibrate PBE/SCAN values for new predictions. However,  our study does not delve into a linear correction for PBE or SCAN because our primary focus is exploring long-period polytypes with multiple atoms in the unit cell. In this context, the pseudo-spin models fitted directly to HSE06 reference values will enable rapid application.

For all five error metrics considered in TABLE~\ref{tab:errormetrics}, Model-4 delivers better performances than PBE and SCAN. The MAD for the model's prediction is about 0.04 eV with an RMSD of about 0.07 eV. These errors are smaller than those of the reference DFT method, HSE06. It is important to note that when applied to other material classes, the DFT methods will retain their transferability with similar errors as seen for the AB polytypes. On the other hand, the models are fitted for a given composition and chemical formula. Hence, even though the pseudo-spin models are very accurate for the polytypes of a given composition, AB, they lose their prediction power when applied to a new composition, such as A$_2$B. For a given composition, the Ising-type additivity model based on the parameters given in TABLE~\ref{tab:five} allows for rapid and accurate estimations of $\varepsilon_g$ of long-period polytypes with multiple bilayers in the unit cell.

\begin{figure}[hpt]
    \centering
    \includegraphics[width=\linewidth]{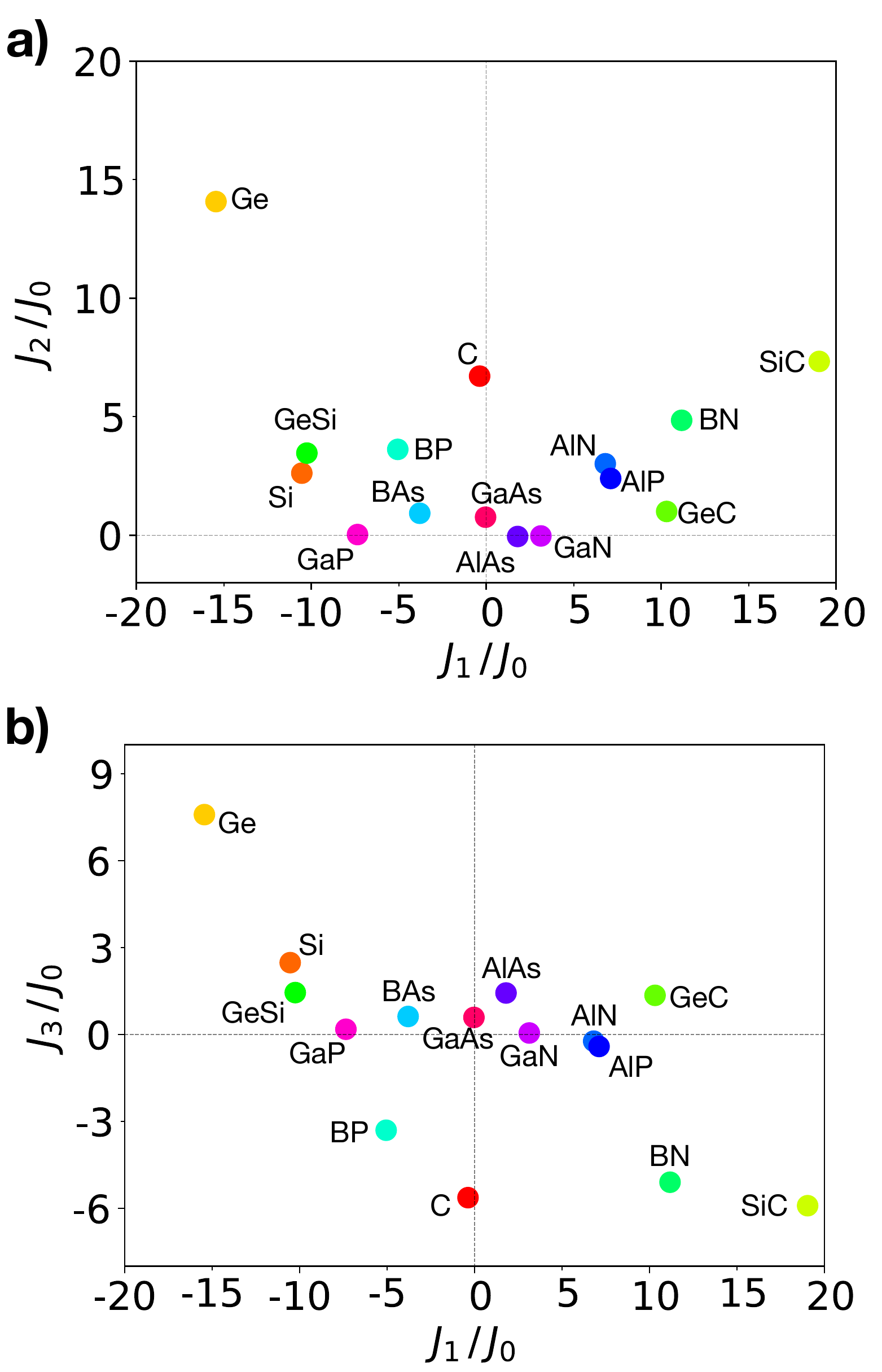}
    \caption{
    Parameter map for the band gap model of semiconductors:
    a) Normalized values of $J_2$ and $J_1$ are plotted. 
    Values on the right (left) side of the plot show an overall increase (decrease) in the band gap
    with hexagonality of the crystal structure. 
    b) Normalized values of $J_3$ and $J_1$ are plotted. 
    Values on the right side show a maximum for \%hex$>50$,
    while those on the left side show a maximum for \%hex$<50$.
    For clarity, both axes are multiplied by 100. 
    }
    \label{fig:three}
\end{figure}

\begin{table}[hpb]
\centering
\caption{
    Error metrics for various approximations corresponding to the
    plot shown in  FIG.~\ref{fig:two}.
    MAE, RMSD, MAX (all in eV) correspond to mean 
    absolute error, root mean square
    deviation and maximum absolute error, respectively. 
    $R^2$ and $\rho$ (both dimensionless) correspond to
    coefficient of determination and Spearman's
    rank correlation coefficient, respectively. 
 }
\small\addtolength{\tabcolsep}{1.2pt}
\begin{tabular}[t]{l rr rr r}
\hline 
\multicolumn{1}{l}{Method} & 
\multicolumn{1}{l}{MAE} & 
\multicolumn{1}{l}{RMSD} & 
\multicolumn{1}{l}{MAX} & 
\multicolumn{1}{l}{$R^2$} & 
\multicolumn{1}{l}{$\rho$} \\
\hline 
PBE &      0.841 &   0.334  &  1.474  &  0.996  &  0.986 \\
SCAN &     0.528 &   0.223  &  0.947  &  0.997  &  0.986 \\ 
Model &  0.039 &   0.069  &  0.348  &  0.999  &  0.997 \\
\hline
\end{tabular}
\label{tab:errormetrics}
\end{table}%

\subsection{Correlation between band gap and hexagonality}
FIG.~\ref{fig:four} presents scatterplots depicting the variation in $\varepsilon_g$ as a function of \%hex for 497 polytypes with up to 15 bilayers. While Model-3 demonstrates improved accuracy for element C (as shown in TABLE\ref{tab:four}), we employ Model-4 uniformly for all compositions to ensure a fair comparison.

As noted by Choyke {\it et al.}\cite{choyke1964optical}, the experimental band gaps of different phases of SiC, namely 2$H$ (3.33 eV), 3$C$ (2.39 eV), 4$H$ (3.26 eV), and 6$H$ (3.02 eV), show an increasing trend from \%hex=0 (3$C$) to \%hex=50 (4$H$), and then remain relatively constant for \%hex values between 50 and 100. This observation has been further discussed by others\cite{presser2008silica,kackell1994electronic,wenzien1995quasiparticle}.
Additionally, we have found 8$H$(2)-SiC with \%hex=25 exhibits a band gap of 2.80 eV following this trend. FIG.~\ref{fig:four} illustrates the relationship between the structure and 
properties of 497 SiC polytypes. The inclusion of a large number 
of polytypes in this study offers examples with \%hex greater than 50, indicating that 
the upper bound for $\varepsilon_g$ continues to increase until 
\%hex=66.67, after which it gradually declines until reaching \%hex=100. 

An explanation of the statistical trends seen in FIG.~\ref{fig:four} can be deduced by the correlations: $J_2$-vs.-$J_1$ and $J_3$-vs.-$J_1$ as depicted in FIG.~\ref{fig:three}. Since the band gaps of materials 
are clustered around different mean values, in order to treat all materials 
on the same footing, we have divided $J_{1-3}$ by $J_0$ in this plot. All materials in a particular region on the configuration map exhibit similar behavior. This analysis effectively elucidates the band gap variation with respect to \%hex while also providing insights into the existence of polytypes with maximum band gaps for structures with less than or greater than 50\% hexagonality. 

The band gap behavior of the C polytypes, in terms of \%hex, resembles the polytypes of BP (refer to FIG.~\ref{fig:four}). However, carefully examining the parameters presented in  TABLE~\ref{tab:five} reveals C and GaAs as exceptions with significantly smaller $J_1$ than $J_2$ and $J_3$. Consequently, C occupies a similar position to GaAs in FIG.~\ref{fig:three}. Moreover, TABLE~\ref{tab:four} highlights that Model-3 outperforms Model-4 for the polytypes of C. Hence, a more robust rationale could have been established by employing the best model for each composition. However, since this study aims to emphasize the applicability of the pseudo-spin model of band gaps for long-period polytypes of SiC, we have refrained from conducting a separate analysis for C polytypes with Model-3.

Furthermore, the relationship between $\varepsilon_g$ and \%hex has not been extensively discussed for several binary
semiconductors. Our findings, as illustrated in FIG.~\ref{fig:four}, reveal consistent patterns in $\varepsilon_g$ for GaN and GaP, albeit with distinct behaviors. Specifically, GaN demonstrates an 
increase in the bandgap with \%hex, while GaP exhibits a reversal of this trend. 

\begin{figure}[hpb]
    \centering
    \includegraphics[width=\linewidth]{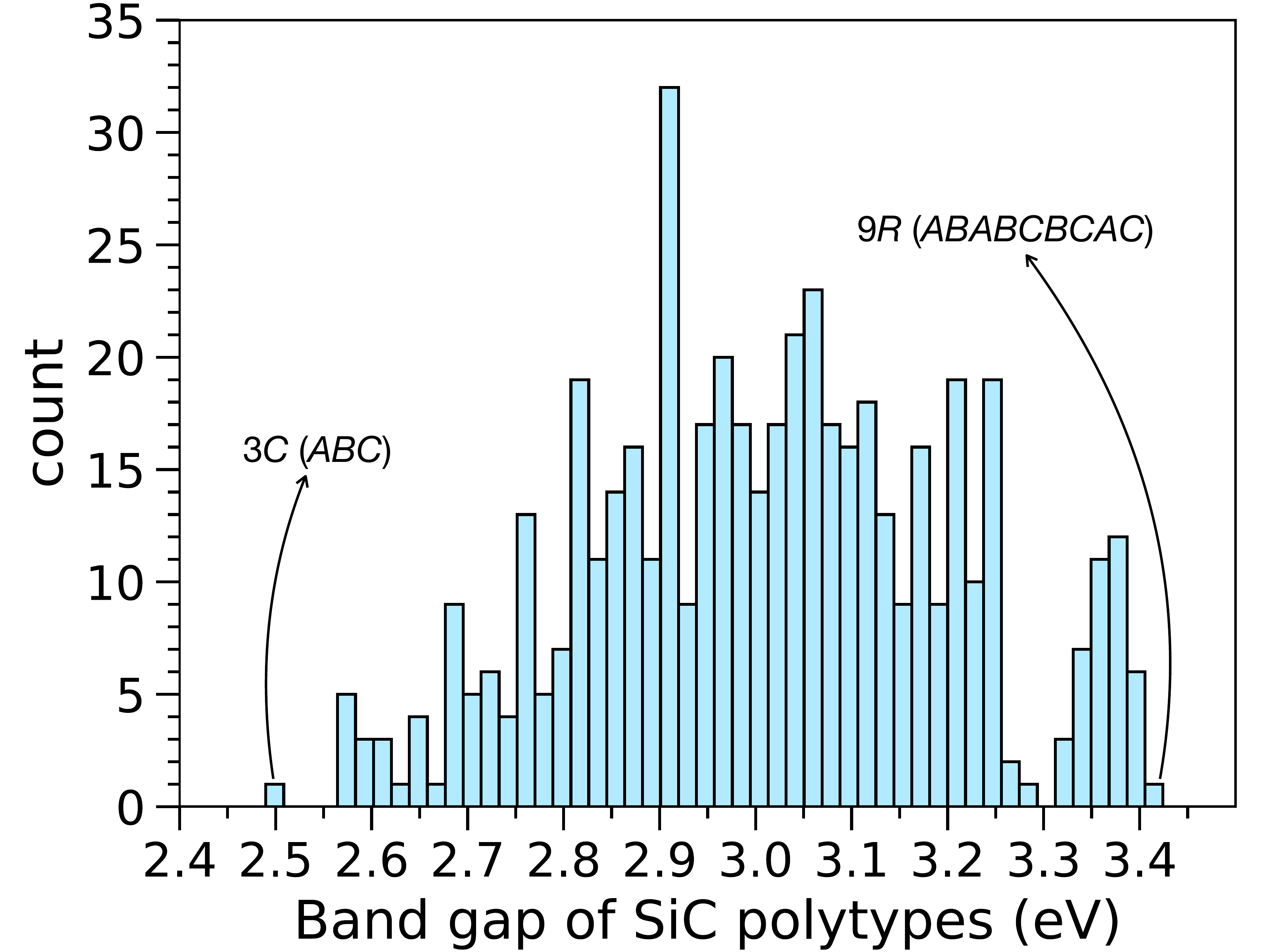}
    \caption{Histogram of 
    model-estimated band gaps of 497 polytypes of SiC 
    with up to 15 bilayers in the unit cell. 
    }
    \label{fig:five}
\end{figure}

\begin{figure*}[hpt]
    \centering
    \includegraphics[width=\linewidth]{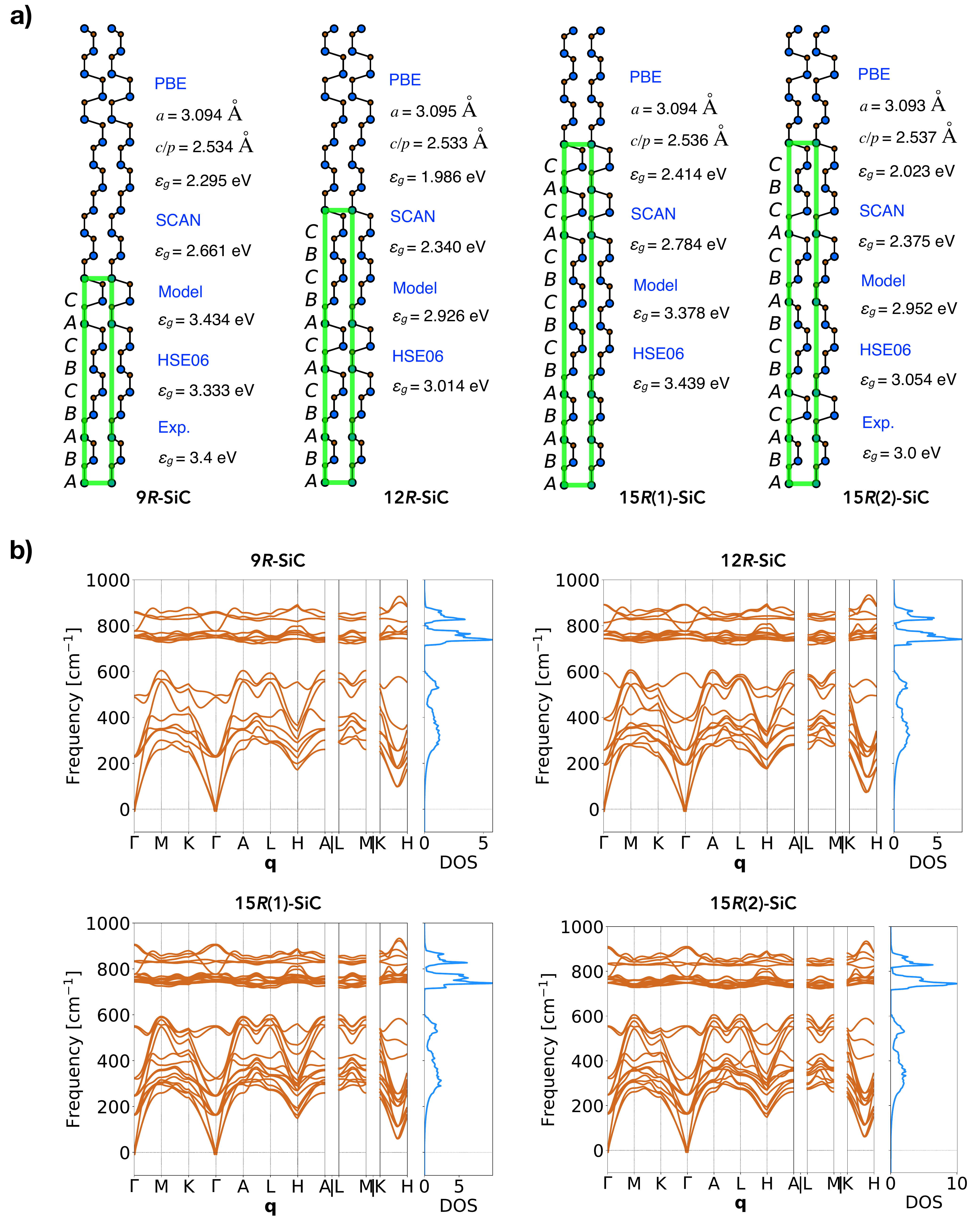}
    \caption{ Details of four rhombohedral 
    long-period polytypes of SiC: 9$R$, 12$R$, 15$R$(1), and
    15$R$(2): a) Schematic representation of the stacking sequence (brown/blue circles correspond to C/Si atoms), 
    calculated values of lattice parameters
    and band gaps are given along with available experimental 
    band gaps\cite{yaghoubi2018predicting}.
     b) GGA-PBE phonon band structures along with the corresponding density of states.}
    \label{fig:six}
\end{figure*}

Notably, GaAs maintains a relatively constant value of $\varepsilon_g$ for all \%hex. This trend can be understood through the vanishing interaction parameters $J_{1-3}$ and $K$ listed in TABLE~\ref{tab:five}. Nevertheless, the structure and cohesive energies of the polytypes of GaAs have been shown to vary with \%hex\cite{panse2011polytypism}. Furthermore,
zero-Kelvin experimental band gaps \cite{kusch2012band} 
of 2$H$ (1.46 eV) and 3$C$ (1.52 eV) imply a small variation with \%hex.
In \ref{ref:IIIA}, we discussed how the HSE06 band gap of 
3$C$-GaAs widens to 1.14 eV when using the crystal structure determined with HSE06. We also performed separate geometry relaxation of 2$H$-GaAs at the HSE06 level, resulting in $a=4.013$ \AA{} and
$c/p=3.307$ \AA{} (see Table S1 for the corresponding PBE values). At this geometry, the HSE06 $\varepsilon_g$ of 2$H$-GaAs turned out to be 1.13 eV, differing from the 3$C$ value by only 0.1 eV. For a given polytype of GaAs, DFT-predicted band gaps are sensitive to small changes 
in lattice parameters. However, across polytypes, they show less variation, resulting in a rather flat profile seen in FIG.~\ref{fig:four}; see also the almost constant values of the band gaps of GaAs at PBE, SCAN and HSE06 levels in TABLE~\ref{tab:sthree}. It is interesting to note that two past studies\cite{zanolli2007model,giorgi2020doping} have noted 
that many-body effects in the wavefunction are necessary to break the `degeneracy' of $\varepsilon_g$ among the wurtzite and zincblende phases of GaAs. In the
more recent study\cite{giorgi2020doping}, the
many-body method G$_0$W$_0$ was applied to predict the band gaps of 2$H$- and 3$C$-GaAs as 1.46 and 1.66 eV, in good qualitative and quantitative agreement with the experimental values: 1.46 and 1.52 eV\cite{kusch2012band}. Hence, even though GaAs is not known to exhibit extensive polytypism as SiC, for the pseudo-spin models to forecast a quantitative trend across hypothetical polytypic phases, it is necessary to consider reference band gaps calculated using post-DFT methods.

\subsection{Case study of rhombohedral SiC}
Over 250 different polytypes of SiC have been observed in experimental studies\cite{krishna1966crystal}. Among them are long-period polytypes containing even more than 50 bilayers in the unit cell, as well as ultralong period polytypes, such as 393$R$ and 594$R$\cite{krishna1966crystal,fal1999strong}. While ZnS also exhibits polytypism, it is a direct band gap semiconductor, implying that the variation of $\varepsilon_g$ across its polytypic forms is relatively limited. Since SiC is an indirect gap material, it can exhibit diverse gaps due to band-folding effects\cite{lambrecht1997electronic} across polytypes
of different unit cell sizes. FIG.~\ref{fig:five} illustrates a histogram depicting the model-predicted $\varepsilon_g$ for 497 polytypes of SiC, demonstrating a range of $\approx$0.9 eV. According to the model, the polytype with the smallest $\varepsilon_g$ is 3$C$, while the largest gap is found in 9$R$. A past study\cite{yaghoubi2018predicting} has reported the HSE06 level band gap of this phase to be 3.233 eV and discussed its dynamic stability through phonon analysis. It is worth noting that several long-period polytypes of SiC observed experimentally exhibit rhombohedral symmetry. Those with experimental band gaps are:
9$R$ (\%hex=66.67, 3.4 eV)\cite{yaghoubi2018predicting}, 
15$R$ (\%hex=40, 3.0 eV)\cite{yaghoubi2018predicting}, 
21$R$ (\%hex=28.57, 2.86 eV)\cite{choyke1964optical}, and 
33$R$ (\%hex=36, 3.01 eV)\cite{choyke1964optical}.

Given the prevalence of polytypes with rhombohedral symmetry in reported studies, we have specifically chosen four phases: 9$R$, 12$R$, and two 15$R$ phases from the 497 set for
 detailed analysis. Only 4 out of 497 polytypes with $\le15$ bilayers are rhombohedral, which highlights the relatively low statistical likelihood of encountering $R$ phases compared to $H$ and $T$. 
The reason behind the low occurrence of an $R$-polytype is that the unit cell must consist of a multiple of three bilayers with an equal number of $A$, $B$, and $C$-type bilayers. Consequently, although the number of polytypes increases significantly with the number of bilayers in the unit cell, rhombohedral phases represent a tiny fraction within the overall landscape of polytype compounds.

We have optimized the structure of four $R$-type polytypes of silicon carbide (SiC), as depicted in FIG.~\ref{fig:six}a. We have confirmed that all four phases are dynamically stable by phonon band structure analysis (see FIG.~\ref{fig:six}b). The stacking sequences, as indicated in List~S1 of the SI, reveal the hexagonal stacking percentages (\%hex) of each polytype: 66.67\% for 9$R$, 50.0\% for 12$R$, 80.0\% for 15$R$(1), and 40.0\% for 15$R$(2). Notably, since the 15$R$(1)-SiC phase exhibits a greater \%hex than the 9$R$ phase, comparing their respective band gaps through HSE06 calculations is intriguing. While our model predicts a larger $\varepsilon_g$ of 3.43 eV for the 9$R$ polytype, the HSE06 reference method suggests that 15$R$(1) possesses a slightly larger band gap than 9$R$. It is important to acknowledge that a minor discrepancy between the model and HSE06 is expected, considering the average error of the model, as reported in TABLE~\ref{tab:four}.

Among the various polytypes of SiC, 9$R$, 12$R$, and 15$R$(2) have been experimentally studied. As of yet, the 15$R$(1) polytype has only been investigated  computationally\cite{limpijumnong1998total}. Hence, to gain further insights, we examined the cohesive energies of all four $R$-SiC polytypes using the HSE06 level of theory. Compared to 3$C$-SiC (-7.056344 eV/atom), 2$H$, 9$R$, 12$R$, and 15$R$(1) are higher in energy by 2.11, 0.29, 0.14, and 1.03 meV/atom, respectively. On the other hand, 4$H$ and 15$R$(2) are lower in energy by  -0.77 and  -0.69 meV/atom, respectively. These findings support previous conclusions\cite{limpijumnong1998total} that 15$R$(1) is thermodynamically less stable than 9$R$, yet more stable than the 2$H$ phase. Overall, 15$R$(1) can be regarded as a wide-gap metastable phase of SiC.

\begin{table*}[hpt]
\centering
\caption{Energies of the highest valence (V) and the lowest 
conduction (C) bands of polytypes of SiC used in the model. 
For each energy, parameters fitted to Eq.~\ref{eq:annni2} are given.
In all cases, VBM is at the $\Gamma$-point
that is set to zero. All values are in eV.}
\small\addtolength{\tabcolsep}{1.2pt}
\begin{tabular}[t]{lll  r r r r r l r r r r r}
\hline 
\multicolumn{2}{l}{${\bf k}$} & 
\multicolumn{1}{l}{band} &
\multicolumn{5}{l}{Polytypes} &
\multicolumn{5}{l}{Parameters}  \\
\cline{4-8} \cline{10-14} 
\multicolumn{3}{l}{} & 
\multicolumn{1}{l}{2$H$} & 
\multicolumn{1}{l}{3$C$} & 
\multicolumn{1}{l}{4$H$} & 
\multicolumn{1}{l}{5$T$} & 
\multicolumn{1}{l}{6$T$} &
\multicolumn{1}{l}{} &
\multicolumn{1}{l}{$J_0$} & 
\multicolumn{1}{l}{$J_1$} & 
\multicolumn{1}{l}{$J_2$} & 
\multicolumn{1}{l}{$J_3$} & 
\multicolumn{1}{l}{$K$}  \\
\hline 
$\Gamma$ & (0,0,0)   & V &      0.000 &      0.000 &      0.000 &      0.000 &      0.000  &  &      0.000 &     0.000 &      0.000 &     0.000 &      0.000  \\
&                    & C &      5.851 &      6.779 &      6.154 &      6.152 &      6.011  &  &      6.115 &     -0.407 &     -0.081 &     -0.057 &     -0.120   \\
$M$ & (1/2,0,0)      & V &     -1.317 &     -1.900 &     -1.233 &     -1.505 &     -1.264  &  &     -1.413 &      0.280 &      0.188 &      0.012 &      0.007   \\
&                    & C &      3.779 &      2.341 &      3.227 &      2.927 &      3.388  &  &      3.108 &      0.720 &      0.084 &     -0.001 &     -0.035  \\
$K$& (1/3,1/3,0)     & V &     -4.115 &     -2.355 &     -1.823 &     -2.567 &     -2.288  &  &     -2.575 &     -0.385 &      0.706 &     -0.494 &     -0.046  \\
&                    & C &      3.294 &      4.913 &      5.063 &      4.020 &      4.093  &  &      4.190 &     -0.986 &      0.480 &      0.177 &     -0.393   \\
$A$ &(0,0,1/2)       & V &     -0.786 &     -0.357 &     -0.226 &     -0.108 &     -0.083  &  &     -0.222 &      0.104 &      0.173 &     -0.318 &      0.176   \\
&                    & C &      7.004 &      6.507 &      6.374 &      6.129 &      6.052  &  &      6.299 &     -0.284 &     -0.191 &      0.533 &     -0.265  \\
$L$ &(1/2,0,1/2)     & V &     -2.527 &     -1.155 &     -1.690 &     -1.224 &     -1.465  &  &     -1.504 &     -0.192 &      0.075 &     -0.494 &      0.262  \\
&                    & C &      4.316 &      3.765 &      3.650 &      2.733 &      3.285  &  &      3.160 &      0.080 &     -0.195 &      0.195 &     -0.685  \\
$H$ &(1/3,1/3,1/2)   & V &     -1.898 &     -3.320 &     -2.640 &     -2.130 &     -1.852  &  &     -2.228 &      1.126 &     -0.015 &     -0.415 &      0.396  \\
&                    & C &      6.163 &      6.288 &      4.299 &      4.465 &      3.813  &  &      4.635 &     -1.096 &     -0.963 &      1.034 &     -0.627 \\
\hline
\end{tabular}
\label{tab:seven}
\end{table*}%

\subsection{Model for band energies at high-symmetry points of the Brillouin zone}
The distinct variation in the band gap across different polytypes of SiC is attributed to the
indirect nature of their band gap\cite{van1997band}. Within the band structure, certain high-symmetry points, namely $\Gamma$, $M$, and $K$, represent the band origin with respect to $k_z$ (with $k_z = 0$), while $A$, $L$, and $H$ denote the corresponding band edges (with $k_z = 1/2$). Across all polytypes examined in this study, the VBM is consistently associated with the downward progression of the $p$-band from $\Gamma$. Only in 2$H$-SiC does the CBM occur at $K$, while in all other polytypes analyzed, the CBM of SiC resides either at $M$ or $L$. Previous studies investigating 2$H$, 3$C$, 4$H$, and 6$H$-SiC have demonstrated 
that the band gap widens when hexagonality increases from 0 to 50, then remains constant
until 100\% hexagonality (as in 2$H$), while the CBM energies vary linearly with the hexagonality of the crystal structure\cite{park1994structural}.

\begin{figure}[hpb]
    \centering
    \includegraphics[width=\linewidth]{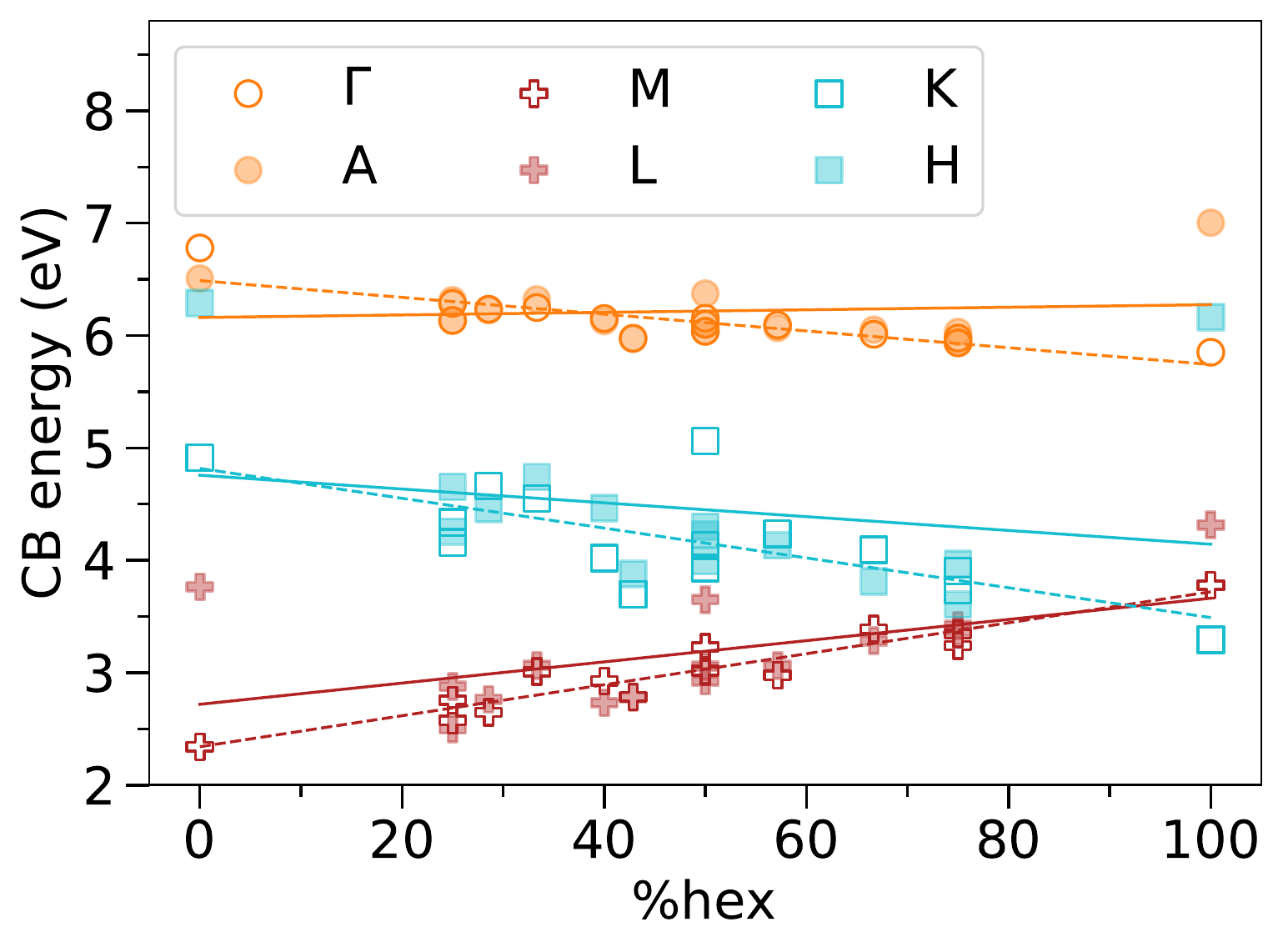}
    \caption{Variation of HSE06 conduction band (CB) energies (in eV)
    of 15 SiC polytypes (listed in TABLE~\ref{tab:one}) with the 
    percentage hexagonality (\%hex) of the crystal structure. 
    Fitted straight lines are also shown;
    dashed lines correspond to $\Gamma$, $M$, and $K$ while
    solid lines correspond to $A$, $L$, and $H$.
    }
    \label{fig:seven}
\end{figure}

In FIG~\ref{fig:seven}, we present the computed energies of the CB at various high-symmetry $k$-points for different crystal structures of SiC, explicitly focusing on the 15 simplest polytypes. The purpose of this figure is to observe how the hexagonality of the crystal structure influences the CB energy. The plot reveals linear trends suggesting a preference for CBM at $M$ or $L$ for hexagonality less than 90\%; beyond this limit, $K$ is a preferred location for CBM. Among the polytypes with up to 15 bilayers, only 2$H$ has \%hex$>90$. 
Other close examples include three 12-bilayer polytypes (12$T$: {\it ABABABABABAC}, 12$T$: {\it ABABABABACAC}, and 12$H$: {\it ABABABACACAC}) with a hexagonality of 83.33\% and three 14-bilayer polytypes (14$T$: {\it ABABABABABABAC}, 14$T$: {\it ABABABABABACAC}, and 14$T$: {\it ABABABABACACAC}) with a hexagonality of 85.71. For a comprehensive list of 497 polytypes with up to 15 bilayers in the unit cell, please refer to List~S1 in the SI.
Among $M$ and $L$ points, the preference for the CBM is not as pronounced. It is expected that in long-period polytypes, the energies of both points would be similar because of band-folding effects. In other words, as the unit cell size increases, the VB and CB along the paths $\Gamma \rightarrow A$, $M \rightarrow L$, and $K \rightarrow H$ tend to exhibit flatter energy profiles with diminishing band dispersions.

\begin{figure*}
    \centering
    \includegraphics[width=\linewidth]{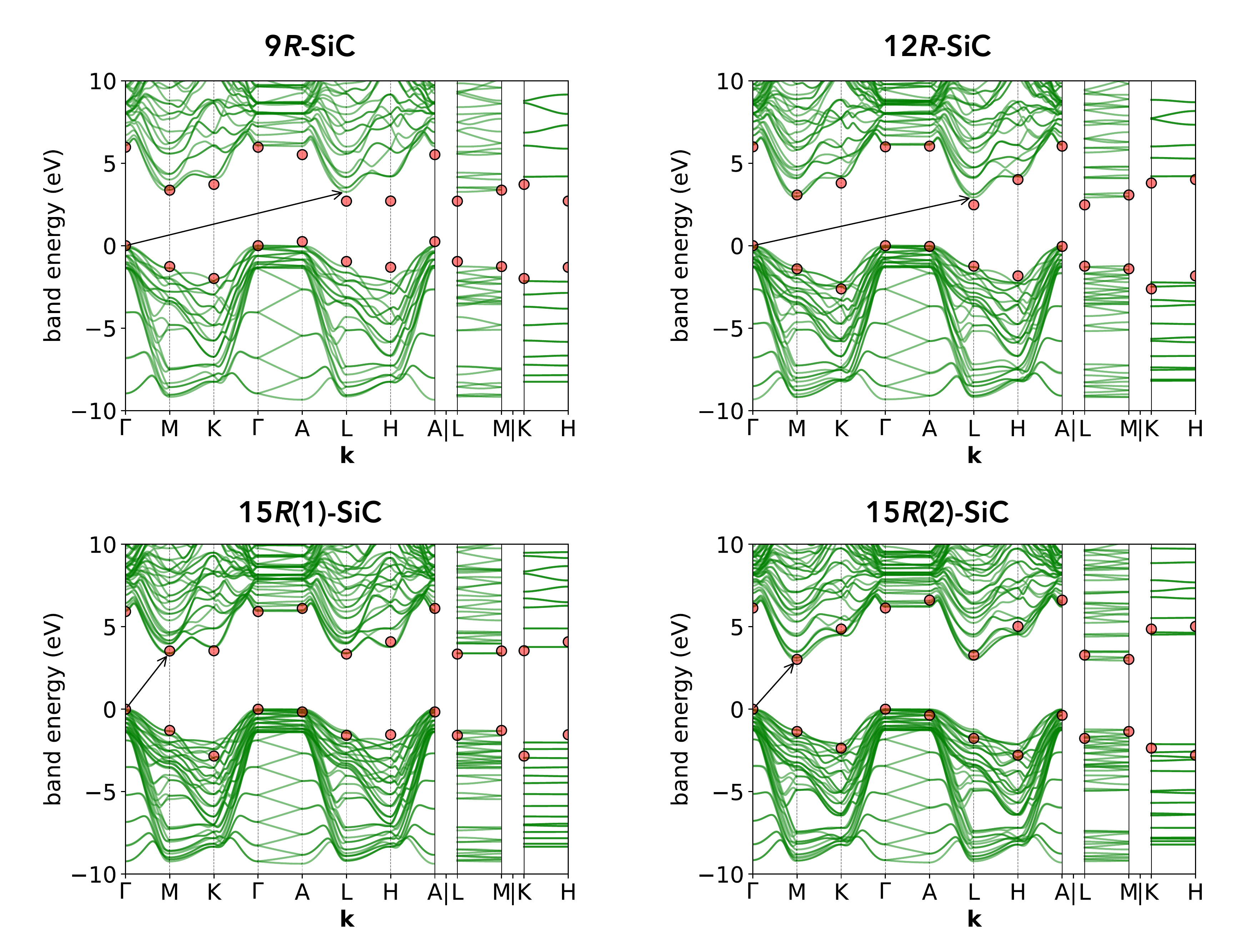}
    \caption{
    HSE06 electronic band structures of the rhombohedral phases
    of SiC along with VB and CB energies at the high-symmetry points
    predicted with a model (red points). 
    The arrow indicates the indirect band gap. 
    In all cases, VBM is set to zero. 
    }
    \label{fig:eight}
\end{figure*}

PBE, SCAN, and HSE06 band structures of the smallest six polytypes are collected in Figures~S1 of SI. It is evident that the underestimation of the band gap is prominent in PBE and SCAN, as observed through the trend: PBE $<$ SCAN $<$ HSE06. Using HSE06 band structures of the 2$H$, 3$C$, 4$H$, 5$T$, and 6$T$ polytypes of SiC, we fitted Model-4 using the energies of VB and CB at each high-symmetry point within the Brillouin zone. The energies and the resulting parameters are provided in TABLE~\ref{tab:seven}. As expected, the model reproduces the target
energies for the polytypes used in the model, as illustrated in
Figures~S2--S6 in the SI.

Additionally, we made predictions for the VB and CB energies at the high-symmetry points for the 6$H$ phase and for phases with 7 and 8 bilayers in the unit cell not included in the model. The SI presents the HSE06 band structures alongside the energies predicted by the model (Figures~S7--S16 in the SI). Overall, the predictions align well with HSE06 for all phases, capturing the indirect band gap nature and the associated symmetry of the CBM. However, the model exhibits discrepancies at the $H$ point for a few polytypes. Notably, the model spuriously widens the gap at $H$ consistently by: (i) overestimating the energy of CB and underestimating the VB energy in 6$H$, 7$T$(3)
(ii) overestimating the energy of CB with no effect on the corresponding VB in 7$T$(1), 8$T$(1), 8$T$(2), 8$T$(3), and 8$H$(2). In contrast, the gap is underestimated at $H$ in 7$T$(2) and  8$H$(1) (Figure~S9 and Figure~S12 in the SI). 8$T$(4) is the only case where the model 
fully agrees with the reference theory, HSE06 (Figure~S15 in the SI). Overall, these trends do not indicate the model's prediction accuracy at the $H$-point to correlate with change in \%hex of these polytypes collected in TABLE~\ref{tab:one}.

To explore the scope to apply the model to predict the positions of the VBM and CBM of long-period polytypes, we applied the models to 9, 12, and 15 $R$ polytypes of SiC. The results are presented in FIG~\ref{fig:eight}. The model's prediction is overall in good agreement for all four polytypes. The best prediction is seen for 12$R$-SiC with \%hex=50. For 15$R$(1) and 15$R$(2), the VB energies are underestimated at $K$ and $H$, respectively. Since 15$R$(1) has \%hex=80, the HSE06 CB energy at $H$ is lowered compared to $L$ or $M$ than in 15$R$(2) with \%hex=40 as expected according to the linear trends revealed in FIG.~\ref{fig:seven}. In the case of 9$R$-SiC with \%hex=66.67, the model significantly underestimates the CB energy at $H$, bringing the energy close to that of $L$ (see FIG~\ref{fig:eight}). This prediction is in disagreement with the linear relationships established using HSE06 band structures shown in FIG.~\ref{fig:seven} according to which the energies of $K$/$H$ pairs should lie above the energy of $M$/$L$ pairs for \%hex$<90$.

In contrast to cohesive energy and the overall band gap, which are size-intensive material properties, the band energies at specific high-symmetry points may be considered size-extensive. In systems with larger unit cell sizes, the energy at high-symmetry points approaches $\Gamma$. Therefore, when building models, it might be helpful to consider the effects of Brillouin zone folding in the band structure for polytypes of different unit cell sizes.

\section{Computational Details}\label{sec:II}

\subsection{Crystal structure generation}\label{subsec:A}
For all polytypes, in four crystal classes---$C$, $H$, $T$, and $R$---we generated initial structures in a hexagonal lattice defined by the lattice vectors:
\begin{eqnarray}
    {\bf a}_1 & = & \hat{x} a / 2 -  \hat{y} \sqrt{3} a / 2 \nonumber  \\
    {\bf a}_2 & = & \hat{x} a / 2 +  \hat{y} \sqrt{3} a / 2  \nonumber \\
    {\bf a}_3 & = & \hat{z} c  \nonumber 
\end{eqnarray}
Notably, the cubic polytype 3$C$ is a specific case of $R$ when the angles $\alpha$, $\beta$, and $\gamma$ are equal to $60^\circ$, along with the condition $a=b=c$ that usually applies to $R$ systems. The fractional coordinates for a given polytype are generated as follows:
\begin{enumerate}
    \item For the $A$, $B$, and $C$ type bilayers, the $x$ and $y$ Wyckoff coordinates (coefficients of ${\bf a}_1$ and ${\bf a}_2$ lattice vectors) are $\left( 0, 0\right)$, $\left( \frac{1}{3}, \frac{2}{3}\right)$, and $\left( \frac{2}{3}, \frac{1}{3}\right)$, respectively.
    \item The $z$-coordinate of atom-1 (e.g., Si in SiC) is assigned as $q/p+u_i$, where $p$ represents the total number of bilayers, $q$ ranges from $0$ to $p-1$, and $u_i$ is a free parameter for atom-1 
    in bilayer $i$.
    \item  For atom-2 (e.g., C in SiC), the $z$-coordinate is assigned as $q/p+d+v_i$, where $d=\frac{3}{4q}$ represents an increment in the $z$-direction, and $v_i$ is a free parameter for atom-2 in bilayer $i$.
\end{enumerate}
Similar structural information are available in \RRefs{kackell1994electronic,bauer2001structure}.

\subsection{Density functional theory calculations}\label{subsec:B}
Lattice vectors and atomic coordinates were fully 
relaxed with the all-electron, numeric atom-centred orbital (NAO) code
FHI-aims\cite{blum2009ab} with the Perdew--Burke--Ernzerhof (PBE)\cite{perdew1996generalized}
GGA XC functional. In all calculations a {\tt tight}
basis set was employed. The quality of the NAO
basis sets used in this study can be assessed 
by comparing them with the $E_{\rm cut}$ parameter in the plane-wave framework. For
determining the energies of 63 binary materials, the accuracy of {\tt light} and {\tt tight} basis sets were shown to be equivalent to that of using an $E_{\rm cut}$ of
400 eV and 800 eV in the plane-wave framework\cite{carbogno2022numerical}. 
During geometry relaxations, the convergence thresholds for
electron density, analytic gradients, and maximum force 
component for gradient-based minimization were set to $10^{-6}$ $e/$\AA$^3$ 
(where $e$ is the charge of an electron),
$5\times10^{-4}$ eV/\AA,
and $5\times10^{-3}$ eV/\AA, respectively. 
Lattice relaxations were performed using FHI-aims'
feature to constrain the crystal structure at the desired space group and 
to constrain the coordinates of the atoms to fixed forms corresponding
to the Wyckoff sites of interest\cite{lenz2019parametrically,lenz2022towards}.
Using this approach, it is possible to symbolically map the deviations in the $z$ Wyckoff coordinates from their ideal positions 
and to explicitly determine the 
corrections $u_i$ and $v_i$ mentioned in \ref{subsec:A}. 

The crystal structures of the 2$H$ phase were optimized for all compositions, and the corresponding
lattice constant $a$ was used as the initial guess for polytypes with larger unit cells. 
 Further, we used 
$p \times c^{2H}/2$, where $c^{2H}$ is the minimum energy value of the $c$-lattice constant of the 2$H$ polytype, 
for other polytypes with $p$ bilayers. 
All lattice parameters reported here are calculated at the GGA-PBE level.
We report band gaps calculated using PBE geometries at three levels:
GGA-PBE, the strongly constrained and appropriately normed (SCAN) mGGA functional  \cite{sun2015strongly} and the Heyd--Scuseria--Ernzerhof (HSE)\cite{krukau2006influence} hybrid XC functional with a screening parameter of $\omega=0.11$ bohr$^{-1}$ (referred to as the HSE06 functional).

For each reciprocal lattice vector (${\bf b}_i$, where $i$ = 1, 2, 3), 
$k$-grids were selected as equally spaced points between $-\pi |{\bf b}_i|$ and
$+\pi |{\bf b}_i|+dk$, with a separation (or $k$-spacking) 
of $dk$ (in \AA$^{-1}$). An additional $dk$ value was 
added to the upper limit 
to ensure a symmetric range, as $\pi |{\bf b}_i|$ may not be divisible by $dk$. 
The $k$-grids size obtained in this manner is directly linked to the unit cell volume, which is determined by the real-space lattice vectors (${\bf a}_i$). 
Accordingly, larger volumes result in a smaller number of $k$-grids. In our calculations, we used $dk=0.3$ \AA$^{-1}$ 
and sampled the Brillouin zone using the Monkhorst--Pack scheme\cite{monkhorst1976special}.
For 2$H$-SiC phase, the number of $k$-grids is $9\times9\times6$ while for 3$C$-SiC and 4$H$-SiC, the number of $k$-grids is $9\times9\times4$. For
SiC polytypes with 5-8 bilayers, $9\times9\times3$ $k$-grids were used to sample
the Brillouin zone. 
In the case of long-period polytypes of SiC---9$R$,
12$R$, 15$R$(1), and 15$R$(2)---our approach yielded initial $k$-grids of size $9\times9\times2$, which were subsequently increased to $9\times9\times3$ in order to improve the band structures.

Phonon spectra were obtained for supercells using 
finite-derivatives of analytic forces
with an atomic displacement of 0.01 \AA{} 
and tighter thresholds ($10^{-7}$ $e/$\AA$^3$ for electron density and
$10^{-6}$ eV/\AA{} for analytic forces)
using the Phonopy code\cite{togo2015first}
interfaced with FHI-aims. For 9$R$, 
12$R$, 15$R$(1), and 15$R$(2), we used a $3\times3\times1$ supercell containing 162, 216, 270, and 270 atoms, respectively. 
Scalar relativistic
effects are accounted for within the atomic zeroth-order regular approximation (atomic ZORA)\cite{blum2009ab}.
All band structures were generated along high-symmetry points 
in the Brillouin zone as defined by Setyawan {\it et al.}\cite{setyawan2010high}.
Optimized crystal 
structures were visually analyzed using the program Vesta\cite{momma2011vesta}.

\section{Conclusions}
We have conducted a systematic study on the polytypes of group IV elements, binary compounds of IV-IV and III-V groups, and employed hybrid-DFT level calculations to model their band gaps. Our approach, inspired by the cluster expansion of the Ising model, incorporates a four-body coupling that results in an overall error of 0.039 eV across the investigated compositions. Remarkably, the model's predictions exhibit minimal systematic errors, aligning well with the HSE06 level band gaps for long-period polytypes not used during the model's development. The mean absolute deviation of our model is an order of magnitude smaller than that of GGA-PBE and mGGA-SCAN DFT methods.

The accuracy of our model enables us to efficiently determine the band gaps of 497 polytypes for each composition containing up to 15 bilayers in the unit cell. Although GGA/mGGA band gaps calibrated using linear fitting could demonstrate better agreement with hybrid-DFT values, implementing such methods necessitates calculating GGA/mGGA band structure, increasing the computational overhead. Therefore, empirical models reported in this study offer a practical solution for the rapid high-throughput screening of electronic properties in the materials space of polytypes. 

We re-examined the empirical relationship between hexagonality and band gap in SiC polytypes and observed that the highest band gaps occur for hexagonality greater than 50\%. We have also established similar correlations for the other compositions: BN, BP, AlN, and AlP. Using a phase-diagram-type analysis based on the model parameters, we have predicted the statistical dependence of band gaps on the hexagonality of the crystal structure. We focussed on SiC polytypes and identified a metastable rhombohedral phase, 15$R$(1), with a higher band gap (3.44 eV at the HSE06 level) than the previously studied 9$R$ phase.

Furthermore, we have explored the feasibility of modeling band gaps at high-symmetry points of the Brillouin zone. While these models exhibit minimal errors for the overall band gap, their accuracy diminishes when predicting individual band energies at high-symmetry points. For long-period polytypes, one expects the bands connecting the band origins ($\Gamma$/$M$/$K$) and their corresponding edges ($A$/$L$/$H$) to be flat. Since our models use the information from small polytypes, significant disparities between the band origins and edges arise when applied to larger polytypes. On the other hand, for the total band gaps, the additivity model presented here delivers better accuracies than that of the reference DFT method. Hence, empirical models applied in this study can be employed fruitfully using more precise reference band gaps that incorporate effects such as many-body correlation or spin-orbit coupling.

\section{Data Availability}
The data that support the findings of this study are
within the article and its supplementary material.
Lattice parameters, electronic band structures, and stacking sequences are provided.

\section{Acknowledgments}
SJ gratefully acknowledges a summer fellowship of the TIFR Visiting Students’ Research Programme (VSRP).
We acknowledge the support of the Department of Atomic Energy, Government
of India, under Project Identification No.~RTI~4007. 
All calculations have been performed using the Helios computer cluster, 
which is an integral part of the MolDis Big Data facility, TIFR Hyderabad \href{http://moldis.tifrh.res.in}{(http://moldis.tifrh.res.in)}.

\section{Author Declarations}

\subsection{Author contributions}
{\bf RR}: Conceptualization; 
Funding acquisition; 
Project administration and supervision; 
Resources; 
Data collection and analysis; 
Writing (main).
{\bf SJ}: Preliminary investigation;
Analysis; 
Writing (supporting).

\subsection{Conflicts of Interest}
The authors have no conflicts of interest to disclose.

\bibliographystyle{apsrev}
\bibliography{aipsamp}

\end{document}


\begin{table*}[ht]
\centering
\caption{GGA-PBE equilibrium lattice constants (in \AA)
of the hexagonal unit cell 
for six polytypes of semiconductors. The polytypes are denoted as 
$pY$, where $p=2,\ldots,6$ and 
$Y$ stands for cubic ($C$), hexagonal ($H$), and trigonal ($T$). 
Note that $p$ 
for the compounds corresponds to the number of 
bilayers; for the elements $2p$ 
corresponds to the number of monolayers in the unit cell. 
For 3$C$, we have considered a
non-primitive rhombohedral unit cell using hexagonal lattice. 
 }
\small\addtolength{\tabcolsep}{1.2pt}
\begin{tabular}[t]{ll llll ll}
\hline
\multicolumn{2}{l}{Parameters}   &
\multicolumn{1}{l}{2$H$} & 
\multicolumn{1}{l}{3$C$} & 
\multicolumn{1}{l}{4$H$} & 
\multicolumn{1}{l}{5$T$} & 
\multicolumn{1}{l}{6$T$} & 
\multicolumn{1}{l}{6$H$} \\
\hline 
C&$a$    & 2.512& 2.526& 2.518& 2.520& 2.516& 2.521\\
&$c/p$   & 2.090& 2.062& 2.076& 2.074& 2.081& 2.072\\
Si&$a$   & 3.854& 3.870& 3.861& 3.863& 3.859& 3.864\\
&$c/p$   & 3.184& 3.159& 3.170& 3.168& 3.175& 3.166\\
Ge&$a$   & 4.062& 4.079& 4.071& 4.074& 4.069& 4.073\\
&$c/p$   & 3.346& 3.322& 3.332& 3.329& 3.335& 3.328\\
SiC&$a$  & 3.093& 3.098& 3.095& 3.095& 3.094& 3.096\\
&$c/p$   & 2.538& 2.529& 2.533& 2.532& 2.534& 2.532\\
GeC&$a$  & 3.248& 3.257& 3.252& 3.253& 3.251& 3.254\\
&$c/p$   & 2.672& 2.659& 2.665& 2.664& 2.667& 2.663\\
GeSi&$a$ & 3.945& 3.962& 3.953& 3.955& 3.950& 3.956\\
&$c/p$   & 3.257& 3.234& 3.244& 3.242& 3.249& 3.240\\
BN&$a$   & 2.555& 2.563& 2.559& 2.560& 2.558& 2.560\\
&$c/p$   & 2.113& 2.093& 2.103& 2.101& 2.106& 2.099\\
BP&$a$   & 3.201& 3.217& 3.208& 3.210& 3.205& 3.211\\
&$c/p$   & 2.652& 2.626& 2.638& 2.637& 2.643& 2.634\\
BAs&$a$  & 3.385& 3.401& 3.392& 3.394& 3.389& 3.395\\
&$c/p$   & 2.800& 2.777& 2.787& 2.786& 2.792& 2.784\\
AlN&$a$  & 3.131& 3.114& 3.121& 3.120& 3.124& 3.119\\
&$c/p$   & 2.509& 2.543& 2.528& 2.531& 2.522& 2.533\\
AlP&$a$  & 3.888& 3.894& 3.891& 3.892& 3.890& 3.892\\
&$c/p$   & 3.191& 3.180& 3.184& 3.184& 3.187& 3.183\\
AlAs&$a$ & 4.043& 4.052& 4.047& 4.048& 4.046& 4.049\\
&$c/p$   & 3.321& 3.307& 3.313& 3.312& 3.316& 3.311\\
GaN&$a$  & 3.218& 3.216& 3.217& 3.217& 3.218& 3.217\\
&$c/p$   & 2.622& 2.626& 2.624& 2.624& 2.623& 2.624\\
GaP&$a$  & 3.883& 3.894& 3.888& 3.890& 3.887& 3.891\\
&$c/p$   & 3.198& 3.181& 3.190& 3.187& 3.193& 3.186\\
GaAs&$a$ & 4.052& 4.065& 4.058& 4.060& 4.056& 4.061\\
&$c/p$   & 3.338& 3.317& 3.327& 3.325& 3.331& 3.324\\
\hline
\end{tabular}
\label{tab:W1}
\end{table*}%

\begin{table*}[ht]
\centering
\caption{GGA-PBE equilibrium lattice constants (in \AA)
of the hexagonal unit cell 
for nine polytypes of semiconductors. The polytypes are denoted as 
$pY$, where $p=7,\&,8$ and 
$Y$ stands for cubic ($C$), hexagonal ($H$), and trigonal ($T$). 
Note that $p$ 
for the compounds corresponds to the number of 
bilayers; for the elements $2p$ 
corresponds to the number of monolayers in the unit cell. 
 }
\small\addtolength{\tabcolsep}{1.2pt}
\begin{tabular}[t]{ll lll lll lll}
\hline
\multicolumn{2}{l}{Parameters}&
\multicolumn{1}{l}{7$T$(1) } & 
\multicolumn{1}{l}{7$T$(2) } & 
\multicolumn{1}{l}{7$T$(3) } & 
\multicolumn{1}{l}{ 8$T$(1)} & 
\multicolumn{1}{l}{ 8$H$(1)} & 
\multicolumn{1}{l}{ 8$T$(2)} & 
\multicolumn{1}{l}{ 8$T$(3)} & 
\multicolumn{1}{l}{ 8$T$(4)} & 
\multicolumn{1}{l}{ 8$H$(2)} \\
\hline 
C&$a$    & 2.518& 2.518& 2.521& 2.515& 2.515& 2.519& 2.522& 2.518& 2.522\\
&$c/p$   & 2.078& 2.078& 2.070& 2.083& 2.083& 2.076& 2.069& 2.076& 2.069\\
Si&$a$   & 3.861& 3.860& 3.865& 3.858& 3.858& 3.862& 3.866& 3.862& 3.866\\
&$c/p$   & 3.173& 3.172& 3.165& 3.177& 3.177& 3.171& 3.165& 3.170& 3.164\\
Ge&$a$   & 4.070& 4.071& 4.075& 4.068& 4.069& 4.072& 4.076& 4.072& 4.075\\
&$c/p$   & 3.334& 3.333& 3.327& 3.337& 3.337& 3.332& 3.326& 3.331& 3.327\\
SiC&$a$  & 3.095& 3.095& 3.096& 3.094& 3.094& 3.095& 3.096& 3.095& 3.096\\
&$c/p$   & 2.534& 2.534& 2.531& 2.535& 2.535& 2.533& 2.531& 2.533& 2.531\\
GeC&$a$  & 3.252& 3.252& 3.254& 3.250& 3.250& 3.252& 3.255& 3.252& 3.255\\
&$c/p$   & 2.666& 2.666& 2.663& 2.668& 2.668& 2.665& 2.662& 2.665& 2.662\\
GeSi&$a$ & 3.952& 3.952& 3.957& 3.949& 3.949& 3.953& 3.957& 3.953& 3.957\\
&$c/p$   & 3.247& 3.246& 3.240& 3.251& 3.251& 3.245& 3.239& 3.244& 3.239\\
BN&$a$   & 2.559& 2.559& 2.561& 2.557& 2.557& 2.559& 2.561& 2.559& 2.561\\
&$c/p$   & 2.104& 2.104& 2.098& 2.108& 2.108& 2.103& 2.098& 2.103& 2.098\\
BP&$a$   & 3.207& 3.207& 3.211& 3.204& 3.204& 3.208& 3.212& 3.208& 3.212\\
&$c/p$   & 2.641& 2.641& 2.633& 2.646& 2.646& 2.639& 2.633& 2.639& 2.632\\
BAs&$a$  & 3.391& 3.391& 3.396& 3.388& 3.388& 3.392& 3.397& 3.392& 3.396\\
&$c/p$   & 2.790& 2.790& 2.783& 2.794& 2.794& 2.788& 2.782& 2.788& 2.782\\
AlN&$a$  & 3.123& 3.123& 3.118& 3.126& 3.126& 3.121& 3.117& 3.121& 3.117\\
&$c/p$   & 2.526& 2.526& 2.535& 2.519& 2.519& 2.528& 2.536& 2.528& 2.536\\
AlP&$a$  & 3.891& 3.890& 3.892& 3.889& 3.889& 3.891& 3.893& 3.891& 3.893\\
&$c/p$   & 3.186& 3.186& 3.183& 3.188& 3.188& 3.185& 3.182& 3.185& 3.182\\
AlAs&$a$ & 4.047& 4.047& 4.049& 4.045& 4.045& 4.047& 4.050& 4.047& 4.050\\
&$c/p$   & 3.315& 3.315& 3.311& 3.317& 3.317& 3.314& 3.310& 3.314& 3.310\\
GaN&$a$  & 3.217& 3.217& 3.217& 3.218& 3.218& 3.217& 3.217& 3.217& 3.217\\
&$c/p$   & 2.623& 2.623& 2.624& 2.622& 2.622& 2.624& 2.625& 2.623& 2.625\\
GaP&$a$  & 3.888& 3.888& 3.891& 3.886& 3.886& 3.889& 3.892& 3.889& 3.892\\
&$c/p$   & 3.190& 3.190& 3.185& 3.194& 3.194& 3.189& 3.184& 3.189& 3.184\\
GaAs&$a$ & 4.058& 4.058& 4.061& 4.056& 4.055& 4.059& 4.062& 4.059& 4.062\\
&$c/p$   & 3.329& 3.329& 3.323& 3.332& 3.333& 3.327& 3.322& 3.327& 3.322\\
\hline
\end{tabular}
\label{tab:W1}
\end{table*}%


\clearpage
\begin{figure*}[!hbtp]
    \centering
    \includegraphics[width=1.0\linewidth]{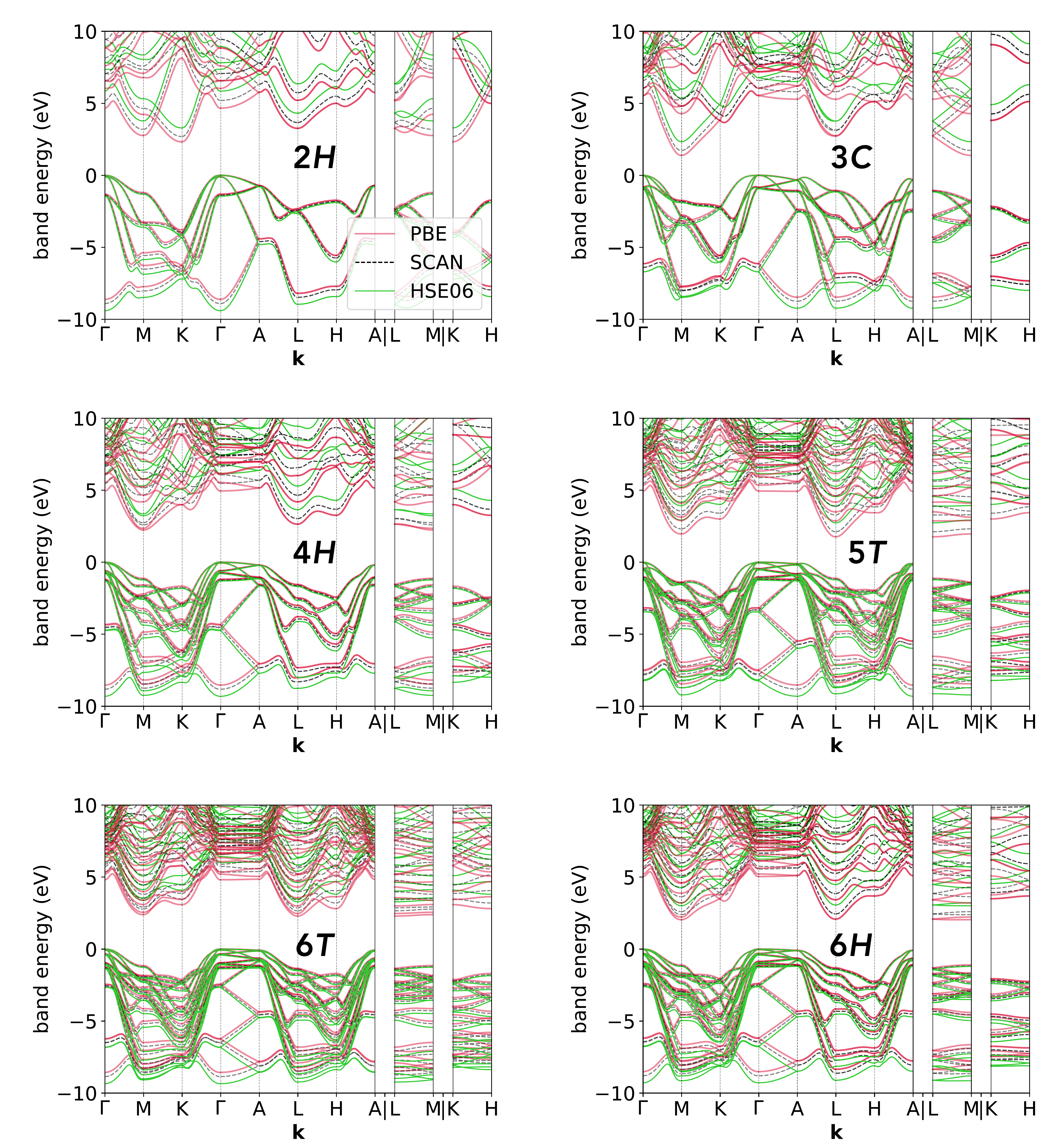}
    \caption{
    Electronic band structures of simplest polytypes of SiC 
    calculated with PBE, SCAN and HSE06 DFT methods.
    }
    \label{fig:predband}
\end{figure*}

\clearpage
\begin{figure*}[!hbtp]
    \centering
    \includegraphics[width=0.7\linewidth]{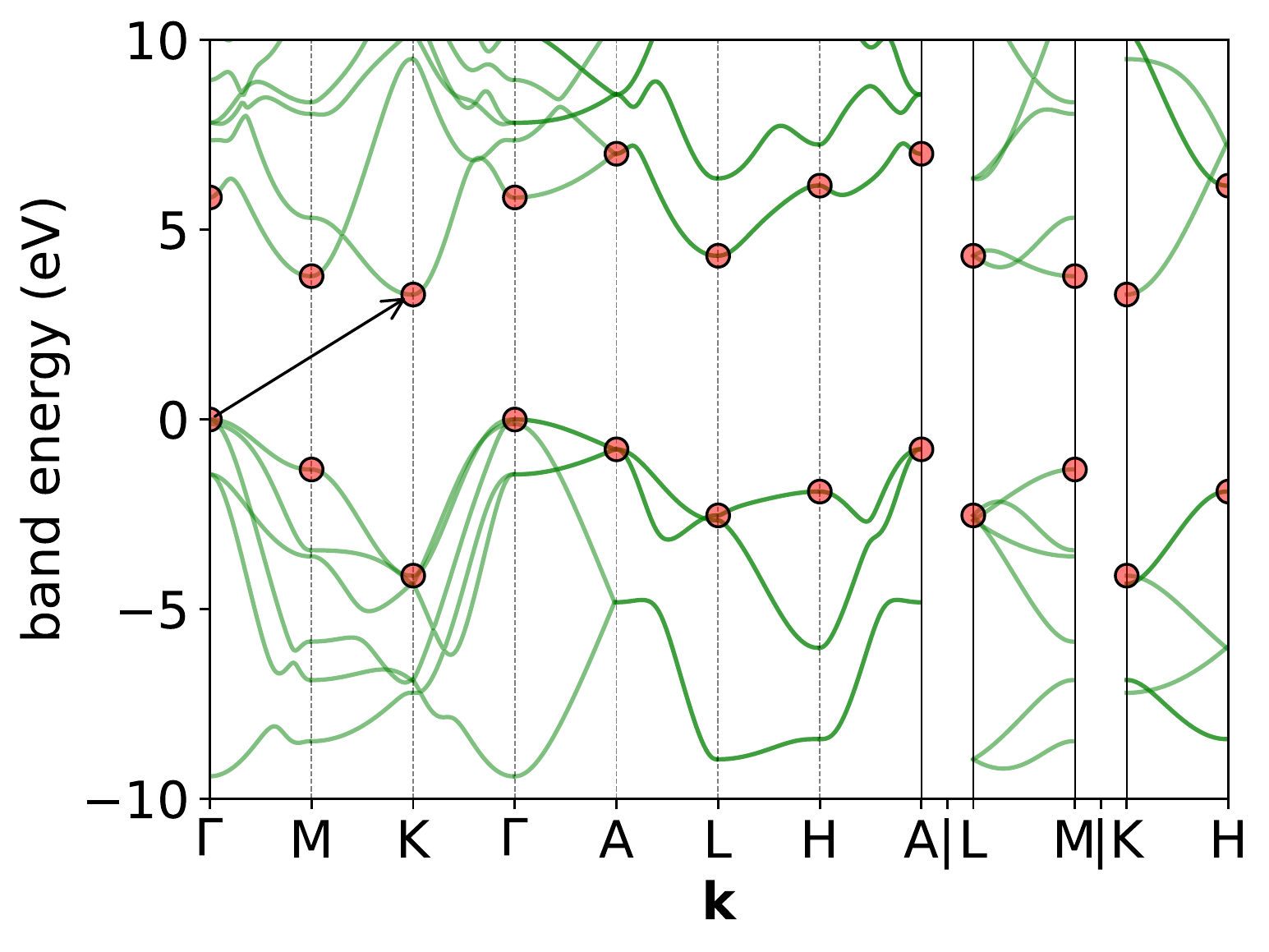}
    \caption{
    HSE06 electronic band structures of the 2$H$-SiC along with VB and CB energies at the high-symmetry points used in a model (red points).    }
    \label{fig:predband}
\end{figure*}

\begin{figure*}[!hbtp]
    \centering
    \includegraphics[width=0.7\linewidth]{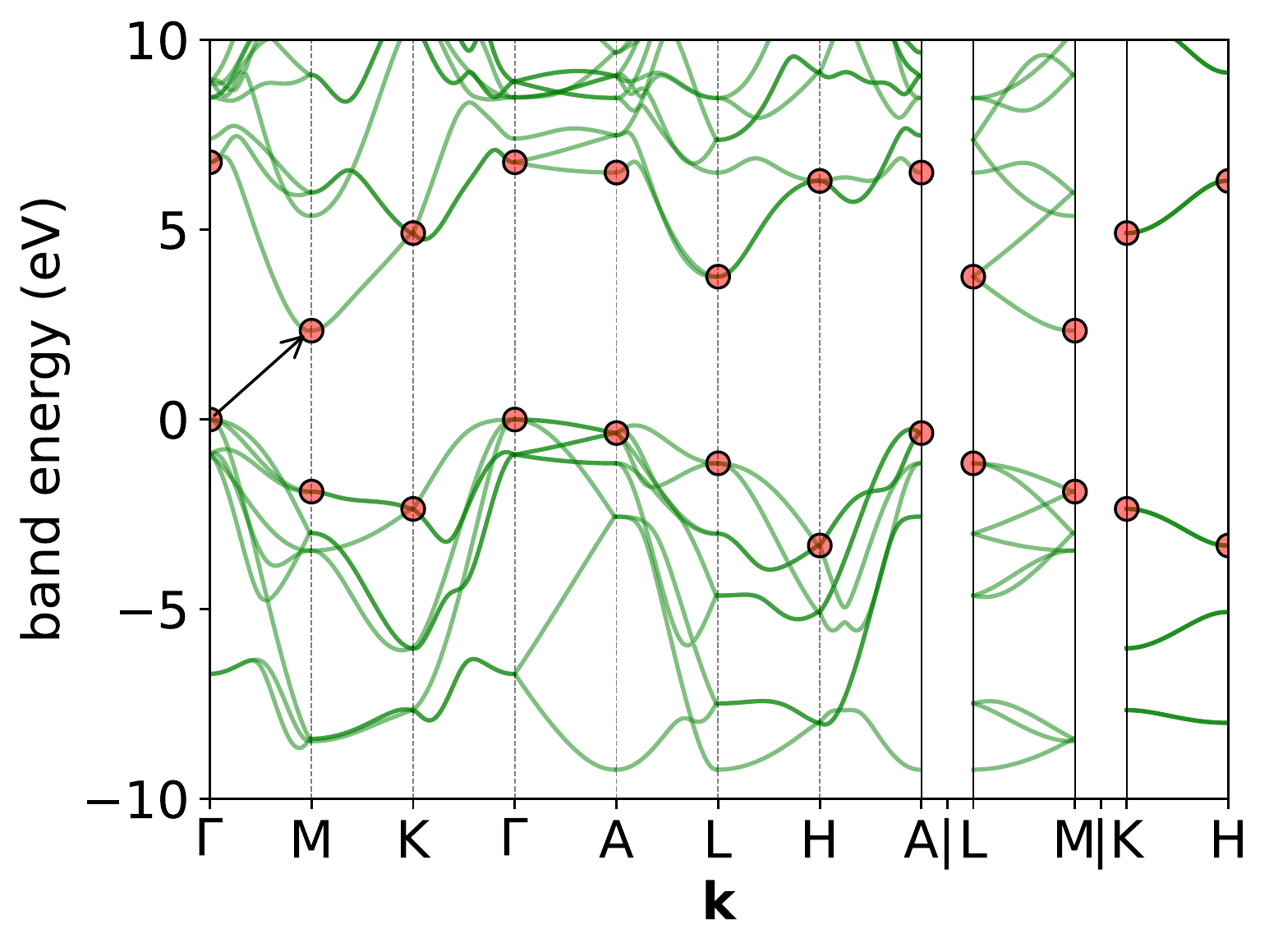}
    \caption{
    HSE06 electronic band structures of the 3$C$-SiC along with VB and CB energies at the high-symmetry points used in a model (red points).    }
    \label{fig:predband}
\end{figure*}

\begin{figure*}[!hbtp]
    \centering
    \includegraphics[width=0.7\linewidth]{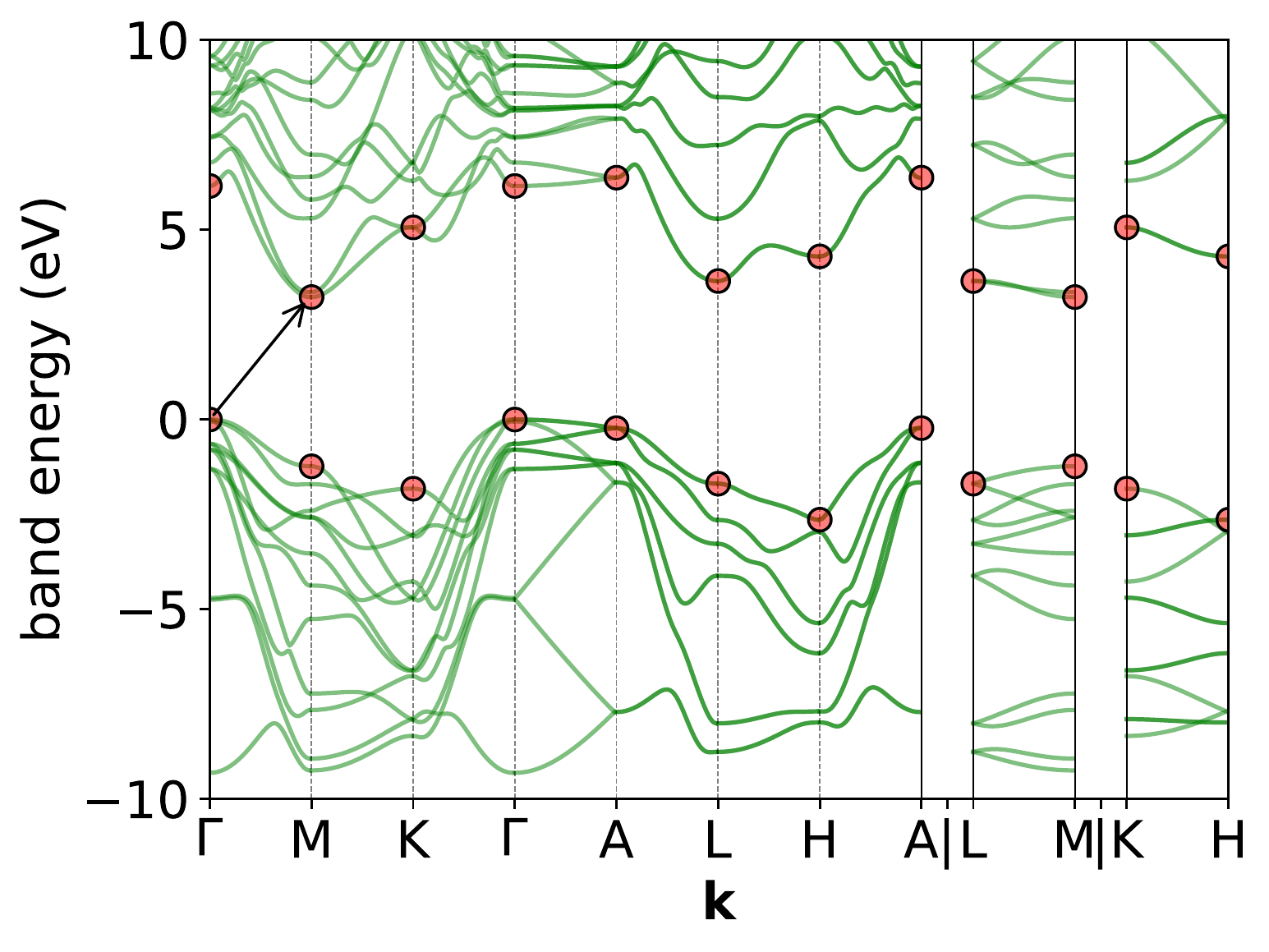}
    \caption{
    HSE06 electronic band structures of the 4$H$-SiC along with VB and CB energies at the high-symmetry points used in a model (red points).    }
    \label{fig:predband}
\end{figure*}

\begin{figure*}[!hbtp]
    \centering
    \includegraphics[width=0.7\linewidth]{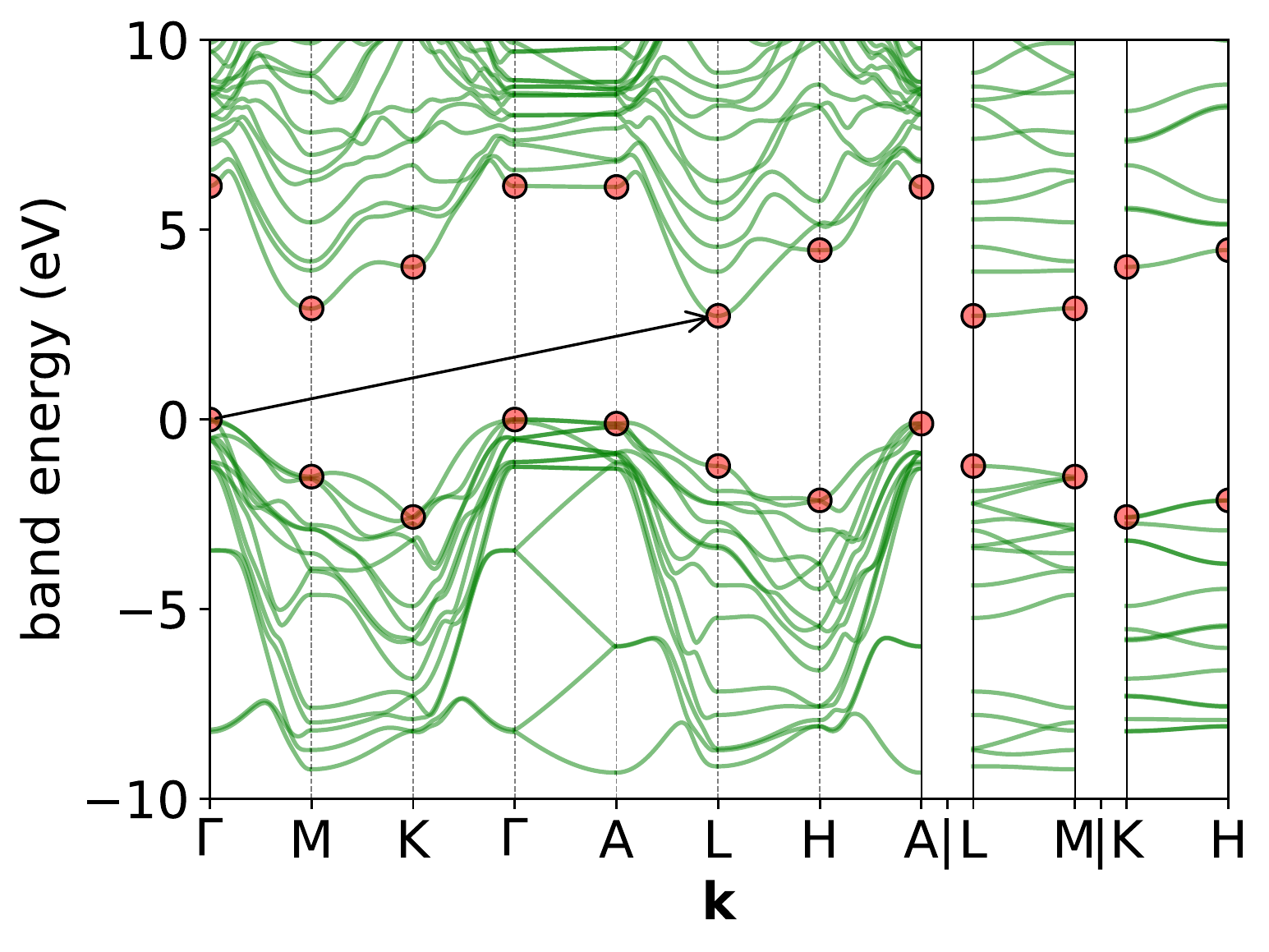}
    \caption{
    HSE06 electronic band structures of the 5$T$-SiC along with VB and CB energies at the high-symmetry points used in a model (red points).    }
    \label{fig:predband}
\end{figure*}

\begin{figure*}[!hbtp]
    \centering
    \includegraphics[width=0.7\linewidth]{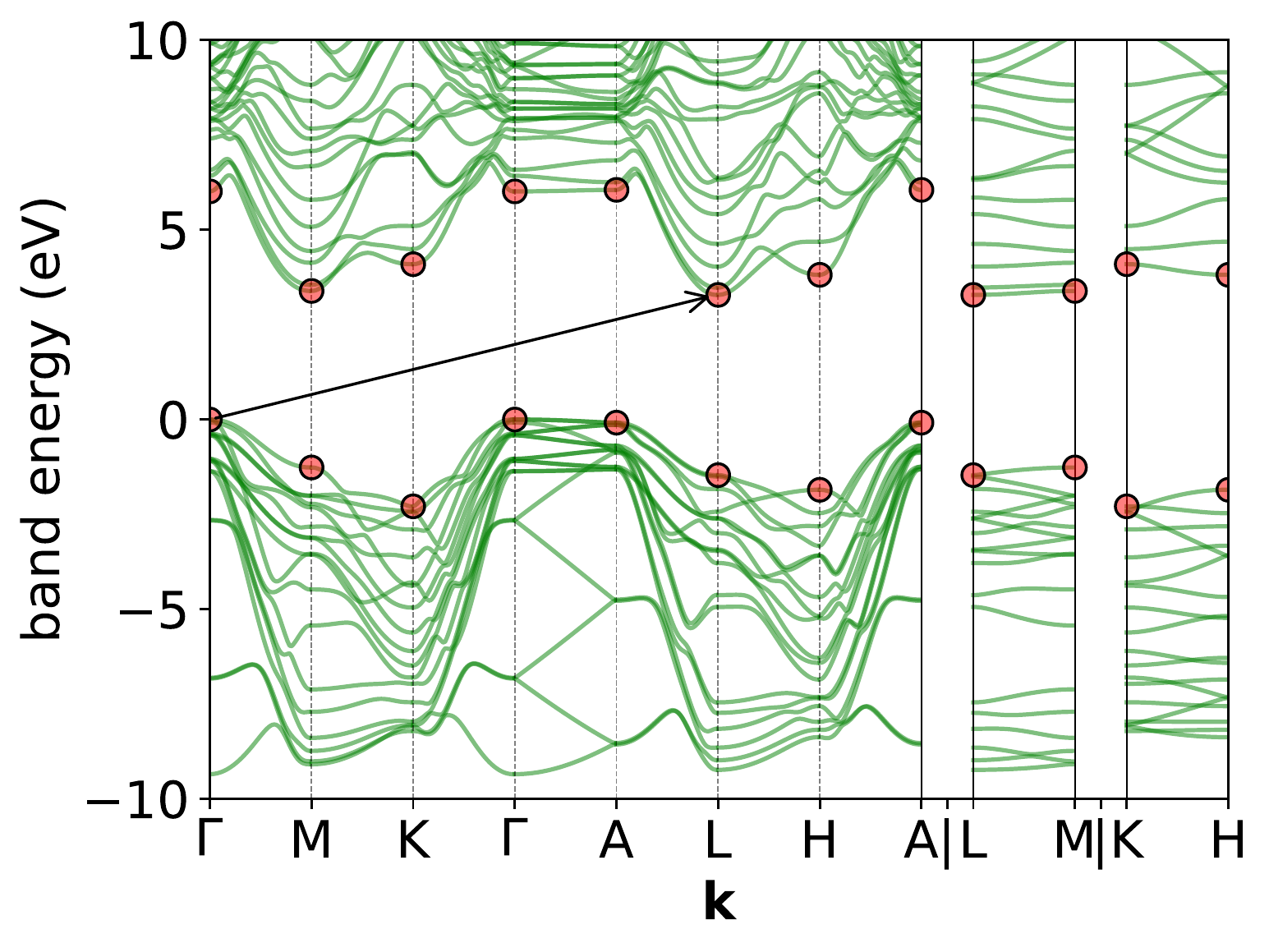}
    \caption{
    HSE06 electronic band structures of the 6$T$-SiC along with VB and CB energies at the high-symmetry points used in a model (red points).
    }
    \label{fig:predband}
\end{figure*}

\clearpage
\begin{figure*}[!hbtp]
    \centering
    \includegraphics[width=0.7\linewidth]{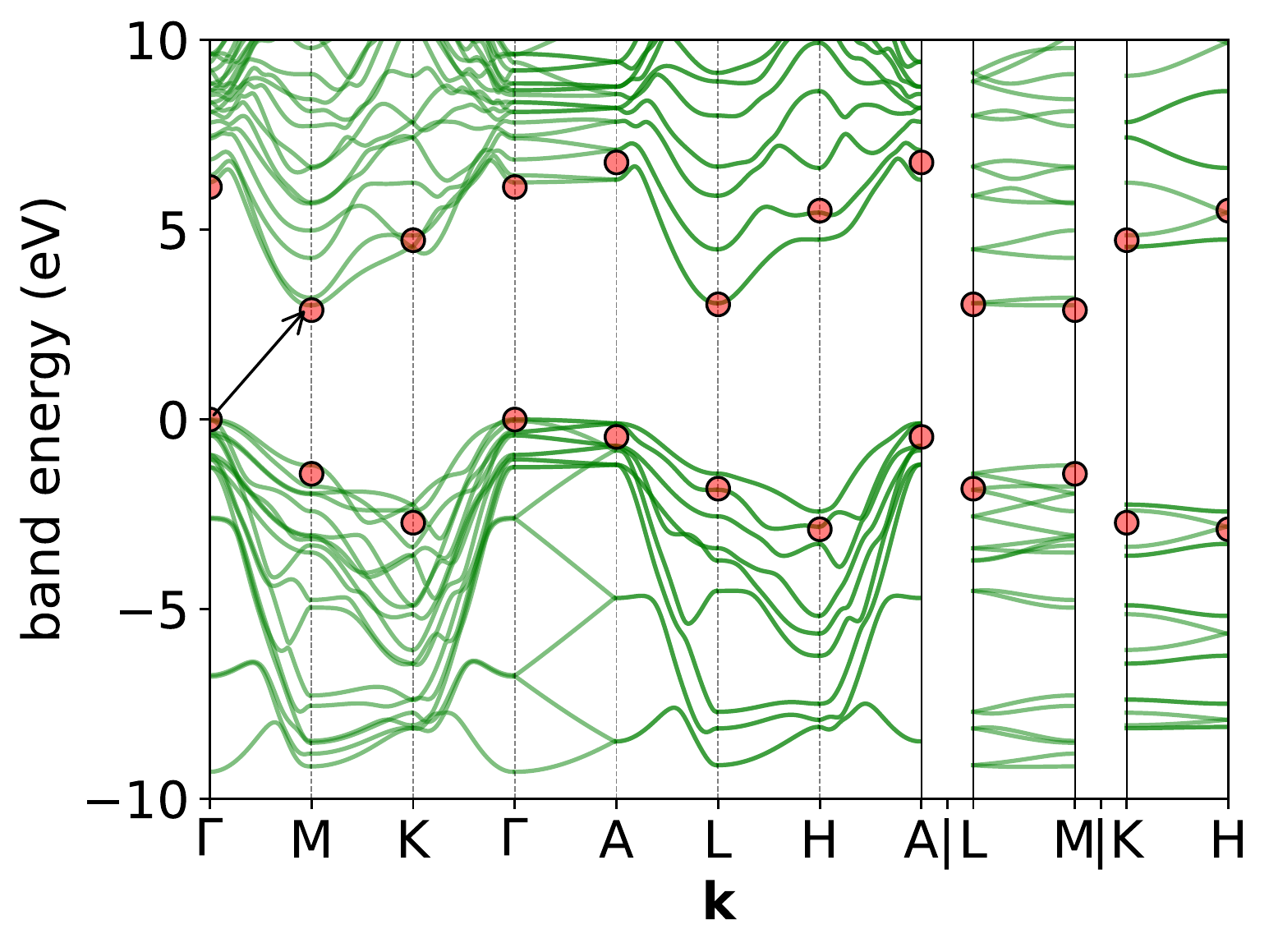}
    \caption{
    HSE06 electronic band structures of the 6$H$-SiC along with VB and CB energies at the high-symmetry points predicted with a model (red points).
    }
    \label{fig:predband}
\end{figure*}

\begin{figure*}[!hbtp]
    \centering
    \includegraphics[width=0.7\linewidth]{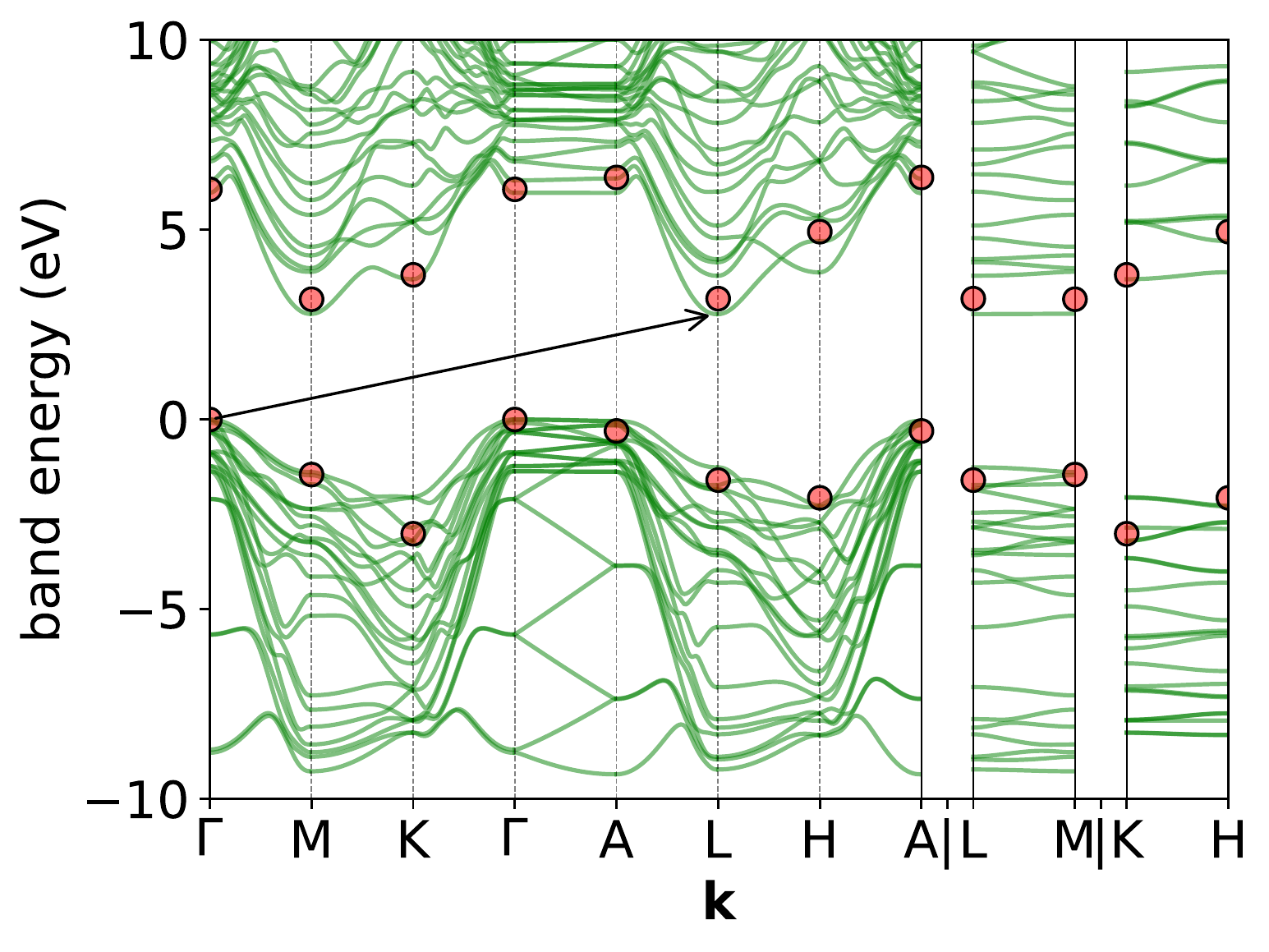}
    \caption{
    HSE06 electronic band structures of the 7$T$(1)-SiC along with VB and CB energies at the high-symmetry points predicted with a model (red points).
    }
    \label{fig:predband}
\end{figure*}

\begin{figure*}[!hbtp]
    \centering
    \includegraphics[width=0.7\linewidth]{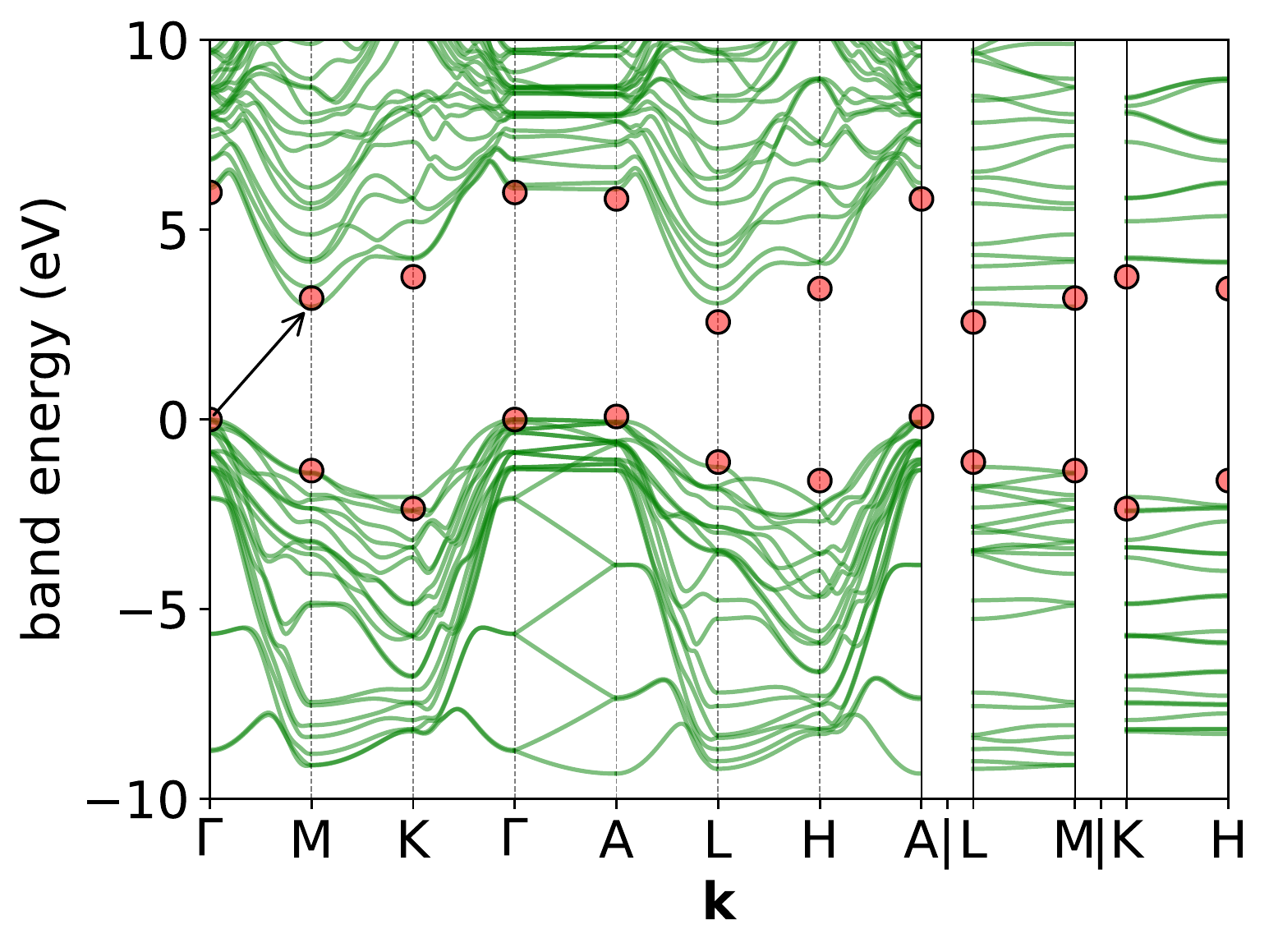}
    \caption{
    HSE06 electronic band structures of the 7$T$(2)-SiC along with VB and CB energies at the high-symmetry points predicted with a model (red points).
    }
    \label{fig:predband}
\end{figure*}

\begin{figure*}[!hbtp]
    \centering
    \includegraphics[width=0.7\linewidth]{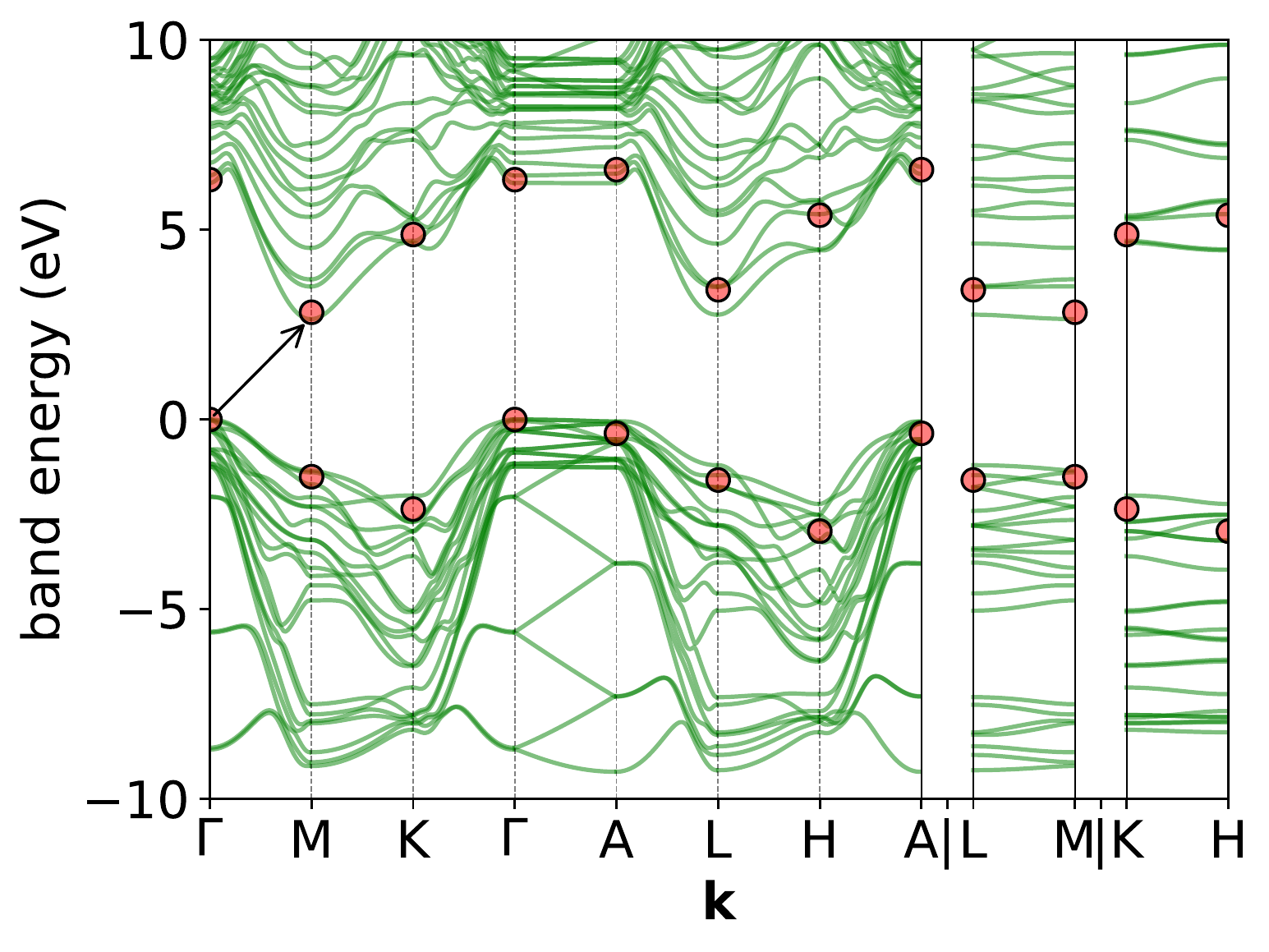}
    \caption{
    HSE06 electronic band structures of the 7$T$(3)-SiC along with VB and CB energies at the high-symmetry points predicted with a model (red points).
    }
    \label{fig:predband}
\end{figure*}

\begin{figure*}[!hbtp]
    \centering
    \includegraphics[width=0.7\linewidth]{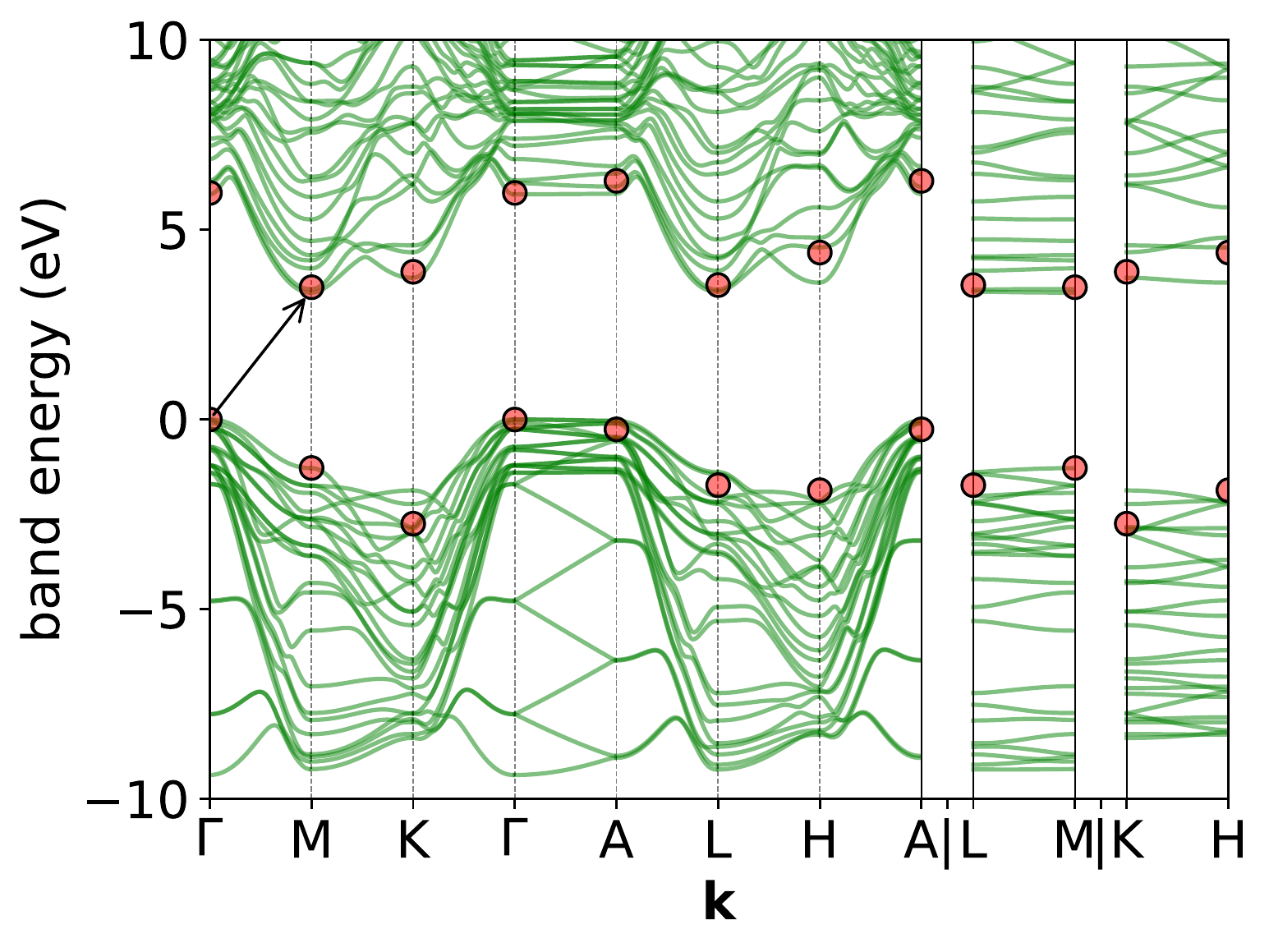}
    \caption{
    HSE06 electronic band structures of the 8$T$(1)-SiC along with VB and CB energies at the high-symmetry points predicted with a model (red points).
    }
    \label{fig:predband}
\end{figure*}

\begin{figure*}[!hbtp]
    \centering
    \includegraphics[width=0.7\linewidth]{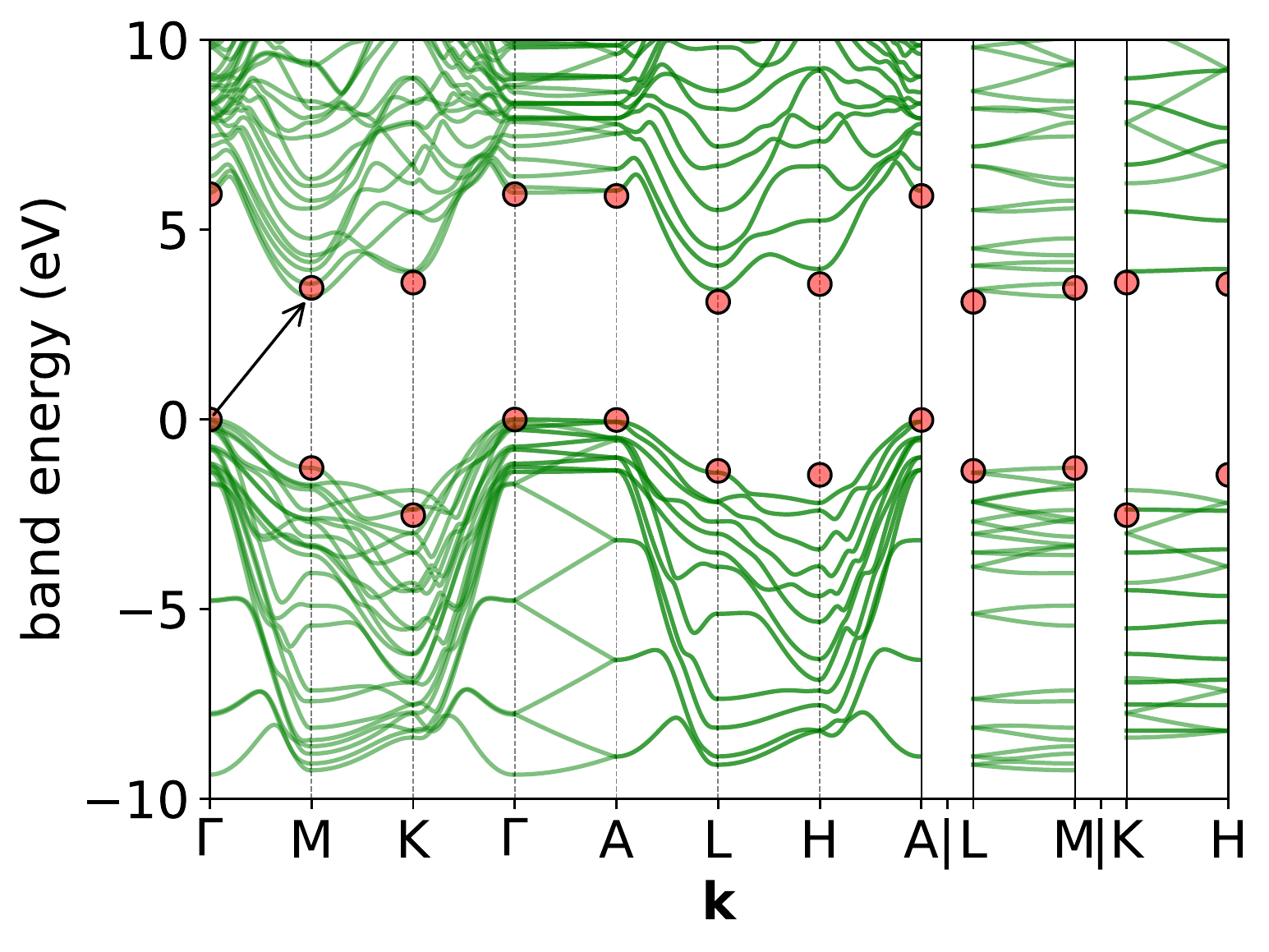}
    \caption{
    HSE06 electronic band structures of the 8$H$(1)-SiC along with VB and CB energies at the high-symmetry points predicted with a model (red points).
    }
    \label{fig:predband}
\end{figure*}

\begin{figure*}[!hbtp]
    \centering
    \includegraphics[width=0.7\linewidth]{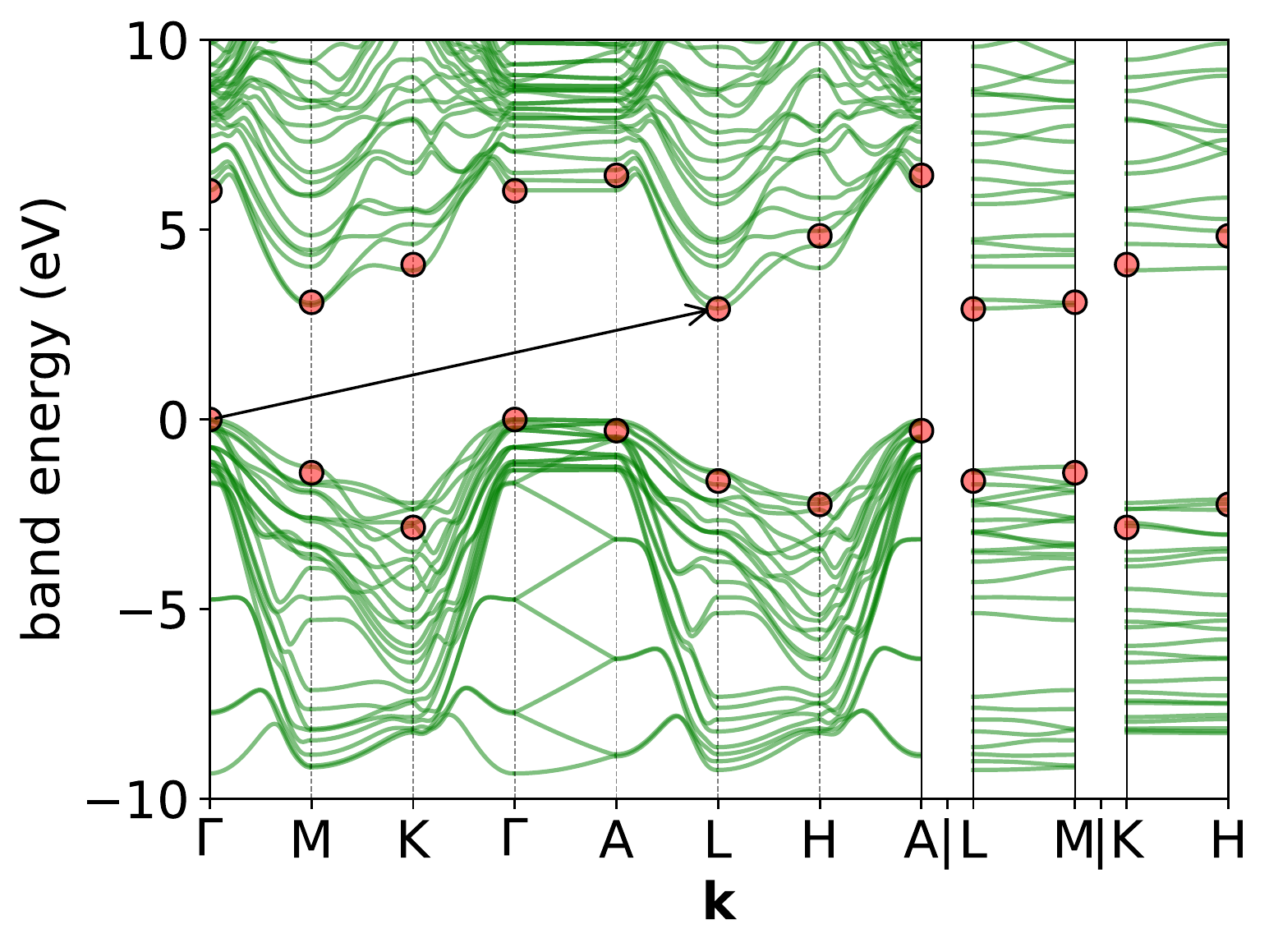}
    \caption{
    HSE06 electronic band structures of the 8$T$(2)-SiC along with VB and CB energies at the high-symmetry points predicted with a model (red points).
    }
    \label{fig:predband}
\end{figure*}

\begin{figure*}[!hbtp]
    \centering
    \includegraphics[width=0.7\linewidth]{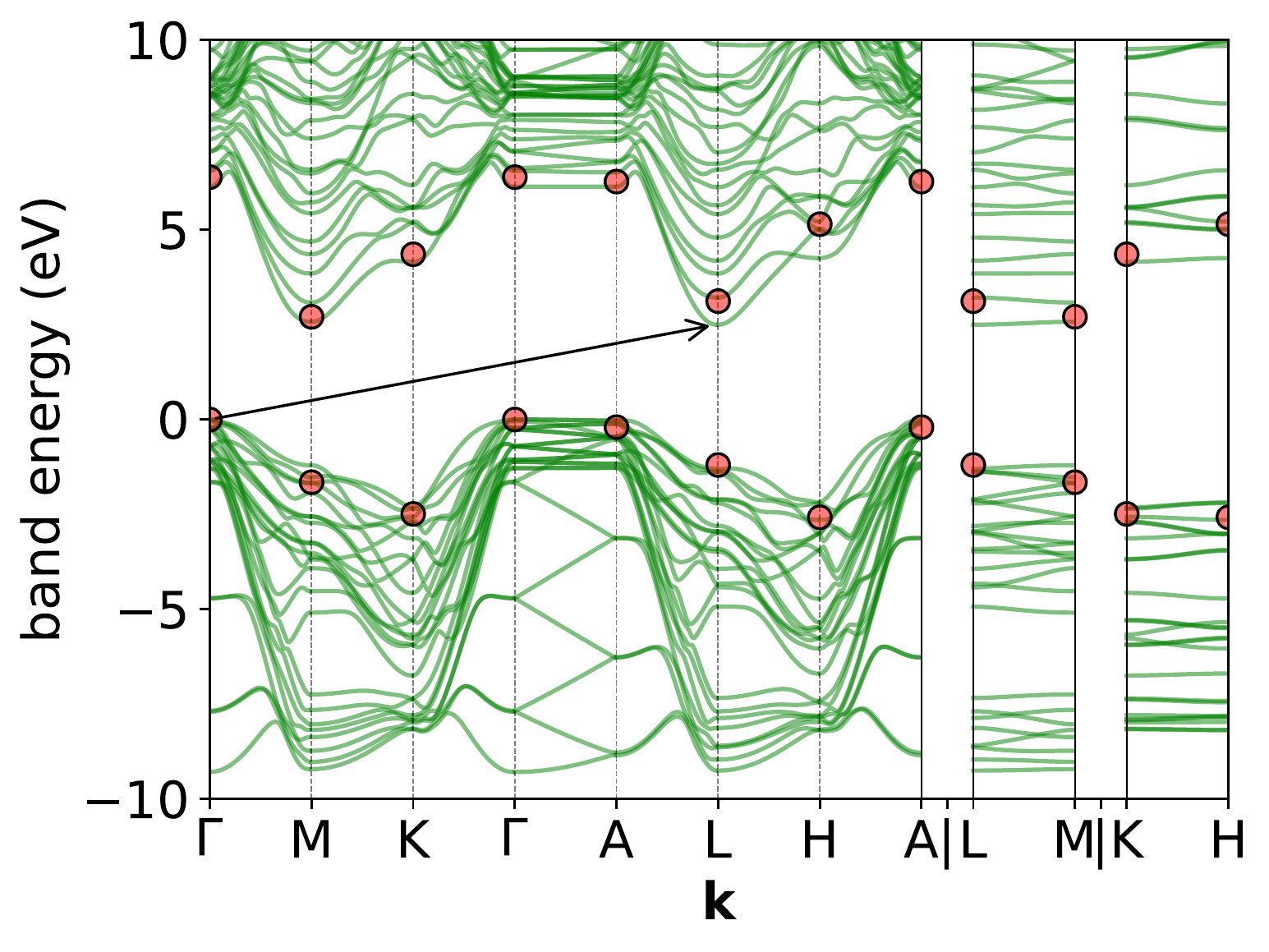}
    \caption{
    HSE06 electronic band structures of the 8$T$(3)-SiC along with VB and CB energies at the high-symmetry points predicted with a model (red points).
    }
    \label{fig:predband}
\end{figure*}

\begin{figure*}[!hbtp]
    \centering
    \includegraphics[width=0.7\linewidth]{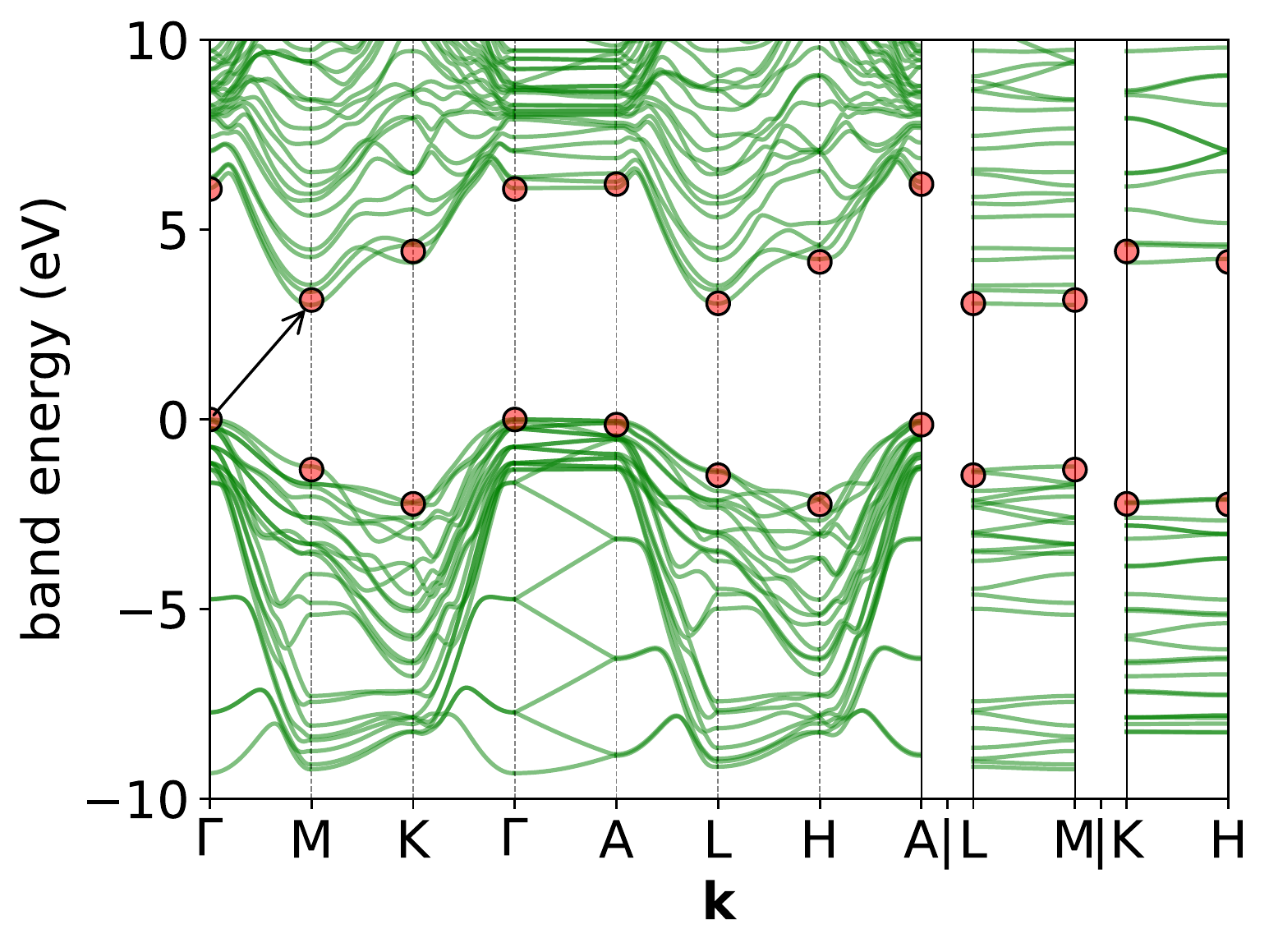}
    \caption{
    HSE06 electronic band structure of 8$T$(4)-SiC
    }
    \label{fig:predband}
\end{figure*}

\begin{figure*}[!hbtp]
    \centering
    \includegraphics[width=0.7\linewidth]{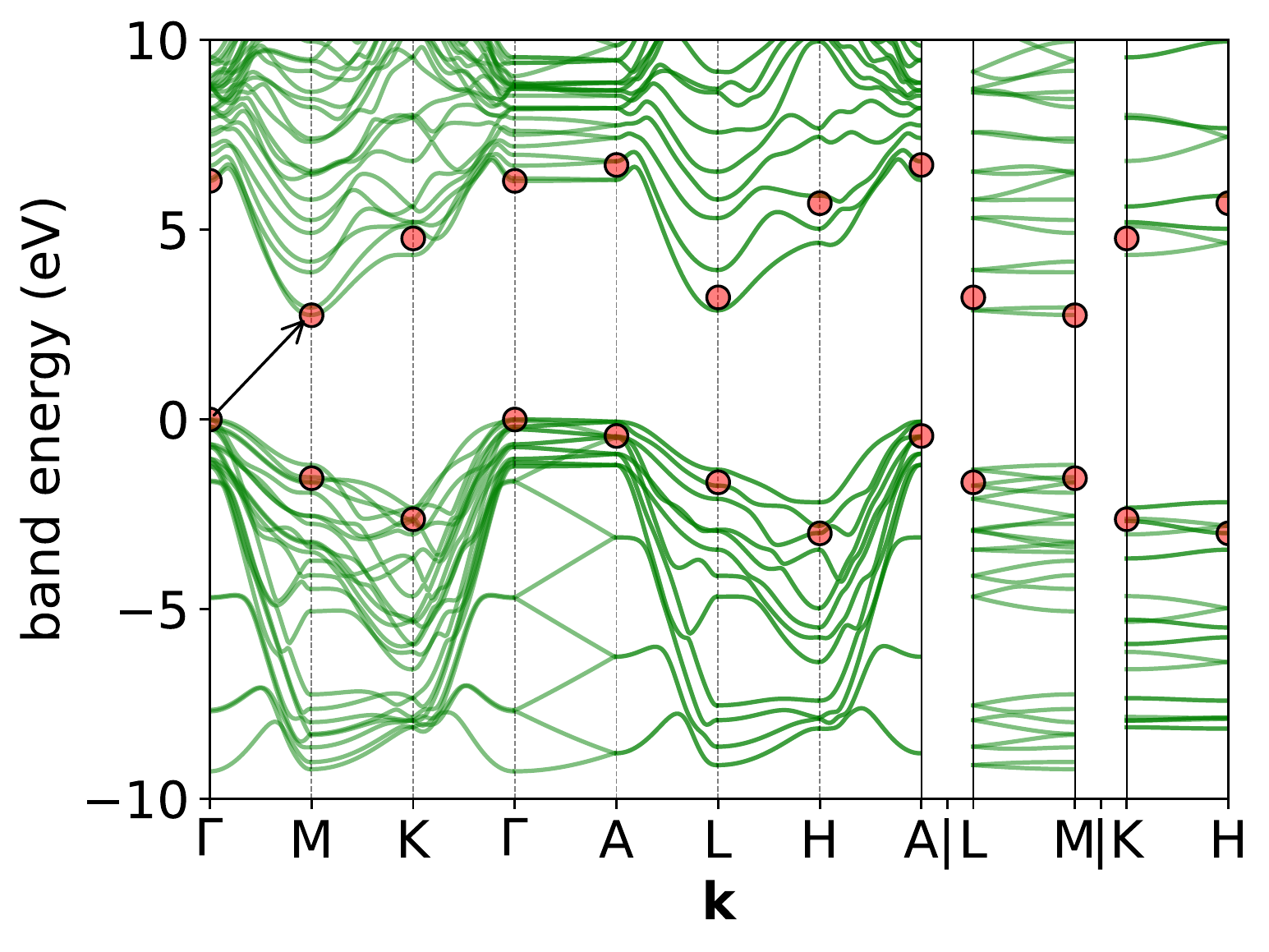}
    \caption{
    HSE06 electronic band structures of the 8$H$(2)-SiC along with VB and CB energies at the high-symmetry points predicted with a model (red points). 
    }
    \label{fig:predband}
\end{figure*}

\clearpage
\noindent{\bf List S1:} Details of polytypes with up to 15 layers 
in the unit cell. Space group names and numbers are given for AB compound
polytypes followed by elemental polytypes. 
\singlespacing
{\tt \noindent 
1    2   AB               hh               2H   P6\_3mc  (186)  2H   P6\_3/mmc  (194)  \\
2    3   ABC              kkk              3C   F4-3m    (216)  3C   Fd-3m      (227)  \\
3    4   ABAC             hkhk             4H   P6\_3mc  (186)  4H   P6\_3/mmc  (194)  \\
4    5   ABABC            hhkkk            5T   P3m1     (156)  5T   P-3m1      (164)  \\
5    6   ABABAC           hhhkhk           6T   P3m1     (156)  6H   P-6m2      (187)  \\
6    6   ABACBC           hkkhkk           6H   P6\_3mc  (186)  6H   P6\_3/mmc  (194)  \\
7    7   ABABABC          hhhhkkk          7T   P3m1     (156)  7T   P-3m1      (164)  \\
8    7   ABABCAC          hhkhhkk          7T   P3m1     (156)  7T   P-3m1      (164)  \\
9    7   ABACABC          hkhkkkk          7T   P3m1     (156)  7T   P-3m1      (164)  \\
10   8   ABABABAC         hhhhhkhk         8T   P3m1     (156)  8H   P-6m2      (187)  \\
11   8   ABABACAC         hhhkhhhk         8H   P6\_3mc  (186)  8H   P6\_3/mmc  (194)  \\
12   8   ABABACBC         hhhkkhkk         8T   P3m1     (156)  8H   P-6m2      (187)  \\
13   8   ABABCABC         hhkkkkkk         8T   P3m1     (156)  8T   P-3m1      (164)  \\
14   8   ABABCBAC         hhkhkkhk         8T   P3m1     (156)  8T   P-3m1      (164)  \\
15   8   ABACBABC         hkkkhkkk         8H   P6\_3mc  (186)  8H   P6\_3/mmc  (194)  \\
16   9   ABABABABC        hhhhhhkkk        9T   P3m1     (156)  9T   P-3m1      (164)  \\
17   9   ABABABCAC        hhhhkhhkk        9T   P3m1     (156)  9T   P3m1       (156)  \\
18   9   ABABACABC        hhhkhkkkk        9T   P3m1     (156)  9T   P3m1       (156)  \\
19   9   ABABCABAC        hhkhkhkkk        9T   P3m1     (156)  9T   P3m1       (156)  \\
20   9   ABABCACBC        hhkkhkhkk        9T   P3m1     (156)  9T   P-3m1      (164)  \\
21   9   ABABCBCAC        hhkhhkhhk        9R   R3m      (160)  9R   R-3m       (166)  \\
22   9   ABACBACBC        hkkhkkkkk        9T   P3m1     (156)  9T   P-3m1      (164)  \\
23   10  ABABABABAC       hhhhhhhkhk       10T  P3m1     (156)  10H  P-6m2      (187)  \\
24   10  ABABABACAC       hhhhhkhhhk       10T  P3m1     (156)  10H  P-6m2      (187)  \\
25   10  ABABABACBC       hhhhhkkhkk       10T  P3m1     (156)  10H  P-6m2      (187)  \\
26   10  ABABABCABC       hhhhkkkkkk       10T  P3m1     (156)  10T  P-3m1      (164)  \\
27   10  ABABABCBAC       hhhhkhkkhk       10T  P3m1     (156)  10T  P-3m1      (164)  \\
28   10  ABABACABAC       hhhkhkhkhk       10T  P3m1     (156)  10H  P-6m2      (187)  \\
29   10  ABABACACBC       hhhkhhkhkk       10T  P3m1     (156)  10T  P3m1       (156)  \\
30   10  ABABACBABC       hhhkkkhkkk       10T  P3m1     (156)  10H  P-6m2      (187)  \\
31   10  ABABACBCBC       hhhkkhhhkk       10H  P6\_3mc  (186)  10H  P6\_3/mmc  (194)  \\
32   10  ABABCABCAC       hhkhhkkkkk       10T  P3m1     (156)  10T  P-3m1      (164)  \\
33   10  ABABCABCBC       hhkkhhkkkk       10T  P3m1     (156)  10T  P-3m1      (164)  \\
34   10  ABABCACBAC       hhkhkkkhkk       10T  P3m1     (156)  10T  P3m1       (156)  \\
35   10  ABABCBABAC       hhkhkhhkhk       10H  P6\_3mc  (186)  10H  P6\_3/mmc  (194)  \\
36   10  ABACABACBC       hkhkhkkhkk       10T  P3m1     (156)  10H  P-6m2      (187)  \\
37   10  ABACABCABC       hkhkkkkkkk       10T  P3m1     (156)  10T  P-3m1      (164)  \\
38   10  ABACBACABC       hkkkkhkkkk       10H  P6\_3mc  (186)  10H  P6\_3/mmc  (194)  \\
39   11  ABABABABABC      hhhhhhhhkkk      11T  P3m1     (156)  11T  P-3m1      (164)  \\
40   11  ABABABABCAC      hhhhhhkhhkk      11T  P3m1     (156)  11T  P3m1       (156)  \\
41   11  ABABABACABC      hhhhhkhkkkk      11T  P3m1     (156)  11T  P3m1       (156)  \\
42   11  ABABABCABAC      hhhhkhkhkkk      11T  P3m1     (156)  11T  P3m1       (156)  \\
43   11  ABABABCACAC      hhhhkhhhhkk      11T  P3m1     (156)  11T  P-3m1      (164)  \\
44   11  ABABABCACBC      hhhhkkhkhkk      11T  P3m1     (156)  11T  P-3m1      (164)  \\
45   11  ABABABCBCAC      hhhhkhhkhhk      11T  P3m1     (156)  11T  P-3m1      (164)  \\
46   11  ABABACABABC      hhhkhkhhkkk      11T  P3m1     (156)  11T  P3m1       (156)  \\
47   11  ABABACABCAC      hhhkhhkkkhk      11T  P3m1     (156)  11T  P3m1       (156)  \\
48   11  ABABACABCBC      hhhkhkkhhkk      11T  P3m1     (156)  11T  P3m1       (156)  \\
49   11  ABABACACABC      hhhkhhhkkkk      11T  P3m1     (156)  11T  P-3m1      (164)  \\
50   11  ABABACBACBC      hhhkkhkkkkk      11T  P3m1     (156)  11T  P3m1       (156)  \\
51   11  ABABCABACBC      hhkkhkkhkkk      11T  P3m1     (156)  11T  P3m1       (156)  \\
52   11  ABABCABCABC      hhkkkkkkkkk      11T  P3m1     (156)  11T  P-3m1      (164)  \\
53   11  ABABCABCBAC      hhkhkkhkkkk      11T  P3m1     (156)  11T  P3m1       (156)  \\
54   11  ABABCACABAC      hhkhkhkhhkk      11T  P3m1     (156)  11T  P-3m1      (164)  \\
55   11  ABABCACBCAC      hhkhhkhkhkk      11T  P3m1     (156)  11T  P3m1       (156)  \\
56   11  ABABCBACBAC      hhkhkkkkkhk      11T  P3m1     (156)  11T  P-3m1      (164)  \\
57   11  ABACABACABC      hkhkhkhkkkk      11T  P3m1     (156)  11T  P-3m1      (164)  \\
58   11  ABACABCACBC      hkhkkhkhkkk      11T  P3m1     (156)  11T  P-3m1      (164)  \\
59   11  ABACBABCABC      hkkkhkkkkkk      11T  P3m1     (156)  11T  P-3m1      (164)  \\
60   12  ABABABABABAC     hhhhhhhhhkhk     12T  P3m1     (156)  12H  P-6m2      (187)  \\
61   12  ABABABABACAC     hhhhhhhkhhhk     12T  P3m1     (156)  12H  P-6m2      (187)  \\
62   12  ABABABABACBC     hhhhhhhkkhkk     12T  P3m1     (156)  12H  P-6m2      (187)  \\
63   12  ABABABABCABC     hhhhhhkkkkkk     12T  P3m1     (156)  12T  P-3m1      (164)  \\
64   12  ABABABABCBAC     hhhhhhkhkkhk     12T  P3m1     (156)  12T  P-3m1      (164)  \\
65   12  ABABABACABAC     hhhhhkhkhkhk     12T  P3m1     (156)  12H  P-6m2      (187)  \\
66   12  ABABABACACAC     hhhhhkhhhhhk     12H  P6\_3mc  (186)  12H  P6\_3/mmc  (194)  \\
67   12  ABABABACACBC     hhhhhkhhkhkk     12T  P3m1     (156)  12T  P3m1       (156)  \\
68   12  ABABABACBABC     hhhhhkkkhkkk     12T  P3m1     (156)  12H  P-6m2      (187)  \\
69   12  ABABABACBCBC     hhhhhkkhhhkk     12T  P3m1     (156)  12H  P-6m2      (187)  \\
70   12  ABABABCABABC     hhhhkkkhhkkk     12T  P3m1     (156)  12T  P-3m1      (164)  \\
71   12  ABABABCABCAC     hhhhkhhkkkkk     12T  P3m1     (156)  12T  P3m1       (156)  \\
72   12  ABABABCABCBC     hhhhkkhhkkkk     12T  P3m1     (156)  12T  P3m1       (156)  \\
73   12  ABABABCACBAC     hhhhkhkkkhkk     12T  P3m1     (156)  12T  P3m1       (156)  \\
74   12  ABABABCBABAC     hhhhkhkhhkhk     12T  P3m1     (156)  12T  P-3m1      (164)  \\
75   12  ABABABCBACAC     hhhhkhhhkkhk     12T  P3m1     (156)  12T  P3m1       (156)  \\
76   12  ABABACABACAC     hhhkhhhkhkhk     12T  P3m1     (156)  12T  P-3m1      (164)  \\
77   12  ABABACABACBC     hhhkhkhkkhkk     12T  P3m1     (156)  12T  P3m1       (156)  \\
78   12  ABABACABCABC     hhhkhkkkkkkk     12T  P3m1     (156)  12T  P3m1       (156)  \\
79   12  ABABACABCBAC     hhhkhkkhkkhk     12T  P3m1     (156)  12H  P-6m2      (187)  \\
80   12  ABABACACBABC     hhhkhhkkhkkk     12T  P3m1     (156)  12T  P3m1       (156)  \\
81   12  ABABACACBCAC     hhhkhhkhkhhk     12T  P3m1     (156)  12H  P-6m2      (187)  \\
82   12  ABABACACBCBC     hhhkhhkhhhkk     12T  P3m1     (156)  12T  P-3m1      (164)  \\
83   12  ABABACBABABC     hhhkkkhhhkkk     12H  P6\_3mc  (186)  12H  P6\_3/mmc  (194)  \\
84   12  ABABACBABCBC     hhhkkhhkhkkk     12T  P3m1     (156)  12T  P3m1       (156)  \\
85   12  ABABACBACABC     hhhkkkkhkkkk     12T  P3m1     (156)  12H  P-6m2      (187)  \\
86   12  ABABACBCACBC     hhhkkhkhkhkk     12T  P3m1     (156)  12H  P-6m2      (187)  \\
87   12  ABABCABABCAC     hhkhhkkhhkkk     12T  P3m1     (156)  12T  P3m1       (156)  \\
88   12  ABABCABACABC     hhkkkhkhkkkk     12T  P3m1     (156)  12T  P3m1       (156)  \\
89   12  ABABCABACBAC     hhkhkkkkhkkk     12T  P3m1     (156)  12T  P3m1       (156)  \\
90   12  ABABCABCABAC     hhkhkhkkkkkk     12T  P3m1     (156)  12T  P3m1       (156)  \\
91   12  ABABCABCACBC     hhkkhkhkkkkk     12T  P3m1     (156)  12T  P3m1       (156)  \\
92   12  ABABCABCBCAC     hhkhhkhhkkkk     12T  P3m1     (156)  12T  P-3m1      (164)  \\
93   12  ABABCACABCBC     hhkkhhkkhhkk     12R  R3m      (160)  12R  R-3m       (166)  \\
94   12  ABABCACBABAC     hhkhkhhkkhkk     12T  P3m1     (156)  12H  P-6m2      (187)  \\
95   12  ABABCACBACBC     hhkkhkkkkhkk     12T  P3m1     (156)  12T  P-3m1      (164)  \\
96   12  ABABCACBCBAC     hhkhkkhhkhkk     12H  P6\_3mc  (186)  12H  P6\_3mc    (186)  \\
97   12  ABABCBABCBAC     hhkhkhkhkkhk     12T  P3m1     (156)  12T  P3m1       (156)  \\
98   12  ABABCBACBCAC     hhkhhkhkkkhk     12T  P3m1     (156)  12T  P-3m1      (164)  \\
99   12  ABACABACBABC     hkhkhkkkhkkk     12T  P3m1     (156)  12H  P-6m2      (187)  \\
100  12  ABACABCBACBC     hkhkkhkkkhkk     12T  P3m1     (156)  12T  P-3m1      (164)  \\
101  12  ABACBACBACBC     hkkhkkkkkkkk     12T  P3m1     (156)  12T  P-3m1      (164)  \\
102  12  ABACBACBCABC     hkkkkkhkkkkk     12H  P6\_3mc  (186)  12H  P6\_3/mmc  (194)  \\
103  13  ABABABABABABC    hhhhhhhhhhkkk    13T  P3m1     (156)  13T  P-3m1      (164)  \\
104  13  ABABABABABCAC    hhhhhhhhkhhkk    13T  P3m1     (156)  13T  P3m1       (156)  \\
105  13  ABABABABACABC    hhhhhhhkhkkkk    13T  P3m1     (156)  13T  P3m1       (156)  \\
106  13  ABABABABCABAC    hhhhhhkhkhkkk    13T  P3m1     (156)  13T  P3m1       (156)  \\
107  13  ABABABABCACAC    hhhhhhkhhhhkk    13T  P3m1     (156)  13T  P3m1       (156)  \\
108  13  ABABABABCACBC    hhhhhhkkhkhkk    13T  P3m1     (156)  13T  P-3m1      (164)  \\
109  13  ABABABABCBCAC    hhhhhhkhhkhhk    13T  P3m1     (156)  13T  P-3m1      (164)  \\
110  13  ABABABACABABC    hhhhhkhkhhkkk    13T  P3m1     (156)  13T  P3m1       (156)  \\
111  13  ABABABACABCAC    hhhhhkhhkkkhk    13T  P3m1     (156)  13T  P3m1       (156)  \\
112  13  ABABABACABCBC    hhhhhkhkkhhkk    13T  P3m1     (156)  13T  P3m1       (156)  \\
113  13  ABABABACACABC    hhhhhkhhhkkkk    13T  P3m1     (156)  13T  P3m1       (156)  \\
114  13  ABABABACBACBC    hhhhhkkhkkkkk    13T  P3m1     (156)  13T  P3m1       (156)  \\
115  13  ABABABCABABAC    hhhhkhkhhhkkk    13T  P3m1     (156)  13T  P3m1       (156)  \\
116  13  ABABABCABACAC    hhhhkhhhkhkkk    13T  P3m1     (156)  13T  P3m1       (156)  \\
117  13  ABABABCABACBC    hhhhkkhkkhkkk    13T  P3m1     (156)  13T  P3m1       (156)  \\
118  13  ABABABCABCABC    hhhhkkkkkkkkk    13T  P3m1     (156)  13T  P-3m1      (164)  \\
119  13  ABABABCABCBAC    hhhhkhkkhkkkk    13T  P3m1     (156)  13T  P3m1       (156)  \\
120  13  ABABABCACABAC    hhhhkhkhkhhkk    13T  P3m1     (156)  13T  P3m1       (156)  \\
121  13  ABABABCACACBC    hhhhkkhhhkhkk    13T  P3m1     (156)  13T  P3m1       (156)  \\
122  13  ABABABCACBCAC    hhhhkhhkhkhkk    13T  P3m1     (156)  13T  P3m1       (156)  \\
123  13  ABABABCBABCAC    hhhhkhhkkhkhk    13T  P3m1     (156)  13T  P3m1       (156)  \\
124  13  ABABABCBACBAC    hhhhkhkkkkkhk    13T  P3m1     (156)  13T  P-3m1      (164)  \\
125  13  ABABABCBCACAC    hhhhkhhhhkhhk    13T  P3m1     (156)  13T  P-3m1      (164)  \\
126  13  ABABACABABCAC    hhhkhhkkhhkhk    13T  P3m1     (156)  13T  P3m1       (156)  \\
127  13  ABABACABABCBC    hhhkhkhhkhhkk    13T  P3m1     (156)  13T  P3m1       (156)  \\
128  13  ABABACABACABC    hhhkhkhkhkkkk    13T  P3m1     (156)  13T  P3m1       (156)  \\
129  13  ABABACABACBAC    hhhkhkhkkkkhk    13T  P3m1     (156)  13T  P3m1       (156)  \\
130  13  ABABACABCACBC    hhhkhkkkhkhkk    13T  P3m1     (156)  13T  P3m1       (156)  \\
131  13  ABABACABCBABC    hhhkhkkhkhkkk    13T  P3m1     (156)  13T  P3m1       (156)  \\
132  13  ABABACABCBCAC    hhhkhhkhhkkhk    13T  P3m1     (156)  13T  P3m1       (156)  \\
133  13  ABABACACABABC    hhhkhhhkhhkkk    13T  P3m1     (156)  13T  P3m1       (156)  \\
134  13  ABABACACABCBC    hhhkhhhkkhhkk    13T  P3m1     (156)  13T  P-3m1      (164)  \\
135  13  ABABACACBACBC    hhhkhhkkkkhkk    13T  P3m1     (156)  13T  P3m1       (156)  \\
136  13  ABABACACBCABC    hhhkhhkhkkkkk    13T  P3m1     (156)  13T  P3m1       (156)  \\
137  13  ABABACBABACBC    hhhkkhkkhhkkk    13T  P3m1     (156)  13T  P3m1       (156)  \\
138  13  ABABACBABCABC    hhhkkkhkkkkkk    13T  P3m1     (156)  13T  P3m1       (156)  \\
139  13  ABABACBACACBC    hhhkkhkhhkkkk    13T  P3m1     (156)  13T  P3m1       (156)  \\
140  13  ABABACBACBCBC    hhhkkhhhkkkkk    13T  P3m1     (156)  13T  P-3m1      (164)  \\
141  13  ABABACBCABCBC    hhhkkhhkkkhkk    13T  P3m1     (156)  13T  P3m1       (156)  \\
142  13  ABABCABABCABC    hhkkkhhkkkkkk    13T  P3m1     (156)  13T  P-3m1      (164)  \\
143  13  ABABCABABCBAC    hhkhkkhkhhkkk    13T  P3m1     (156)  13T  P-3m1      (164)  \\
144  13  ABABCABACABAC    hhkhkhkhkhkkk    13T  P3m1     (156)  13T  P3m1       (156)  \\
145  13  ABABCABACACBC    hhkhkkhhkkkhk    13T  P3m1     (156)  13T  P3m1       (156)  \\
146  13  ABABCABACBABC    hhkkkhkkkhkkk    13T  P3m1     (156)  13T  P-3m1      (164)  \\
147  13  ABABCABACBCAC    hhkhhkhkkhkkk    13T  P3m1     (156)  13T  P3m1       (156)  \\
148  13  ABABCABCABCAC    hhkhhkkkkkkkk    13T  P3m1     (156)  13T  P-3m1      (164)  \\
149  13  ABABCABCABCBC    hhkkhhkkkkkkk    13T  P3m1     (156)  13T  P-3m1      (164)  \\
150  13  ABABCABCACABC    hhkkkkhhkkkkk    13T  P3m1     (156)  13T  P-3m1      (164)  \\
151  13  ABABCABCACBAC    hhkhkkkhkkkkk    13T  P3m1     (156)  13T  P3m1       (156)  \\
152  13  ABABCABCBABAC    hhkhkhhkhkkkk    13T  P3m1     (156)  13T  P3m1       (156)  \\
153  13  ABABCABCBACBC    hhkkhkkkhkkkk    13T  P3m1     (156)  13T  P3m1       (156)  \\
154  13  ABABCACABACBC    hhkhkkhkkhhkk    13T  P3m1     (156)  13T  P3m1       (156)  \\
155  13  ABABCACBABCAC    hhkhhkkhkkhkk    13T  P3m1     (156)  13T  P-3m1      (164)  \\
156  13  ABABCACBACBAC    hhkhkkkkkkhkk    13T  P3m1     (156)  13T  P3m1       (156)  \\
157  13  ABABCACBCABAC    hhkhkhkkhkhkk    13T  P3m1     (156)  13T  P3m1       (156)  \\
158  13  ABABCACBCACBC    hhkkhkhkhkhkk    13T  P3m1     (156)  13T  P-3m1      (164)  \\
159  13  ABABCBABCABAC    hhkhkhkkkhkhk    13T  P3m1     (156)  13T  P-3m1      (164)  \\
160  13  ABABCBABCBCAC    hhkhhkhhkhkhk    13T  P3m1     (156)  13T  P-3m1      (164)  \\
161  13  ABACABACBACBC    hkhkhkkhkkkkk    13T  P3m1     (156)  13T  P3m1       (156)  \\
162  13  ABACABCABACBC    hkhkkhkkhkkkk    13T  P3m1     (156)  13T  P3m1       (156)  \\
163  13  ABACABCABCABC    hkhkkkkkkkkkk    13T  P3m1     (156)  13T  P-3m1      (164)  \\
164  13  ABACABCACBABC    hkhkkkhkkhkkk    13T  P3m1     (156)  13T  P-3m1      (164)  \\
165  13  ABACBACABCABC    hkkkkhkkkkkkk    13T  P3m1     (156)  13T  P-3m1      (164)  \\
166  14  ABABABABABABAC   hhhhhhhhhhhkhk   14T  P3m1     (156)  14H  P-6m2      (187)  \\
167  14  ABABABABABACAC   hhhhhhhhhkhhhk   14T  P3m1     (156)  14H  P-6m2      (187)  \\
168  14  ABABABABABACBC   hhhhhhhhhkkhkk   14T  P3m1     (156)  14H  P-6m2      (187)  \\
169  14  ABABABABABCABC   hhhhhhhhkkkkkk   14T  P3m1     (156)  14T  P-3m1      (164)  \\
170  14  ABABABABABCBAC   hhhhhhhhkhkkhk   14T  P3m1     (156)  14T  P-3m1      (164)  \\
171  14  ABABABABACABAC   hhhhhhhkhkhkhk   14T  P3m1     (156)  14H  P-6m2      (187)  \\
172  14  ABABABABACACAC   hhhhhhhkhhhhhk   14T  P3m1     (156)  14H  P-6m2      (187)  \\
173  14  ABABABABACACBC   hhhhhhhkhhkhkk   14T  P3m1     (156)  14T  P3m1       (156)  \\
174  14  ABABABABACBABC   hhhhhhhkkkhkkk   14T  P3m1     (156)  14H  P-6m2      (187)  \\
175  14  ABABABABACBCBC   hhhhhhhkkhhhkk   14T  P3m1     (156)  14H  P-6m2      (187)  \\
176  14  ABABABABCABABC   hhhhhhkkkhhkkk   14T  P3m1     (156)  14T  P-3m1      (164)  \\
177  14  ABABABABCABCAC   hhhhhhkhhkkkkk   14T  P3m1     (156)  14T  P3m1       (156)  \\
178  14  ABABABABCABCBC   hhhhhhkkhhkkkk   14T  P3m1     (156)  14T  P3m1       (156)  \\
179  14  ABABABABCACBAC   hhhhhhkhkkkhkk   14T  P3m1     (156)  14T  P3m1       (156)  \\
180  14  ABABABABCBABAC   hhhhhhkhkhhkhk   14T  P3m1     (156)  14T  P-3m1      (164)  \\
181  14  ABABABABCBACAC   hhhhhhkhhhkkhk   14T  P3m1     (156)  14T  P3m1       (156)  \\
182  14  ABABABACABABAC   hhhhhkhkhhhkhk   14T  P3m1     (156)  14H  P-6m2      (187)  \\
183  14  ABABABACABACAC   hhhhhkhhhkhkhk   14T  P3m1     (156)  14T  P3m1       (156)  \\
184  14  ABABABACABACBC   hhhhhkhkhkkhkk   14T  P3m1     (156)  14T  P3m1       (156)  \\
185  14  ABABABACABCABC   hhhhhkhkkkkkkk   14T  P3m1     (156)  14T  P3m1       (156)  \\
186  14  ABABABACABCBAC   hhhhhkhkkhkkhk   14T  P3m1     (156)  14H  P-6m2      (187)  \\
187  14  ABABABACACACBC   hhhhhkhhhhkhkk   14T  P3m1     (156)  14T  P3m1       (156)  \\
188  14  ABABABACACBABC   hhhhhkhhkkhkkk   14T  P3m1     (156)  14T  P3m1       (156)  \\
189  14  ABABABACACBCAC   hhhhhkhhkhkhhk   14T  P3m1     (156)  14H  P-6m2      (187)  \\
190  14  ABABABACACBCBC   hhhhhkhhkhhhkk   14T  P3m1     (156)  14T  P3m1       (156)  \\
191  14  ABABABACBABABC   hhhhhkkkhhhkkk   14T  P3m1     (156)  14H  P-6m2      (187)  \\
192  14  ABABABACBABCBC   hhhhhkkhhkhkkk   14T  P3m1     (156)  14T  P3m1       (156)  \\
193  14  ABABABACBACABC   hhhhhkkkkhkkkk   14T  P3m1     (156)  14H  P-6m2      (187)  \\
194  14  ABABABACBCACBC   hhhhhkkhkhkhkk   14T  P3m1     (156)  14H  P-6m2      (187)  \\
195  14  ABABABACBCBCBC   hhhhhkkhhhhhkk   14H  P6\_3mc  (186)  14H  P6\_3/mmc  (194)  \\
196  14  ABABABCABABCAC   hhhhkhhkkhhkkk   14T  P3m1     (156)  14T  P3m1       (156)  \\
197  14  ABABABCABABCBC   hhhhkkhhkhhkkk   14T  P3m1     (156)  14T  P3m1       (156)  \\
198  14  ABABABCABACABC   hhhhkkkhkhkkkk   14T  P3m1     (156)  14T  P3m1       (156)  \\
199  14  ABABABCABACBAC   hhhhkhkkkkhkkk   14T  P3m1     (156)  14T  P3m1       (156)  \\
200  14  ABABABCABCABAC   hhhhkhkhkkkkkk   14T  P3m1     (156)  14T  P3m1       (156)  \\
201  14  ABABABCABCACAC   hhhhkhhhhkkkkk   14T  P3m1     (156)  14T  P-3m1      (164)  \\
202  14  ABABABCABCACBC   hhhhkkhkhkkkkk   14T  P3m1     (156)  14T  P3m1       (156)  \\
203  14  ABABABCABCBCAC   hhhhkhhkhhkkkk   14T  P3m1     (156)  14T  P3m1       (156)  \\
204  14  ABABABCABCBCBC   hhhhkkhhhhkkkk   14T  P3m1     (156)  14T  P-3m1      (164)  \\
205  14  ABABABCACABCAC   hhhhkhhkkkhhkk   14T  P3m1     (156)  14T  P3m1       (156)  \\
206  14  ABABABCACABCBC   hhhhkkhhkkhhkk   14T  P3m1     (156)  14T  P-3m1      (164)  \\
207  14  ABABABCACACBAC   hhhhkhkkkhhhkk   14T  P3m1     (156)  14T  P3m1       (156)  \\
208  14  ABABABCACBABAC   hhhhkhkhhkkhkk   14T  P3m1     (156)  14T  P3m1       (156)  \\
209  14  ABABABCACBACAC   hhhhkhhhkkkhkk   14T  P3m1     (156)  14T  P3m1       (156)  \\
210  14  ABABABCACBACBC   hhhhkkhkkkkhkk   14T  P3m1     (156)  14T  P-3m1      (164)  \\
211  14  ABABABCACBCBAC   hhhhkhkkhhkhkk   14T  P3m1     (156)  14T  P3m1       (156)  \\
212  14  ABABABCBABABAC   hhhhkhkhhhhkhk   14H  P6\_3mc  (186)  14H  P6\_3/mmc  (194)  \\
213  14  ABABABCBABACAC   hhhhkhhhkhhkhk   14T  P3m1     (156)  14T  P3m1       (156)  \\
214  14  ABABABCBABCBAC   hhhhkhkhkhkkhk   14T  P3m1     (156)  14T  P3m1       (156)  \\
215  14  ABABABCBACBCAC   hhhhkhhkhkkkhk   14T  P3m1     (156)  14T  P3m1       (156)  \\
216  14  ABABABCBCABCAC   hhhhkhhkkkkhhk   14T  P3m1     (156)  14T  P-3m1      (164)  \\
217  14  ABABABCBCBACAC   hhhhkhhhkkhhhk   14T  P3m1     (156)  14T  P-3m1      (164)  \\
218  14  ABABACABABACAC   hhhkhhhkhhhkhk   14T  P3m1     (156)  14H  P-6m2      (187)  \\
219  14  ABABACABABACBC   hhhkhkhhhkkhkk   14T  P3m1     (156)  14H  P-6m2      (187)  \\
220  14  ABABACABABCABC   hhhkhkhhkkkkkk   14T  P3m1     (156)  14T  P3m1       (156)  \\
221  14  ABABACABABCBAC   hhhkhkhhkhkkhk   14T  P3m1     (156)  14T  P3m1       (156)  \\
222  14  ABABACABACABAC   hhhkhkhkhkhkhk   14T  P3m1     (156)  14H  P-6m2      (187)  \\
223  14  ABABACABACACBC   hhhkhkhkhhkhkk   14T  P3m1     (156)  14T  P3m1       (156)  \\
224  14  ABABACABACBABC   hhhkhkhkkkhkkk   14T  P3m1     (156)  14T  P3m1       (156)  \\
225  14  ABABACABACBCAC   hhhkhhkhkkhkhk   14T  P3m1     (156)  14T  P3m1       (156)  \\
226  14  ABABACABACBCBC   hhhkhkhkkhhhkk   14T  P3m1     (156)  14T  P3m1       (156)  \\
227  14  ABABACABCABABC   hhhkhkkkkhhkkk   14T  P3m1     (156)  14T  P3m1       (156)  \\
228  14  ABABACABCABCAC   hhhkhhkkkkkkhk   14T  P3m1     (156)  14T  P3m1       (156)  \\
229  14  ABABACABCABCBC   hhhkhkkkkkhhkk   14T  P3m1     (156)  14T  P3m1       (156)  \\
230  14  ABABACABCACABC   hhhkhkkkhhkkkk   14T  P3m1     (156)  14T  P3m1       (156)  \\
231  14  ABABACABCACBAC   hhhkhkkkhkkkhk   14T  P3m1     (156)  14H  P-6m2      (187)  \\
232  14  ABABACABCBACAC   hhhkhhhkhkkhkk   14T  P3m1     (156)  14T  P3m1       (156)  \\
233  14  ABABACABCBACBC   hhhkhkkhkkkhkk   14T  P3m1     (156)  14T  P3m1       (156)  \\
234  14  ABABACABCBCABC   hhhkhkkhhkkkkk   14T  P3m1     (156)  14T  P3m1       (156)  \\
235  14  ABABACABCBCBAC   hhhkhkkhhhkkhk   14T  P3m1     (156)  14H  P-6m2      (187)  \\
236  14  ABABACACABCABC   hhhkhhhkkkkkkk   14T  P3m1     (156)  14T  P-3m1      (164)  \\
237  14  ABABACACBABABC   hhhkhhkkhhhkkk   14T  P3m1     (156)  14T  P3m1       (156)  \\
238  14  ABABACACBABCAC   hhhkhhkkhkkhhk   14T  P3m1     (156)  14H  P-6m2      (187)  \\
239  14  ABABACACBABCBC   hhhkhhkkhkhhkk   14T  P3m1     (156)  14T  P3m1       (156)  \\
240  14  ABABACACBACABC   hhhkhhkkkhkkkk   14T  P3m1     (156)  14T  P3m1       (156)  \\
241  14  ABABACACBCACBC   hhhkhhkhkhkhkk   14T  P3m1     (156)  14T  P3m1       (156)  \\
242  14  ABABACACBCBABC   hhhkhhkhhkhkkk   14T  P3m1     (156)  14T  P3m1       (156)  \\
243  14  ABABACACBCBCAC   hhhkhhkhhhkhhk   14H  P6\_3mc  (186)  14H  P6\_3/mmc  (194)  \\
244  14  ABABACBABACABC   hhhkkkhhkhkkkk   14T  P3m1     (156)  14T  P3m1       (156)  \\
245  14  ABABACBABCACBC   hhhkkhkhkkhkkk   14T  P3m1     (156)  14T  P3m1       (156)  \\
246  14  ABABACBABCBABC   hhhkkkhkhkhkkk   14T  P3m1     (156)  14H  P-6m2      (187)  \\
247  14  ABABACBACABCBC   hhhkkhhkkhkkkk   14T  P3m1     (156)  14T  P3m1       (156)  \\
248  14  ABABACBACACABC   hhhkkkkhhhkkkk   14H  P6\_3mc  (186)  14H  P6\_3/mmc  (194)  \\
249  14  ABABACBACBACBC   hhhkkhkkkkkkkk   14T  P3m1     (156)  14T  P3m1       (156)  \\
250  14  ABABACBACBCABC   hhhkkkkkhkkkkk   14T  P3m1     (156)  14H  P-6m2      (187)  \\
251  14  ABABACBCABACBC   hhhkkhkkhkkhkk   14T  P3m1     (156)  14H  P-6m2      (187)  \\
252  14  ABABACBCBABCBC   hhhkkhhkhkhhkk   14T  P3m1     (156)  14H  P-6m2      (187)  \\
253  14  ABABCABABCABAC   hhkhkhkkkhhkkk   14T  P3m1     (156)  14T  P3m1       (156)  \\
254  14  ABABCABABCACBC   hhkkhkhkkhhkkk   14T  P3m1     (156)  14T  P-3m1      (164)  \\
255  14  ABABCABABCBCAC   hhkhhkhhkhhkkk   14T  P3m1     (156)  14T  P-3m1      (164)  \\
256  14  ABABCABACABCAC   hhkhhkkkhkhkkk   14T  P3m1     (156)  14T  P-3m1      (164)  \\
257  14  ABABCABACABCBC   hhkkhhkkhkhkkk   14T  P3m1     (156)  14T  P3m1       (156)  \\
258  14  ABABCABACACBAC   hhkhkkkhhkhkkk   14H  P6\_3mc  (186)  14H  P6\_3mc    (186)  \\
259  14  ABABCABACBABAC   hhkhkhhkkkhkkk   14T  P3m1     (156)  14H  P-6m2      (187)  \\
260  14  ABABCABACBACBC   hhkkhkkkkkhkkk   14T  P3m1     (156)  14T  P3m1       (156)  \\
261  14  ABABCABACBCABC   hhkkkhkkhkkkkk   14T  P3m1     (156)  14T  P3m1       (156)  \\
262  14  ABABCABACBCBAC   hhkhkkhhkkhkkk   14T  P3m1     (156)  14T  P3m1       (156)  \\
263  14  ABABCABCABACBC   hhkkhkkhkkkkkk   14T  P3m1     (156)  14T  P3m1       (156)  \\
264  14  ABABCABCABCABC   hhkkkkkkkkkkkk   14T  P3m1     (156)  14T  P-3m1      (164)  \\
265  14  ABABCABCABCBAC   hhkhkkhkkkkkkk   14T  P3m1     (156)  14T  P3m1       (156)  \\
266  14  ABABCABCACABAC   hhkhkhkhhkkkkk   14T  P3m1     (156)  14T  P-3m1      (164)  \\
267  14  ABABCABCACBCAC   hhkhhkhkhkkkkk   14T  P3m1     (156)  14T  P3m1       (156)  \\
268  14  ABABCABCBABCAC   hhkhhkkhkhkkkk   14T  P3m1     (156)  14T  P3m1       (156)  \\
269  14  ABABCABCBACABC   hhkkkkhkkhkkkk   14T  P3m1     (156)  14T  P-3m1      (164)  \\
270  14  ABABCABCBACBAC   hhkhkkkkkhkkkk   14T  P3m1     (156)  14T  P3m1       (156)  \\
271  14  ABABCABCBCABAC   hhkhkhkkhhkkkk   14T  P3m1     (156)  14T  P3m1       (156)  \\
272  14  ABABCACABABCBC   hhkhhkhhkkhhkk   14T  P3m1     (156)  14T  P-3m1      (164)  \\
273  14  ABABCACABACABC   hhkhkhkkkkhhkk   14T  P3m1     (156)  14T  P3m1       (156)  \\
274  14  ABABCACABACBAC   hhkhkkkkhkhhkk   14T  P3m1     (156)  14T  P-3m1      (164)  \\
275  14  ABABCACBABACBC   hhkkhkkhhkkhkk   14H  P6\_3mc  (186)  14H  P6\_3/mmc  (194)  \\
276  14  ABABCACBABCBAC   hhkhkkhkhkkhkk   14T  P3m1     (156)  14T  P3m1       (156)  \\
277  14  ABABCACBACABAC   hhkhkhkhkkkhkk   14T  P3m1     (156)  14T  P3m1       (156)  \\
278  14  ABABCACBACBCAC   hhkhhkhkkkkhkk   14T  P3m1     (156)  14T  P3m1       (156)  \\
279  14  ABABCACBCACBAC   hhkhkkkhkhkhkk   14T  P3m1     (156)  14T  P3m1       (156)  \\
280  14  ABABCBABACABAC   hhkhkhhkhkhkhk   14T  P3m1     (156)  14H  P-6m2      (187)  \\
281  14  ABABCBABACACBC   hhkhhkhkhhkkhk   14T  P3m1     (156)  14T  P3m1       (156)  \\
282  14  ABABCBABCACBAC   hhkhkhkkhkkkhk   14T  P3m1     (156)  14T  P3m1       (156)  \\
283  14  ABABCBACABCBAC   hhkhkkhkkhkkhk   14T  P3m1     (156)  14T  P-3m1      (164)  \\
284  14  ABABCBACBACBAC   hhkhkkkkkkkkhk   14T  P3m1     (156)  14T  P-3m1      (164)  \\
285  14  ABACABACABACBC   hkhkhkhkhkkhkk   14T  P3m1     (156)  14H  P-6m2      (187)  \\
286  14  ABACABACABCABC   hkhkhkhkkkkkkk   14T  P3m1     (156)  14T  P-3m1      (164)  \\
287  14  ABACABACBACABC   hkhkhkkkkhkkkk   14T  P3m1     (156)  14H  P-6m2      (187)  \\
288  14  ABACABACBCACBC   hkhkhkkhkhkhkk   14H  P6\_3mc  (186)  14H  P6\_3/mmc  (194)  \\
289  14  ABACABCABCACBC   hkhkkhkhkkkkkk   14T  P3m1     (156)  14T  P-3m1      (164)  \\
290  14  ABACABCABCBABC   hkhkkkhkhkkkkk   14T  P3m1     (156)  14T  P-3m1      (164)  \\
291  14  ABACABCACBACBC   hkhkkhkkkkhkkk   14T  P3m1     (156)  14T  P3m1       (156)  \\
292  14  ABACBABCABACBC   hkkhkkhkkkhkkk   14T  P3m1     (156)  14H  P-6m2      (187)  \\
293  14  ABACBABCABCABC   hkkkhkkkkkkkkk   14T  P3m1     (156)  14T  P-3m1      (164)  \\
294  14  ABACBACBABCABC   hkkkkkkhkkkkkk   14H  P6\_3mc  (186)  14H  P6\_3/mmc  (194)  \\
295  15  ABABABABABABABC  hhhhhhhhhhhhkkk  15T  P3m1     (156)  15T  P-3m1      (164)  \\
296  15  ABABABABABABCAC  hhhhhhhhhhkhhkk  15T  P3m1     (156)  15T  P3m1       (156)  \\
297  15  ABABABABABACABC  hhhhhhhhhkhkkkk  15T  P3m1     (156)  15T  P3m1       (156)  \\
298  15  ABABABABABCABAC  hhhhhhhhkhkhkkk  15T  P3m1     (156)  15T  P3m1       (156)  \\
299  15  ABABABABABCACAC  hhhhhhhhkhhhhkk  15T  P3m1     (156)  15T  P3m1       (156)  \\
300  15  ABABABABABCACBC  hhhhhhhhkkhkhkk  15T  P3m1     (156)  15T  P-3m1      (164)  \\
301  15  ABABABABABCBCAC  hhhhhhhhkhhkhhk  15T  P3m1     (156)  15T  P-3m1      (164)  \\
302  15  ABABABABACABABC  hhhhhhhkhkhhkkk  15T  P3m1     (156)  15T  P3m1       (156)  \\
303  15  ABABABABACABCAC  hhhhhhhkhhkkkhk  15T  P3m1     (156)  15T  P3m1       (156)  \\
304  15  ABABABABACABCBC  hhhhhhhkhkkhhkk  15T  P3m1     (156)  15T  P3m1       (156)  \\
305  15  ABABABABACACABC  hhhhhhhkhhhkkkk  15T  P3m1     (156)  15T  P3m1       (156)  \\
306  15  ABABABABACBACBC  hhhhhhhkkhkkkkk  15T  P3m1     (156)  15T  P3m1       (156)  \\
307  15  ABABABABCABABAC  hhhhhhkhkhhhkkk  15T  P3m1     (156)  15T  P3m1       (156)  \\
308  15  ABABABABCABACAC  hhhhhhkhhhkhkkk  15T  P3m1     (156)  15T  P3m1       (156)  \\
309  15  ABABABABCABACBC  hhhhhhkkhkkhkkk  15T  P3m1     (156)  15T  P3m1       (156)  \\
310  15  ABABABABCABCABC  hhhhhhkkkkkkkkk  15T  P3m1     (156)  15T  P-3m1      (164)  \\
311  15  ABABABABCABCBAC  hhhhhhkhkkhkkkk  15T  P3m1     (156)  15T  P3m1       (156)  \\
312  15  ABABABABCACABAC  hhhhhhkhkhkhhkk  15T  P3m1     (156)  15T  P3m1       (156)  \\
313  15  ABABABABCACACAC  hhhhhhkhhhhhhkk  15T  P3m1     (156)  15T  P-3m1      (164)  \\
314  15  ABABABABCACACBC  hhhhhhkkhhhkhkk  15T  P3m1     (156)  15T  P3m1       (156)  \\
315  15  ABABABABCACBCAC  hhhhhhkhhkhkhkk  15T  P3m1     (156)  15T  P3m1       (156)  \\
316  15  ABABABABCBABCAC  hhhhhhkhhkkhkhk  15T  P3m1     (156)  15T  P3m1       (156)  \\
317  15  ABABABABCBACBAC  hhhhhhkhkkkkkhk  15T  P3m1     (156)  15T  P-3m1      (164)  \\
318  15  ABABABABCBCACAC  hhhhhhkhhhhkhhk  15T  P3m1     (156)  15T  P3m1       (156)  \\
319  15  ABABABACABABABC  hhhhhkhkhhhhkkk  15T  P3m1     (156)  15T  P3m1       (156)  \\
320  15  ABABABACABABCAC  hhhhhkhhkkhhkhk  15T  P3m1     (156)  15T  P3m1       (156)  \\
321  15  ABABABACABABCBC  hhhhhkhkhhkhhkk  15T  P3m1     (156)  15T  P3m1       (156)  \\
322  15  ABABABACABACABC  hhhhhkhkhkhkkkk  15T  P3m1     (156)  15T  P3m1       (156)  \\
323  15  ABABABACABACBAC  hhhhhkhkhkkkkhk  15T  P3m1     (156)  15T  P3m1       (156)  \\
324  15  ABABABACABCACAC  hhhhhkhhhhkkkhk  15T  P3m1     (156)  15T  P3m1       (156)  \\
325  15  ABABABACABCACBC  hhhhhkhkkkhkhkk  15T  P3m1     (156)  15T  P3m1       (156)  \\
326  15  ABABABACABCBABC  hhhhhkhkkhkhkkk  15T  P3m1     (156)  15T  P3m1       (156)  \\
327  15  ABABABACABCBCAC  hhhhhkhhkhhkkhk  15T  P3m1     (156)  15T  P3m1       (156)  \\
328  15  ABABABACABCBCBC  hhhhhkhkkhhhhkk  15T  P3m1     (156)  15T  P3m1       (156)  \\
329  15  ABABABACACABABC  hhhhhkhhhkhhkkk  15T  P3m1     (156)  15T  P3m1       (156)  \\
330  15  ABABABACACABCAC  hhhhhkhhhkkkhhk  15T  P3m1     (156)  15T  P3m1       (156)  \\
331  15  ABABABACACABCBC  hhhhhkhhhkkhhkk  15T  P3m1     (156)  15T  P3m1       (156)  \\
332  15  ABABABACACACABC  hhhhhkhhhhhkkkk  15T  P3m1     (156)  15T  P-3m1      (164)  \\
333  15  ABABABACACBACBC  hhhhhkhhkkkkhkk  15T  P3m1     (156)  15T  P3m1       (156)  \\
334  15  ABABABACACBCABC  hhhhhkhhkhkkkkk  15T  P3m1     (156)  15T  P3m1       (156)  \\
335  15  ABABABACBABACBC  hhhhhkkhkkhhkkk  15T  P3m1     (156)  15T  P3m1       (156)  \\
336  15  ABABABACBABCABC  hhhhhkkkhkkkkkk  15T  P3m1     (156)  15T  P3m1       (156)  \\
337  15  ABABABACBACACBC  hhhhhkkhkhhkkkk  15T  P3m1     (156)  15T  P3m1       (156)  \\
338  15  ABABABACBACBCBC  hhhhhkkhhhkkkkk  15T  P3m1     (156)  15T  P3m1       (156)  \\
339  15  ABABABACBCABCBC  hhhhhkkhhkkkhkk  15T  P3m1     (156)  15T  P3m1       (156)  \\
340  15  ABABABCABABACAC  hhhhkhhhkhhhkkk  15T  P3m1     (156)  15T  P3m1       (156)  \\
341  15  ABABABCABABACBC  hhhhkkhkkhhhkkk  15T  P3m1     (156)  15T  P3m1       (156)  \\
342  15  ABABABCABABCABC  hhhhkkkhhkkkkkk  15T  P3m1     (156)  15T  P3m1       (156)  \\
343  15  ABABABCABABCBAC  hhhhkhkkhkhhkkk  15T  P3m1     (156)  15T  P3m1       (156)  \\
344  15  ABABABCABACABAC  hhhhkhkhkhkhkkk  15T  P3m1     (156)  15T  P3m1       (156)  \\
345  15  ABABABCABACACBC  hhhhkkhkhhkhkkk  15T  P3m1     (156)  15T  P3m1       (156)  \\
346  15  ABABABCABACBABC  hhhhkkkhkkkhkkk  15T  P3m1     (156)  15T  P-3m1      (164)  \\
347  15  ABABABCABACBCAC  hhhhkhhkhkkhkkk  15T  P3m1     (156)  15T  P3m1       (156)  \\
348  15  ABABABCABACBCBC  hhhhkkhhhkkhkkk  15T  P3m1     (156)  15T  P3m1       (156)  \\
349  15  ABABABCABCABCAC  hhhhkhhkkkkkkkk  15T  P3m1     (156)  15T  P3m1       (156)  \\
350  15  ABABABCABCABCBC  hhhhkkhhkkkkkkk  15T  P3m1     (156)  15T  P3m1       (156)  \\
351  15  ABABABCABCACABC  hhhhkkkkhhkkkkk  15T  P3m1     (156)  15T  P3m1       (156)  \\
352  15  ABABABCABCACBAC  hhhhkhkkkhkkkkk  15T  P3m1     (156)  15T  P3m1       (156)  \\
353  15  ABABABCABCBABAC  hhhhkhkhhkhkkkk  15T  P3m1     (156)  15T  P3m1       (156)  \\
354  15  ABABABCABCBACAC  hhhhkhhhkkhkkkk  15T  P3m1     (156)  15T  P3m1       (156)  \\
355  15  ABABABCABCBACBC  hhhhkkhkkkhkkkk  15T  P3m1     (156)  15T  P3m1       (156)  \\
356  15  ABABABCABCBCBAC  hhhhkhkkhhhkkkk  15T  P3m1     (156)  15T  P3m1       (156)  \\
357  15  ABABABCACABABAC  hhhhkhkhhhkhhkk  15T  P3m1     (156)  15T  P3m1       (156)  \\
358  15  ABABABCACABACAC  hhhhkhhhkhkhhkk  15T  P3m1     (156)  15T  P3m1       (156)  \\
359  15  ABABABCACABACBC  hhhhkkhhkhkkhkk  15T  P3m1     (156)  15T  P3m1       (156)  \\
360  15  ABABABCACABCBAC  hhhhkhkkhkkhhkk  15T  P3m1     (156)  15T  P3m1       (156)  \\
361  15  ABABABCACACABAC  hhhhkhkhkhhhhkk  15T  P3m1     (156)  15T  P-3m1      (164)  \\
362  15  ABABABCACACBCAC  hhhhkhhkhkhhhkk  15T  P3m1     (156)  15T  P3m1       (156)  \\
363  15  ABABABCACACBCBC  hhhhkkhhhkhhhkk  15T  P3m1     (156)  15T  P-3m1      (164)  \\
364  15  ABABABCACBABCAC  hhhhkhhkkhkkhkk  15T  P3m1     (156)  15T  P3m1       (156)  \\
365  15  ABABABCACBACBAC  hhhhkhkkkkkkhkk  15T  P3m1     (156)  15T  P3m1       (156)  \\
366  15  ABABABCACBCABAC  hhhhkhkhkkhkhkk  15T  P3m1     (156)  15T  P3m1       (156)  \\
367  15  ABABABCACBCACAC  hhhhkhhhhkhkhkk  15T  P3m1     (156)  15T  P3m1       (156)  \\
368  15  ABABABCACBCACBC  hhhhkkhkhkhkhkk  15T  P3m1     (156)  15T  P-3m1      (164)  \\
369  15  ABABABCACBCBCAC  hhhhkhhkhhhkhkk  15T  P3m1     (156)  15T  P3m1       (156)  \\
370  15  ABABABCBABABCAC  hhhhkhhkkhhhkhk  15T  P3m1     (156)  15T  P3m1       (156)  \\
371  15  ABABABCBABACBAC  hhhhkhkhhkkkkhk  15T  P3m1     (156)  15T  P3m1       (156)  \\
372  15  ABABABCBABCABAC  hhhhkhkhkkkhkhk  15T  P3m1     (156)  15T  P-3m1      (164)  \\
373  15  ABABABCBABCBCAC  hhhhkhhkhhkhkhk  15T  P3m1     (156)  15T  P3m1       (156)  \\
374  15  ABABABCBACABCAC  hhhhkhhkkkhkkhk  15T  P3m1     (156)  15T  P3m1       (156)  \\
375  15  ABABABCBACACBAC  hhhhkhkkhhkkkhk  15T  P3m1     (156)  15T  P3m1       (156)  \\
376  15  ABABABCBACBACAC  hhhhkhhhkkkkkhk  15T  P3m1     (156)  15T  P3m1       (156)  \\
377  15  ABABABCBCABACAC  hhhhkhhhkhkkhhk  15T  P3m1     (156)  15T  P3m1       (156)  \\
378  15  ABABABCBCACBCAC  hhhhkhhkhkhkhhk  15T  P3m1     (156)  15T  P-3m1      (164)  \\
379  15  ABABABCBCBCACAC  hhhhkhhhhkhhhhk  15R  R3m      (160)  15R  R-3m       (166)  \\
380  15  ABABACABABACABC  hhhkhkhhhkhkkkk  15T  P3m1     (156)  15T  P3m1       (156)  \\
381  15  ABABACABABCABAC  hhhkhkhhkkkhkhk  15T  P3m1     (156)  15T  P3m1       (156)  \\
382  15  ABABACABABCACBC  hhhkhkhhkkhkhkk  15T  P3m1     (156)  15T  P3m1       (156)  \\
383  15  ABABACABABCBABC  hhhkhkhhkhkhkkk  15T  P3m1     (156)  15T  P3m1       (156)  \\
384  15  ABABACABABCBCAC  hhhkhhkhhkhhkhk  15T  P3m1     (156)  15T  P3m1       (156)  \\
385  15  ABABACABACABABC  hhhkhkhkhkhhkkk  15T  P3m1     (156)  15T  P3m1       (156)  \\
386  15  ABABACABACABCAC  hhhkhhkkkhkhkhk  15T  P3m1     (156)  15T  P3m1       (156)  \\
387  15  ABABACABACABCBC  hhhkhkhkhkkhhkk  15T  P3m1     (156)  15T  P3m1       (156)  \\
388  15  ABABACABACACABC  hhhkhkhkhhhkkkk  15T  P3m1     (156)  15T  P-3m1      (164)  \\
389  15  ABABACABACACBAC  hhhkhkhkhhkkkhk  15T  P3m1     (156)  15T  P3m1       (156)  \\
390  15  ABABACABACBACAC  hhhkhhhkhkhkkkk  15T  P3m1     (156)  15T  P3m1       (156)  \\
391  15  ABABACABACBACBC  hhhkhkhkkkkkhkk  15T  P3m1     (156)  15T  P3m1       (156)  \\
392  15  ABABACABACBCABC  hhhkhkhkkhkkkkk  15T  P3m1     (156)  15T  P3m1       (156)  \\
393  15  ABABACABACBCBAC  hhhkhkhkkhhkkhk  15T  P3m1     (156)  15T  P3m1       (156)  \\
394  15  ABABACABCABACAC  hhhkhhhkhkkkkhk  15T  P3m1     (156)  15T  P-3m1      (164)  \\
395  15  ABABACABCABACBC  hhhkhkkkkhkkhkk  15T  P3m1     (156)  15T  P3m1       (156)  \\
396  15  ABABACABCABCABC  hhhkhkkkkkkkkkk  15T  P3m1     (156)  15T  P3m1       (156)  \\
397  15  ABABACABCABCBAC  hhhkhkkhkkkkkhk  15T  P3m1     (156)  15T  P3m1       (156)  \\
398  15  ABABACABCACACBC  hhhkhkkhhhkhkkk  15T  P3m1     (156)  15T  P3m1       (156)  \\
399  15  ABABACABCACBABC  hhhkhkkkhkkhkkk  15T  P3m1     (156)  15T  P3m1       (156)  \\
400  15  ABABACABCACBCAC  hhhkhhkhkhkkkhk  15T  P3m1     (156)  15T  P3m1       (156)  \\
401  15  ABABACABCACBCBC  hhhkhkkkhkhhhkk  15T  P3m1     (156)  15T  P-3m1      (164)  \\
402  15  ABABACABCBABABC  hhhkhkkhkhhhkkk  15T  P3m1     (156)  15T  P-3m1      (164)  \\
403  15  ABABACABCBABCAC  hhhkhhkkhkhkkhk  15T  P3m1     (156)  15T  P3m1       (156)  \\
404  15  ABABACABCBABCBC  hhhkhkkhkhkhhkk  15T  P3m1     (156)  15T  P3m1       (156)  \\
405  15  ABABACABCBACABC  hhhkhkkhkkhkkkk  15T  P3m1     (156)  15T  P3m1       (156)  \\
406  15  ABABACABCBCACBC  hhhkhkkhhkhkhkk  15T  P3m1     (156)  15T  P3m1       (156)  \\
407  15  ABABACACABABCAC  hhhkhhhkhhkkhhk  15T  P3m1     (156)  15T  P-3m1      (164)  \\
408  15  ABABACACABABCBC  hhhkhhhkhhkhhkk  15T  P3m1     (156)  15T  P3m1       (156)  \\
409  15  ABABACACABCACBC  hhhkhhhkkhkhkkk  15T  P3m1     (156)  15T  P3m1       (156)  \\
410  15  ABABACACBABACBC  hhhkhhkkhhkkhkk  15T  P3m1     (156)  15T  P3m1       (156)  \\
411  15  ABABACACBABCABC  hhhkhhkkhkkkkkk  15T  P3m1     (156)  15T  P3m1       (156)  \\
412  15  ABABACACBACACBC  hhhkhhkkkhhkhkk  15T  P3m1     (156)  15T  P3m1       (156)  \\
413  15  ABABACACBACBABC  hhhkhhkkkkkhkkk  15T  P3m1     (156)  15T  P3m1       (156)  \\
414  15  ABABACACBACBCAC  hhhkhhkhkkkkhhk  15T  P3m1     (156)  15T  P3m1       (156)  \\
415  15  ABABACACBACBCBC  hhhkhhkkkkhhhkk  15T  P3m1     (156)  15T  P3m1       (156)  \\
416  15  ABABACACBCABABC  hhhkhhkhkkhhkkk  15T  P3m1     (156)  15T  P3m1       (156)  \\
417  15  ABABACACBCABCBC  hhhkhhkhkkkhhkk  15T  P3m1     (156)  15T  P3m1       (156)  \\
418  15  ABABACACBCACABC  hhhkhhkhkhhkkkk  15T  P3m1     (156)  15T  P3m1       (156)  \\
419  15  ABABACACBCBACBC  hhhkhhkhhkkkhkk  15T  P3m1     (156)  15T  P3m1       (156)  \\
420  15  ABABACACBCBCABC  hhhkhhkhhhkkkkk  15T  P3m1     (156)  15T  P-3m1      (164)  \\
421  15  ABABACBABABCABC  hhhkkkhhhkkkkkk  15T  P3m1     (156)  15T  P-3m1      (164)  \\
422  15  ABABACBABACACBC  hhhkkhkhhkhhkkk  15T  P3m1     (156)  15T  P3m1       (156)  \\
423  15  ABABACBABACBABC  hhhkkkhhkkkhkkk  15T  P3m1     (156)  15T  P3m1       (156)  \\
424  15  ABABACBABACBCBC  hhhkkhhhkkhhkkk  15T  P3m1     (156)  15T  P3m1       (156)  \\
425  15  ABABACBABCABCBC  hhhkkhhkkkkhkkk  15T  P3m1     (156)  15T  P3m1       (156)  \\
426  15  ABABACBABCACABC  hhhkkkhkkhhkkkk  15T  P3m1     (156)  15T  P3m1       (156)  \\
427  15  ABABACBABCBACBC  hhhkkhkkkhkhkkk  15T  P3m1     (156)  15T  P3m1       (156)  \\
428  15  ABABACBABCBCABC  hhhkkkhkhhkkkkk  15T  P3m1     (156)  15T  P3m1       (156)  \\
429  15  ABABACBACABACBC  hhhkkhkkhkhkkkk  15T  P3m1     (156)  15T  P3m1       (156)  \\
430  15  ABABACBACABCABC  hhhkkkkhkkkkkkk  15T  P3m1     (156)  15T  P3m1       (156)  \\
431  15  ABABACBACBABCBC  hhhkkhhkhkkkkkk  15T  P3m1     (156)  15T  P3m1       (156)  \\
432  15  ABABACBACBCACBC  hhhkkhkhkhkkkkk  15T  P3m1     (156)  15T  P3m1       (156)  \\
433  15  ABABACBCABABCBC  hhhkkhhkhhkkhkk  15T  P3m1     (156)  15T  P3m1       (156)  \\
434  15  ABABACBCABCACBC  hhhkkhkhkkkkhkk  15T  P3m1     (156)  15T  P3m1       (156)  \\
435  15  ABABACBCACABCBC  hhhkkhhkkhhkhkk  15T  P3m1     (156)  15T  P3m1       (156)  \\
436  15  ABABCABABCABCAC  hhkhhkkkhhkkkkk  15T  P3m1     (156)  15T  P3m1       (156)  \\
437  15  ABABCABABCACABC  hhkkhhkkkhhkkkk  15T  P3m1     (156)  15T  P3m1       (156)  \\
438  15  ABABCABABCACBAC  hhkhkkkhkkhhkkk  15T  P3m1     (156)  15T  P3m1       (156)  \\
439  15  ABABCABABCBABAC  hhkhkhhkhkhhkkk  15T  P3m1     (156)  15T  P-3m1      (164)  \\
440  15  ABABCABACABACBC  hhkkhkkhkhkhkkk  15T  P3m1     (156)  15T  P3m1       (156)  \\
441  15  ABABCABACABCABC  hhkkkhkhkkkkkkk  15T  P3m1     (156)  15T  P3m1       (156)  \\
442  15  ABABCABACABCBAC  hhkhkkhkkhkhkkk  15T  P3m1     (156)  15T  P3m1       (156)  \\
443  15  ABABCABACACBABC  hhkhkkkhhkkkhkk  15T  P3m1     (156)  15T  P3m1       (156)  \\
444  15  ABABCABACACBCAC  hhkhhkhkhhkhkkk  15T  P3m1     (156)  15T  P3m1       (156)  \\
445  15  ABABCABACBABCAC  hhkhhkkhkkkhkkk  15T  P3m1     (156)  15T  P3m1       (156)  \\
446  15  ABABCABACBACABC  hhkkkhkkkkhkkkk  15T  P3m1     (156)  15T  P3m1       (156)  \\
447  15  ABABCABACBACBAC  hhkhkkkkkkkhkkk  15T  P3m1     (156)  15T  P3m1       (156)  \\
448  15  ABABCABACBCABAC  hhkhkhkkhkkhkkk  15T  P3m1     (156)  15T  P3m1       (156)  \\
449  15  ABABCABACBCACBC  hhkkhkhkhkkhkkk  15T  P3m1     (156)  15T  P3m1       (156)  \\
450  15  ABABCABCABABCAC  hhkhhkkhhkkkkkk  15T  P3m1     (156)  15T  P3m1       (156)  \\
451  15  ABABCABCABACABC  hhkkkkhkhkkkkkk  15T  P3m1     (156)  15T  P3m1       (156)  \\
452  15  ABABCABCABACBAC  hhkhkkkkhkkkkkk  15T  P3m1     (156)  15T  P3m1       (156)  \\
453  15  ABABCABCABCABAC  hhkhkhkkkkkkkkk  15T  P3m1     (156)  15T  P3m1       (156)  \\
454  15  ABABCABCABCACBC  hhkkhkhkkkkkkkk  15T  P3m1     (156)  15T  P3m1       (156)  \\
455  15  ABABCABCABCBCAC  hhkhhkhhkkkkkkk  15T  P3m1     (156)  15T  P-3m1      (164)  \\
456  15  ABABCABCACABCBC  hhkkhhkkhhkkkkk  15T  P3m1     (156)  15T  P-3m1      (164)  \\
457  15  ABABCABCACBABAC  hhkhkhhkkhkkkkk  15T  P3m1     (156)  15T  P3m1       (156)  \\
458  15  ABABCABCACBACBC  hhkkhkkkkhkkkkk  15T  P3m1     (156)  15T  P3m1       (156)  \\
459  15  ABABCABCACBCABC  hhkkkkkhkhkkkkk  15T  P3m1     (156)  15T  P-3m1      (164)  \\
460  15  ABABCABCACBCBAC  hhkhkkhhkhkkkkk  15T  P3m1     (156)  15T  P3m1       (156)  \\
461  15  ABABCABCBABACBC  hhkhkkkkhhkkhkk  15T  P3m1     (156)  15T  P3m1       (156)  \\
462  15  ABABCABCBABCBAC  hhkhkkhkhkhkkkk  15T  P3m1     (156)  15T  P3m1       (156)  \\
463  15  ABABCABCBACABAC  hhkhkhkhkkhkkkk  15T  P3m1     (156)  15T  P3m1       (156)  \\
464  15  ABABCABCBACACBC  hhkhkkhhkkkkhkk  15T  P3m1     (156)  15T  P3m1       (156)  \\
465  15  ABABCABCBACBCAC  hhkhhkhkkkhkkkk  15T  P3m1     (156)  15T  P3m1       (156)  \\
466  15  ABABCABCBCABCAC  hhkhhkkkkhhkkkk  15T  P3m1     (156)  15T  P-3m1      (164)  \\
467  15  ABABCABCBCACBAC  hhkhkkkhkhhkkkk  15T  P3m1     (156)  15T  P-3m1      (164)  \\
468  15  ABABCACABABCBAC  hhkhhkhkkhkhhkk  15T  P3m1     (156)  15T  P3m1       (156)  \\
469  15  ABABCACABACABAC  hhkhkhkhkhkhhkk  15T  P3m1     (156)  15T  P-3m1      (164)  \\
470  15  ABABCACABACACBC  hhkhkhhkhkkhhkk  15T  P3m1     (156)  15T  P3m1       (156)  \\
471  15  ABABCACABACBABC  hhkhkkkhkkkhhkk  15T  P3m1     (156)  15T  P3m1       (156)  \\
472  15  ABABCACABCBACBC  hhkkhhkkhkkkhkk  15T  P3m1     (156)  15T  P-3m1      (164)  \\
473  15  ABABCACBABCABAC  hhkhkhkkkhkkhkk  15T  P3m1     (156)  15T  P3m1       (156)  \\
474  15  ABABCACBABCACBC  hhkkhkhkkhkkhkk  15T  P3m1     (156)  15T  P3m1       (156)  \\
475  15  ABABCACBABCBCAC  hhkhhkhhkhkkhkk  15T  P3m1     (156)  15T  P3m1       (156)  \\
476  15  ABABCACBACBACBC  hhkkhkkkkkkkhkk  15T  P3m1     (156)  15T  P-3m1      (164)  \\
477  15  ABABCACBCABCBAC  hhkhkkhkkkhkhkk  15T  P3m1     (156)  15T  P3m1       (156)  \\
478  15  ABABCACBCACABAC  hhkhkhkhhkhkhkk  15T  P3m1     (156)  15T  P3m1       (156)  \\
479  15  ABABCACBCACBCAC  hhkhhkhkhkhkhkk  15T  P3m1     (156)  15T  P3m1       (156)  \\
480  15  ABABCACBCBABCAC  hhkhhkkhkhhkhkk  15T  P3m1     (156)  15T  P-3m1      (164)  \\
481  15  ABABCBABCABCBAC  hhkhkhkkkkhkkhk  15T  P3m1     (156)  15T  P3m1       (156)  \\
482  15  ABABCBABCACBCAC  hhkhhkhkhkkhkhk  15T  P3m1     (156)  15T  P-3m1      (164)  \\
483  15  ABABCBABCBACBAC  hhkhkhkhkkkkkhk  15T  P3m1     (156)  15T  P3m1       (156)  \\
484  15  ABABCBACABACBAC  hhkhkkhkhkkkkhk  15T  P3m1     (156)  15T  P3m1       (156)  \\
485  15  ABABCBACBACBCAC  hhkhhkhkkkkkkhk  15T  P3m1     (156)  15T  P-3m1      (164)  \\
486  15  ABABCBACBCACBAC  hhkhkkkhkhkkkhk  15T  P3m1     (156)  15T  P-3m1      (164)  \\
487  15  ABACABACABACABC  hkhkhkhkhkhkkkk  15T  P3m1     (156)  15T  P-3m1      (164)  \\
488  15  ABACABACABCACBC  hkhkhkhkkhkhkkk  15T  P3m1     (156)  15T  P3m1       (156)  \\
489  15  ABACABACBABCABC  hkhkhkkkhkkkkkk  15T  P3m1     (156)  15T  P3m1       (156)  \\
490  15  ABACABCABACBABC  hkhkkkhkkkhkkkk  15T  P3m1     (156)  15T  P3m1       (156)  \\
491  15  ABACABCABCBACBC  hkhkkhkkkhkkkkk  15T  P3m1     (156)  15T  P3m1       (156)  \\
492  15  ABACABCBABCACBC  hkhkkhkhkkhkhkk  15R  R3m      (160)  15R  R-3m       (166)  \\
493  15  ABACABCBACBACBC  hkhkkhkkkkkkhkk  15T  P3m1     (156)  15T  P-3m1      (164)  \\
494  15  ABACBABCACBACBC  hkkhkkkhkkhkkkk  15T  P3m1     (156)  15T  P-3m1      (164)  \\
495  15  ABACBACBACBACBC  hkkhkkkkkkkkkkk  15T  P3m1     (156)  15T  P-3m1      (164)  \\
496  15  ABACBACBACBCABC  hkkkkkhkkkkkkkk  15T  P3m1     (156)  15T  P-3m1      (164)  \\
497  15  ABACBACBCABACBC  hkkhkkhkkhkkkkk  15T  P3m1     (156)  15T  P-3m1      (164)  \\
}
\onehalfspacing